\pdfoutput=1
\documentclass{aa}  
\usepackage{graphicx}
\usepackage{txfonts}
\usepackage{xcolor}
\usepackage{pdflscape}
\newcommand{\TotalNumberDetections}{135,118}  %
\newcommand{\TotalNumberSources}{129,192}     %
\newcommand{\TotalNumberVoges}{124,730}       %
\newcommand{\TotalNumberFscVoges}{105,924}
\newcommand{\TotalNumberBscVoges}{18,806}
\newcommand{\TotalNumberScreened}{5926}       %

\begin{document}

\title{Second ROSAT all-sky survey (2RXS) source catalogue
      \thanks{The catalogue is available at the CDS via anonymous ftp to
              cdsarc.u-strasbg.fr (130.79.128.5) or via
              http://cdsarc.u-strasbg.fr/viz-bin/qcat?J/A+A/???/A???
              and on the catalogue web site http://www.mpe.mpg.de/ROSAT/2RXS
              }
      }

  \author{Th. Boller, M.J. Freyberg, J. Tr\"umper, F. Haberl, W. Voges, K. Nandra}

  \titlerunning{The 2RXS source catalogue}
  \authorrunning{Boller et al.}

  \institute{Max-Planck-Institut f\"ur extraterrestrische Physik, 
             85748 Garching, Giessenbachstra{\ss}e, Germany \\
             \email{bol@mpe.mpg.de}
            }

  \date{Received January 12, 2015; revised November 20, 2015; accepted February 29, 2016}   %

\abstract
  {}
{We present the second ROSAT all-sky survey source catalogue, hereafter referred to 
as the 2RXS catalogue. This is the second publicly released ROSAT catalogue of point-like 
sources obtained from the ROSAT all-sky survey (RASS) observations performed with 
the position-sensitive proportional counter (PSPC) between 
June 1990 and August 1991, and is an extended and revised version 
of the bright and faint source catalogues. 
  }
{We used the latest version of the RASS processing to produce overlapping 
X-ray images of $6.4 \degr \times 6.4 \degr$ sky regions. 
To create a source catalogue, a likelihood-based detection algorithm was
applied  to these, which accounts for the variable point-spread function (PSF)
across the PSPC field of view. 
Improvements in the background determination compared to 1RXS were also implemented. 
X-ray control images showing the source and background extraction regions were  
generated, which were visually inspected. 
Simulations were performed to assess the spurious source content of the 2RXS catalogue. 
X-ray spectra and light curves were extracted for the 2RXS sources, with spectral and variability 
parameters derived from these products.  
}
{We obtained about 135,000 X-ray detections in the $0.1-2.4$ keV energy band down
to a likelihood threshold of 6.5, as adopted in the 1RXS faint source catalogue. 
Our simulations show that the expected spurious content of the catalogue is a 
strong function of detection likelihood, and the full catalogue is expected to 
contain about $30$~\% spurious detections. 
A more conservative likelihood threshold of %
9,
on the other hand, yields %
about 71,000
detections with a $5$~\% spurious fraction.
We 
recommend
thresholds appropriate to the scientific application. 
X-ray images and overlaid X-ray contour lines provide an additional user 
product to evaluate the detections visually, 
and we performed our own visual inspections to flag uncertain detections. 
Intra-day variability in the X-ray light curves was quantified based on the 
normalised excess variance and a maximum amplitude variability analysis. 
X-ray spectral fits were performed using three basic models, 
a power law, a thermal plasma emission model, and black-body emission. 
Thirty-two large extended regions with diffuse emission and embedded point  
sources were identified and excluded from the present analysis.   
  }
{The 2RXS catalogue provides the deepest and cleanest X-ray all-sky survey 
catalogue in advance of eROSITA.}

  \maketitle

\section{Introduction}\label{sec:Introduction}

The ROSAT all-sky survey (RASS) was the first to scan the whole sky with
a powerful imaging X-ray telescope operating in the $0.1 - 2.4$\,keV band 
\citep{Truemper1982}.
The Wolter type I mirror system \citep{Aschenbach1988} was exceptionally 
well suited for the sky survey operation because of the very low micro-roughness of the mirrors ($< 0.3$\,nm), which was 
responsible for the excellent contrast of the X-ray images. 
The focal plane detector used for the sky survey, the position-sensitive 
proportional counter (PSPC), 
had a five-sided anti-coincidence system that
reduced the particle background with an efficiency of $99.85\,\%$ 
\citep{Pfeffermann1986,Pfeffermann2003}. 
This efficient anti-coincidence veto design resulted in a low, non-X-ray (particle) background. 
Another reason for the exceptionally low particle background of 
ROSAT was the low Earth orbit with orbital height of $\sim 580$\,km and 
inclination $53^\circ$ (orbital period 96 minutes).

The ROSAT survey observations were performed in scanning mode, where the 
field of view (FOV) of the PSPC detector scanned a two-degrees-wide strip along 
a great circle over the ecliptic poles
within 96 minutes. With a shift of about one degree per day, an all-sky 
survey was completed within half a year. Because of periods of very high 
background or poor attitude values, some parts of the sky were missed, but 
re-observed during the February survey in 1991 
and the August survey in 1991.
Before the main survey between August 1990 and January 1991, the 
July mini-survey in 1990 was performed for testing 
and is part of the ROSAT all-sky survey.
Still some parts of the sky remained unobserved, which were later 
(February 1997) covered by pointing observations. 
The analysis of these will be reported in a separate paper \citep{Freyberg2016a}.

The RASS sensitivity \citep{Truemper1993} surpassed that of the Uhuru 
\citep{1978ApJS...38..357F} and 
HEAO-1 \citep{1984ApJS...56..507W} surveys by a factor >\,100 in the soft X-ray band.
The RASS bright source catalogue (RASS BSC), containing \TotalNumberBscVoges\ 
sources, was first published in electronic form \citep{Voges1996} and later in 
a printed version \citep{Voges1999}. 
This catalogue has served a very large scientific community 
working in different fields  - from solar system objects (Moon, comets, and 
planets) out to clusters of galaxies and quasars at large cosmological distances. 
The faint part of the ROSAT all-sky survey (RASS FSC) with \TotalNumberFscVoges\ sources 
down to a detection likelihood limit of 6.5
was published only in an electronic version 
\citep{Voges2000}. 
Both RASS BSC and FSC, which taken together constitute the ROSAT 1RXS catalogue, 
were based on the so-called RASS-2 processing \citep{Voges1999}. 
An updated version of the processing (RASS-3) was performed subsequently with
event files being made public, but without further documentation.

After the all-sky survey, ROSAT performed an extended program of pointed 
observations that covered a significant part of the sky ($\sim 18\,\%$)
with deeper observations.  
Several other satellite missions with imaging telescopes
have gathered data over large areas of the sky, producing 
large catalogues of X-ray point sources. 
Of particular note are the 
ROSAT PSPC pointed catalogue \citep{2RXP}, the 
XMM-Newton catalogue of pointed observations \citep{3XMM}, the 
XMM-Newton slew survey \citep{XMMSL1}, and the
deep Swift X-ray telescope point source 
catalogue \citep{2014ApJS..210....8E}.
\begin{figure*}
  \centering
  \includegraphics[angle=-90,width=180mm,clip=]{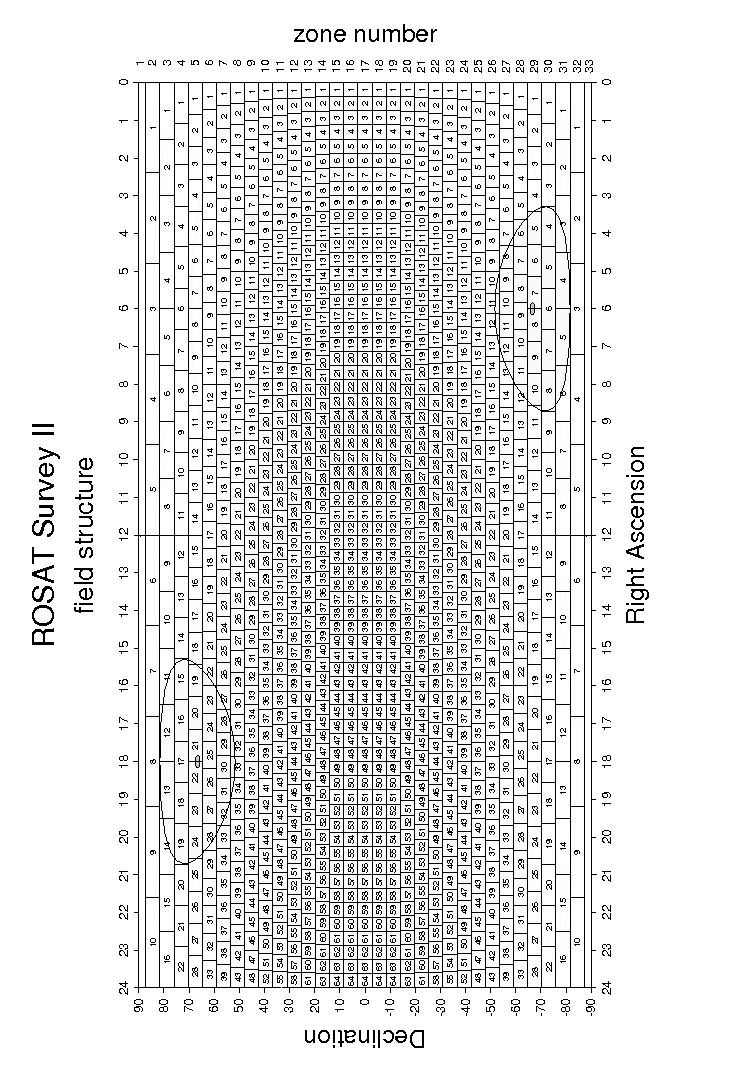}
  \caption{
                    Structure and numbering scheme of the sky
fields in the ROSAT all-sky survey
          in equatorial coordinates (courtesy K.\,Dennerl,
          available at the 2RXS web site.) 
          Areas around the North and South ecliptic poles with 1$\degr$ and $15\degr$ radius 
          are indicated at the upper left and lower right corner.
          }
   \label{Fig0_field_structure}
   \end{figure*}

The aim of this paper is to present a revised point source 
data base of the ROSAT all-sky survey (ROSAT 2RXS). 
The main points of improvement are as follows:
\begin{enumerate}
\item Use of an improved detection algorithm.
\item Reduction of systematic time delays between star tracker and photon arrival time.
\item A significantly improved reduction of spurious detections by a careful visual 
 screening of each catalogue entry and the exclusion of large, extended emission regions,
 in particular from the background-map 
creation process.
\item The provision of X-ray images of 1378 sky fields ($6.4\degr \times 6.4\degr$) 
 covering the whole sky.
\item The provision of local maps ($40\arcmin \times 40\arcmin$) for each detected X-ray source. 
\item The creation of source spectra and light curves and deduction of characteristic parameters.
\item The creation of new photon event tables through correcting astrometric errors that are
present in the publicly available event files (originating from the RASS-3 processing).
\item The delivery of a documented and reproducible point source catalogue.
\item Performing extensive simulations to estimate the amount of spurious detections 
in the 2RXS catalogue as a function of the detection likelihood and other source parameters.
\end{enumerate}

The total number of 
entries 
listed in the 2RXS catalogue is \TotalNumberDetections. 
Of these, \TotalNumberScreened\  have been flagged as uncertain detections 
(cf.\ Sect.\ref{sec:ScreeningIndividual}).
We have also provided the results of cross correlations of the catalogue 
with major source catalogues from X-rays and other wavelength bands 
(see Sect.~\ref{sec:CrossMatches}).

\begin{figure*}[thp]
  \centering
  \includegraphics[clip=]{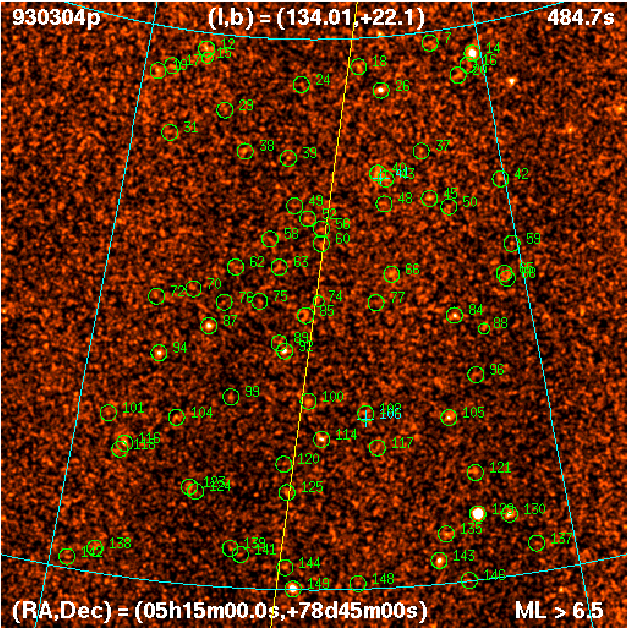}
  \caption{
Example for a source count image of sky field 930304 in the way
it is used for source detection.
For the colour representation in this and all following images the ESO/MIDAS colour table 
{\tt heat} is used
with linear intensity scaling.
The cyan solid lines confine the $6.4\degr \times 6.4\degr$ field and 
the scan direction is marked with the yellow line
through the centre of the field. 
2RXS sources are indicated with green circles. 
Detections that were manually flagged as being uncertain (see Sect.~\ref{sec:Screening})
are marked with cyan crosses. 
Coordinates refer to the image centre, 
the exposure time is the median over the whole image.
          }
   \label{Fig_sky_field}
   \end{figure*}

\section{2RXS - data analysis}\label{sec:2RXS}

In the following we describe how the 2RXS catalogue was created 
(Sect.~\ref{sec:SourceDetectionGeneral} and Appendix A) and how 
new data products were produced (Sect.~\ref{sec:DataProducts}). 
\subsection{Detection algorithm}\label{sec:SourceDetectionGeneral}

We used the data from the third processing of the ROSAT All-Sky Survey 
(RASS-3) as input, which is publicly available from the MPE ROSAT archival 
FTP site\footnote{\tt ftp://ftp.mpe.mpg.de/rosat/archive/900000/}:
sequences {\tt 93nnmmp},
with {\tt nn} ranging from {\tt 01} to {\tt 33}, from the equatorial North Pole
in 
$5.625\degr$
steps to the equatorial South Pole, that is,\ {\tt 17} containing the 
equatorial plane), and {\tt mm} from 1 to a maximum of 64 (depending on Declination), 
dividing the equatorial rings along Right Ascension, and prefix {\tt 93} being the 
year of the RASS-2 processing.\footnote{This scheme has been kept also for the next RASS-3 processing.}
The field structure and numbering scheme are illustrated in Fig.~\ref{Fig0_field_structure}.
Data are thus organised into 1378 overlapping equatorial sky regions of size
$6.4\degr \times 6.4\degr$ , with event lists originating from the RASS-3 processing.

The RASS-3 processing differs in two points from the RASS-2 processing, which was used to create the 
1RXS catalogue. 
Firstly, the time resolution for the interpolation of the attitude solution 
(which is given with 1\,s accuracy) was refined from 1/64 s to 1/32000 s.
A time delay exists between attitude time and photon time, which is given with a
precision of 10 ms. Therefore, a systematic uncertainty of $\pm5$\,ms translates into
a systematic positional error 
(still)
in the RASS-3 processing of at least $\pm 1\arcsec$
along scan direction.
Secondly, in the RASS-3 processing the required number of identified guide stars 
in the star tracker field  
to accept photons was relaxed with respect to the RASS-2 processing.
This resulted in a larger sky coverage in the RASS-3 processing.

The several phases of the ROSAT all-sky survey (and its completion)
are described in a separate paper \citep{Freyberg2016a}.
Here we only used data that had been taken in scanning mode during four phases:
the mini survey,  the main survey, the February survey 91, 
and the August survey 91 \citep[see Table~2 of][]{Voges1999}.

Periods where the PSPC FOV passed over the Moon 
\citep{Freyberg1994}
were almost completely removed during RASS-3 processing (remaining periods were
excluded in this paper). 
The photons were rejected from the event lists but kept in separate files. 
As for 1RXS, the source detection scheme was optimised to a 
survey scanning mode with the PSPC
and was performed using a three-step approach.
The first two steps are based on a sliding cell method, estimating the 
background locally, and then using background from a model fit. 
Finally, the resulting source candidate lists are used as input to a 
maximum likelihood algorithm (for more details see Appendix A). 
Similar techniques are used for the XMM-Newton catalogues 
\citep{Watson2009} %
and the 1SXPS Swift-XRT catalogue \citep{2014ApJS..210....8E}. %
However, for the 
ROSAT survey, likelihood values are determined on an event basis, while 
the XMM-Newton and Swift methods use binned images. 
In the ROSAT method the appropriate point-spread function (PSF) 
and vignetting corrections are
assigned to each event: 
in scanning mode photons from an individual source are detected at
very different places in the detector in contrast to a pointed observation,
while in the latter (image approach) only an average PSF can be used.
In contrast to CCD-type observations, the PSPC does not suffer from so-called out-of-time events such as 
hot pixels, bad columns, or read-out streaks.
Additionally, the highly effective anti-coincidence procedure resulted in a very low
particle background.
Moreover, the scanning mode smears out remaining telescope artefacts
such as stray-light that is due to single reflected X-rays in a Wolter-I telescope.
Therefore, we continued with using the ROSAT source 
detection software, but included several improvements.

Primary source parameters \citep[see][]{Boese2004}
resulting from the source detection procedure 
are for instance the detection likelihood, exposure time, 
source counts, count rate, and source extent (with corresponding uncertainties).
Derived source parameters include hardness ratios, spectral fit, and 
variability parameters (with corresponding uncertainties).
Cross-correlations with selected catalogues are described in Sect.~\ref{sec:CrossMatches}.

The main improvement in the new detection procedure was to use nine smaller
overlapping images ($2.27\degr \times 2.27\degr$) 
within the $6.4\degr \times 6.4\degr$ sky field
instead of only one large image, as used in the second 
processing of the ROSAT all-sky survey data (see Fig.\ref{plot_2rxs_subimages}).
As shown by \citet{Freyberg1994}, variations in the background 
and the exposure create intensity structures in the count intensities with
typical sizes of about two degrees perpendicular to the scan paths. 
These variations can be seen in the $6.4\degr \times 6.4\degr$ images used for
source detection in the 1RXS processing. 
The background maps produced from spline 
fits often did not follow these variations, leading to over- or underestimated
background in different areas. 
To overcome this problem, we divided the images 
into nine (three by three) sub-images. With this 
smaller sky-field map structure we can better account for the count 
intensity variations and can better constrain the background for sources. 
This results in a more precise determination of the detection likelihood 
values, especially for faint 2RXS sources. 
In the RASS FSC catalogue about 12,000 detections have artificially high
likelihood values that are due to an underestimation of the background resulting from a 
background map that is not adequately adapted to the count intensity variations 
present in the ROSAT all-sky survey. 
In the 2RXS processing these are either absent from the seed source lists 
or assigned likelihoods below the acceptance threshold
and therefore absent from the 2RXS catalogue. 
This is described in detail in Sect.~\ref{sec:detection}.

\subsection{Data Products}\label{sec:DataProducts}

In addition to the source properties obtained by the detection algorithm,
higher-level data products were created to allow a more detailed scientific
exploration of the 2RXS catalogue.
We wrote shell scripts, which call the necessary MIDAS/EXSAS programs in 
order to produce X-ray images, light curves, and spectra. 
The data products were produced for all detections above a likelihood of 6.5 
identified in the 1378 individual fields covering the ROSAT all-sky survey.
The screening procedures applied to all 2RXS sources and larger complex regions 
are described in Sect. \ref{sec:Screening}.

\subsubsection{X-ray images of the 1378 sky fields}\label{sec:FieldImages}

We have produced X-ray images in the 0.1-2.4 keV band for each individual 
sky field with a size of $6.4\degr \times 6.4\degr$. 
Figure \ref{Fig_sky_field} shows the source count image for field 930304. 
2RXS sources are marked with green circles, and sources that were manually flagged (see Sect.~\ref{sec:Screening})
are shown as cyan crosses. 
The yellow line indicates the scan path (ecliptic great circle) 
across the centre of the field. 
The variations in the count intensities are caused by variations in the 
background intensity and in the exposure time  (\cite{Freyberg1994}). 
The 1378 sky fields with the 2RXS sources overlaid,  as well as all the other
data products listed below, can be accessed at the 
catalogue web site \footnote{http://www.mpe.mpg.de/ROSAT/2RXS}.

\subsubsection{X-ray source images}\label{sec:Images}

\begin{figure*}

\centerline{%
\includegraphics[angle=0.0,width=0.415\textwidth,clip=true]{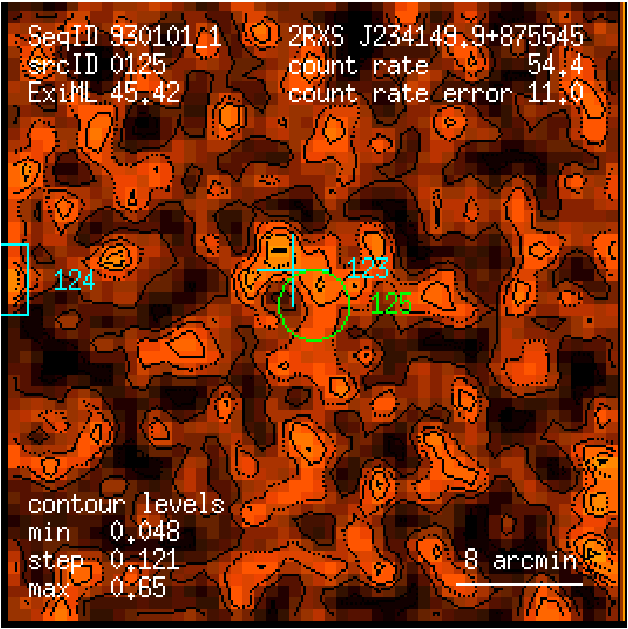}
\includegraphics[angle=0.0,width=0.415\textwidth,clip=true]{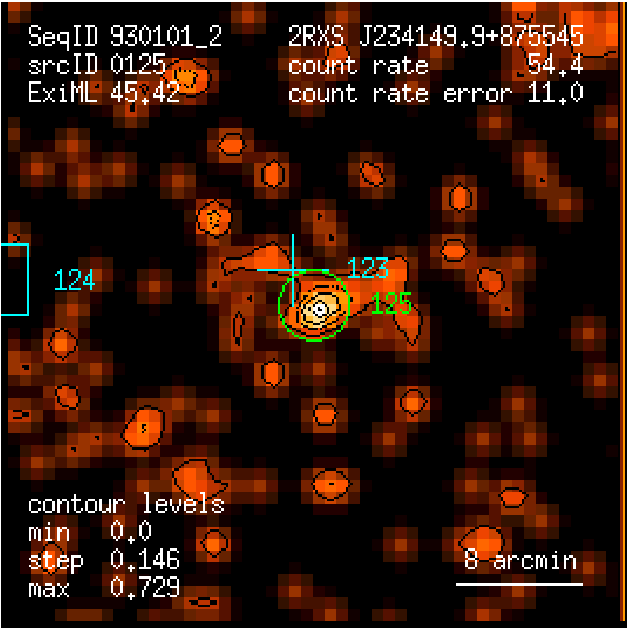}
}%
\centerline{%
\includegraphics[angle=0.0,width=0.415\textwidth,clip=true]{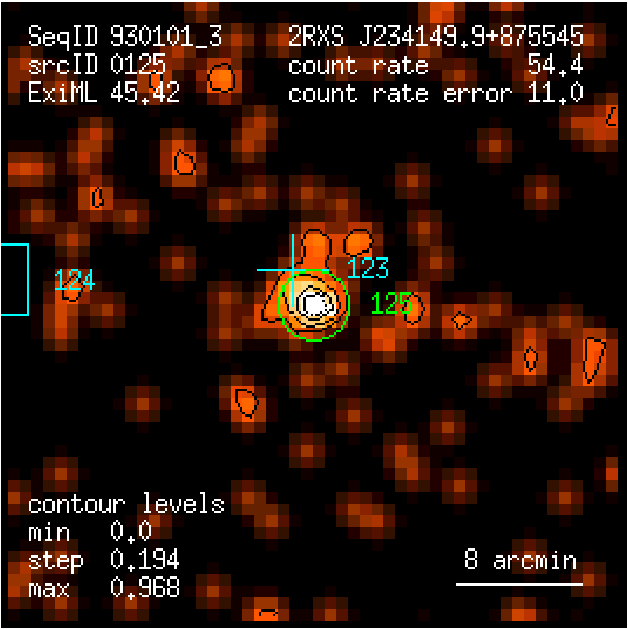}
\includegraphics[angle=0.0,width=0.415\textwidth,clip=true]{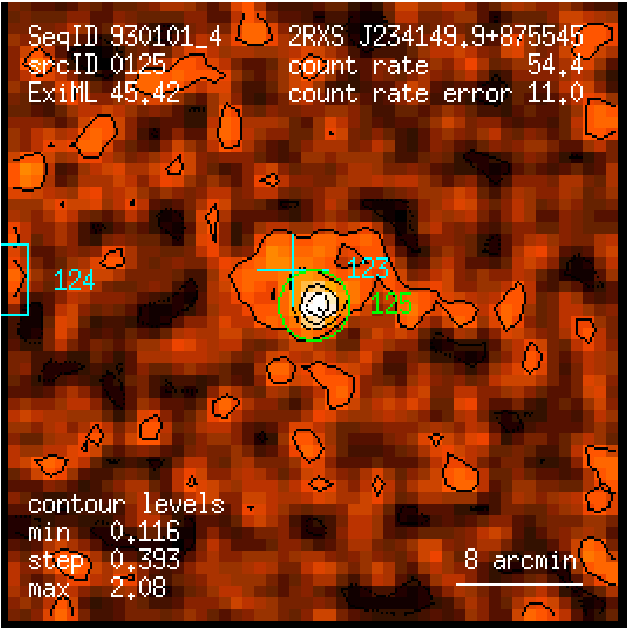}
}%
\centerline{%
\includegraphics[angle=0.0,width=0.415\textwidth,clip=true]{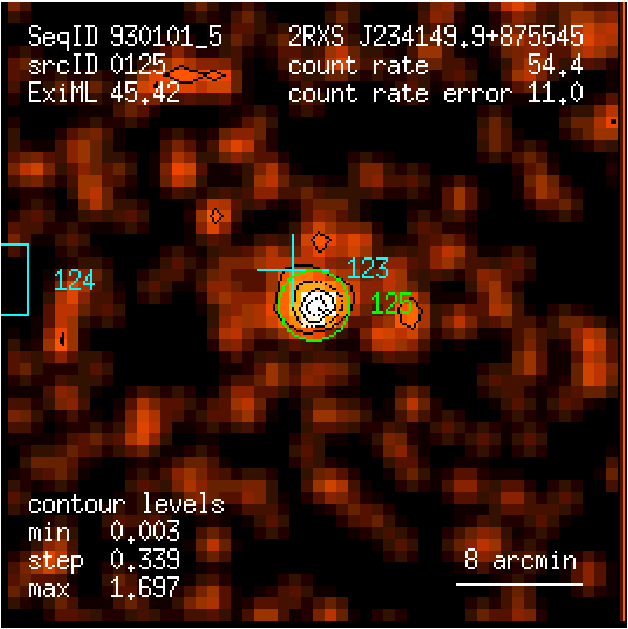}
\includegraphics[angle=0.0,width=0.415\textwidth,clip=true]{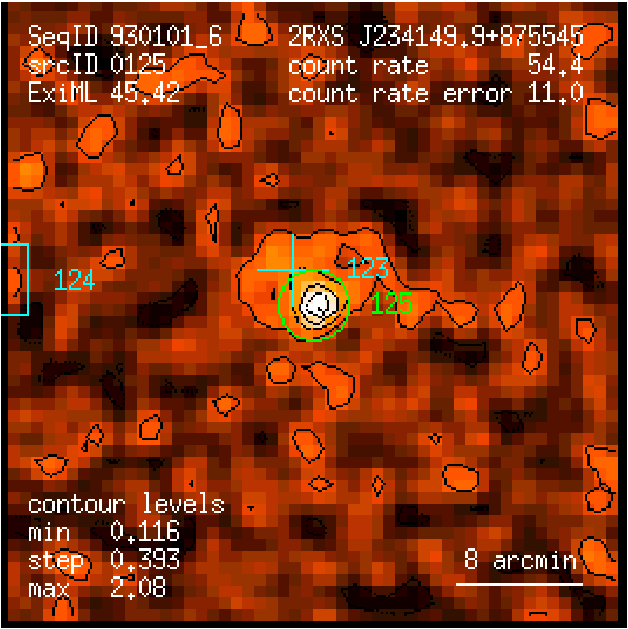}
}
  \caption{
  Example of X-ray count rate images for our six energy bands with overlaid X-ray
  intensity contours (black lines),
  bands 1 to 6 are defined as PSPC channels $11-41$, $52-90$, $91-201$, $11-235$, 
  $52-201$, $11-201$, respectively (with one channel $\sim 10$\,eV).
Each
image is centred on source 125 for which the 2RXS IAU name is provided. 
The broad-band source count rate 
(PSPC channels $11-235$)
with error as determined from the detection 
algorithm is in units of $\rm counts\ ks^{-1}$. 
Contour levels 
are in units of $\rm counts\ ks^{-1}\ arcmin^{-2}$.
At the top of the image the field number (SeqID) and the 
source identification within the field (srcID) are given.
The green circles gives the catalogue entries, and the cyan cross marks a source
that was visually flagged in the screening process.
          }
   \label{Fig1_example_image_6bands}
   \end{figure*}

In addition to the 1378 sky field images, we produced zoomed X-ray images
with a size of $40\arcmin \times 40\arcmin$ centred on each of the point sources
for each of the six energy bands we used in our analysis (see the Appendix for details).
Five equally spaced contour lines were determined linearly between 
the minimum and maximum photon surface density and overlaid on the X-ray images. 
Figure~\ref{Fig1_example_image_6bands} gives an example for the source 125 in field 930101. 
The 2RXS sources are marked with green circles,
and cyan crosses  denote detections that were visually screened and flagged
(cf. Fig~\ref{Fig1_example_image_6bands}).

\subsubsection{X-ray light curves}\label{sec:lightcurves}

To create light curves for ROSAT observations in scanning mode, we 
developed a set of scripts 
that take  all necessary functionality into account
and appropriately handle the detector and instrument behaviour, for example,
the removal of source scans when the detector was switched off, 
counting photons when the source is completely in the FOV.

Background extraction regions were taken from two circles along the scan
direction with radii of 5\arcmin, and separated by a distance
of $+30\arcmin$ and $-30\arcmin$ from the source position.
This results in a time offset between the source and background events of eight seconds. 
On this relatively short time scale, variations in the background are expected to be small,
and this approach minimises the effect of these variations. 
Incorrect background subtraction can occur because 
previous or following scans contribute, separated by the scan period of $5760$\,s.
When background regions overlap with other sources, we corrected for the 
fraction of overlap, making sure to use source-free background regions.

\begin{figure}
  \centering
    \includegraphics[angle=-90,width=90mm,clip=]{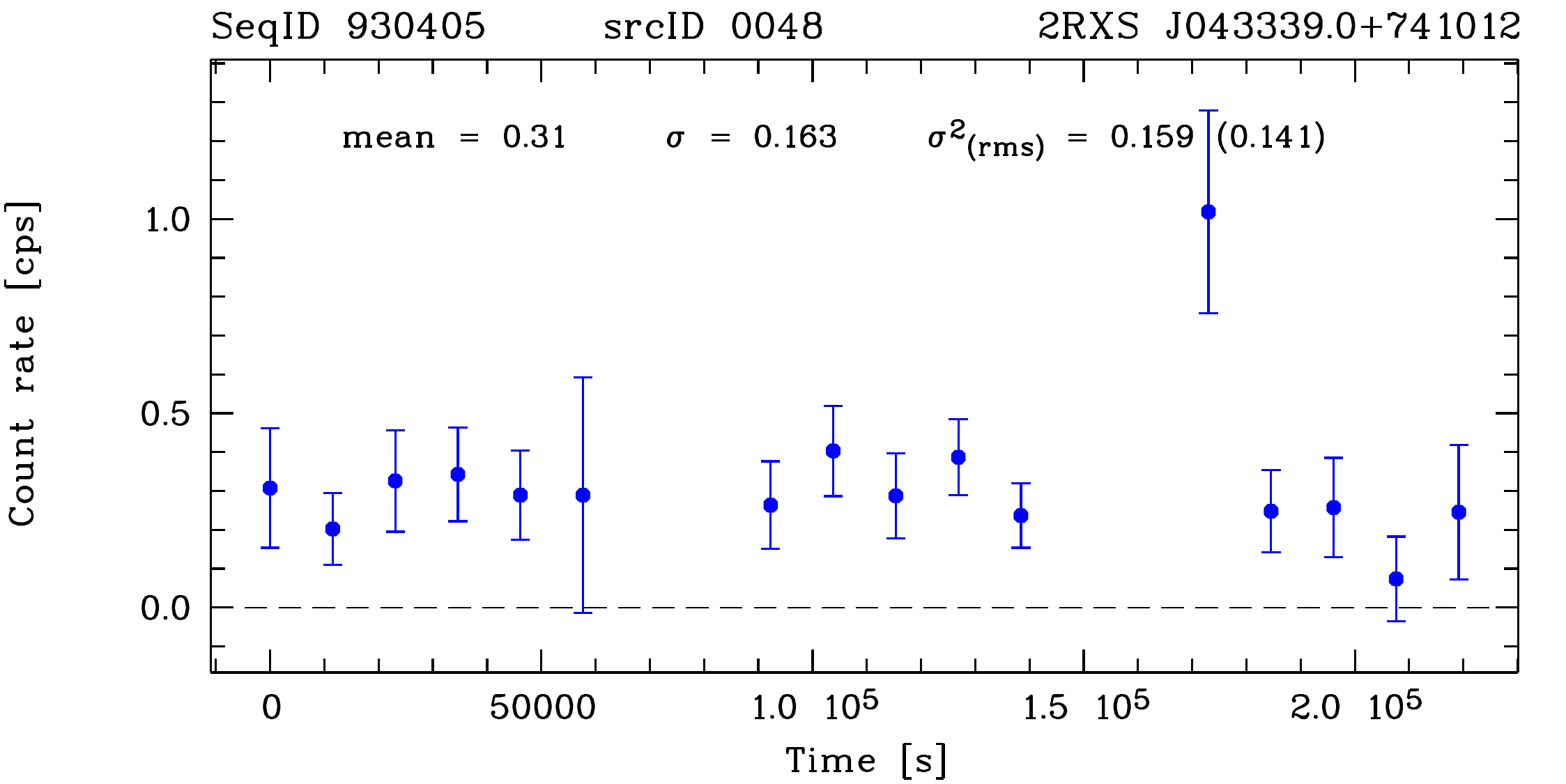}
  \caption{
X-ray light curve of 2RXS\,J043339.0$+$741012. 
The mean count rate, the corresponding standard deviation 
(both in counts s$^{-1}$), and the normalised excess variance 
with 1\,$\sigma$ uncertainty are given. 
           }
   \label{Fig2_example_lightcurve}
   \end{figure}

\begin{figure}
  \centering
  \includegraphics[width=90mm,clip=]{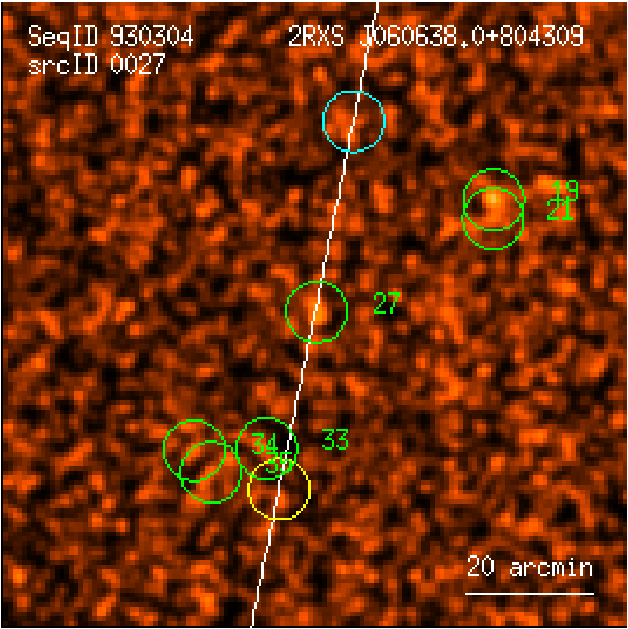}
  \caption{%
Example of a control image demonstrating how the background region is selected.
The background regions 
for source 27 are shown in cyan (at the top) and yellow circles (at the bottom). 
Since the yellow circle is contaminated
with another source (33), the background in the cyan circle 
is used in the background subtraction.
}
             \label{Fig_example_controlimage}%
   \end{figure}

In some of the ROSAT observations (0.9 per cent of the catalogue entries) individual bins have very low exposure values, below 6 seconds. 
Such data points are associated with an extreme large error and the calculation of the standard deviations
leads to improper values. We have therefore flagged such light curves. 

We finally selected the background from the region 
that is 
spatially
less contaminated by other sources, 
for an example see Fig.~\ref{Fig_example_controlimage}.

In the control images 
the selected background regions are marked with circles in cyan.
The source and background light curves were created with a time binning of 
11520 seconds (corresponding to two orbits). 
The light curves are provided as plots  for each source
showing the background-subtracted count rates versus time. 
We also produced a graph with the number of source and background counts, respectively. 

From the light curves we extracted basic parameters, such as the mean count rate, 
standard deviation, and the minimum and maximum count rate, together with their corresponding errors.
To characterise variability, we computed the excess variance with its uncertainty 
and the maximum amplitude variability as described in Sect.~\ref{sec:variab}. 
In Sect.~\ref{sec:Timing} we discuss the  general timing properties of the 2RXS objects. 
Figure~\ref{Fig2_example_lightcurve} shows an example light curve for the source with 
number 48 in the field 930405.

\subsubsection{X-ray spectra}\label{sec:SpectralFits}

We extracted spectra using the same source and background regions as 
for the light curves. The spectral analysis with three standard models
is described in Sect.~\ref{sec:SpectralProperties}.
In Fig.~\ref{Fig_powl_unique} an example for a source spectrum with the
three model fits is presented.

\section{Screening of the second RASS catalogue}\label{sec:Screening}

\subsection{Screening for large extended regions}\label{sec:ScreeningMasks}

First we inspected all 1378 ROSAT sky fields for large extended regions 
with diffuse emission and embedded point sources. 
In Appendix~\ref{sec:AppExt} we list the sky 
fields in Table~\ref{tab:AppExttab},  where we have identified these regions. 
We give the equatorial coordinates of the source we attribute to large extended 
regions, and we supply a source identification whenever 
possible. 
Sources within the masked regions were excluded from the 
present
2RXS catalogue. 
The analysis of point-like X-ray sources within these masked regions will be 
performed in a subsequent paper \citep{Freyberg2016b} 
as further 2RXS sources.

\subsection{Visual screening of the 2RXS sources}\label{sec:ScreeningIndividual}

\begin{figure}
  \centering
  \includegraphics[width=90mm,clip=]{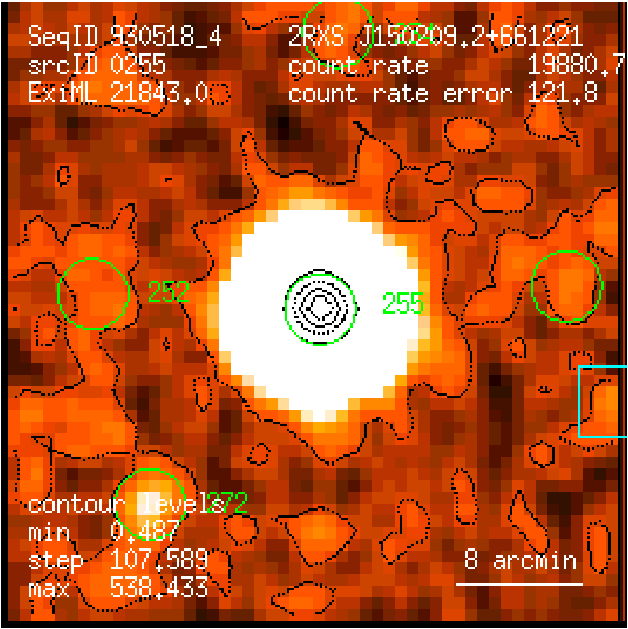}
\includegraphics[width=90mm,clip=]{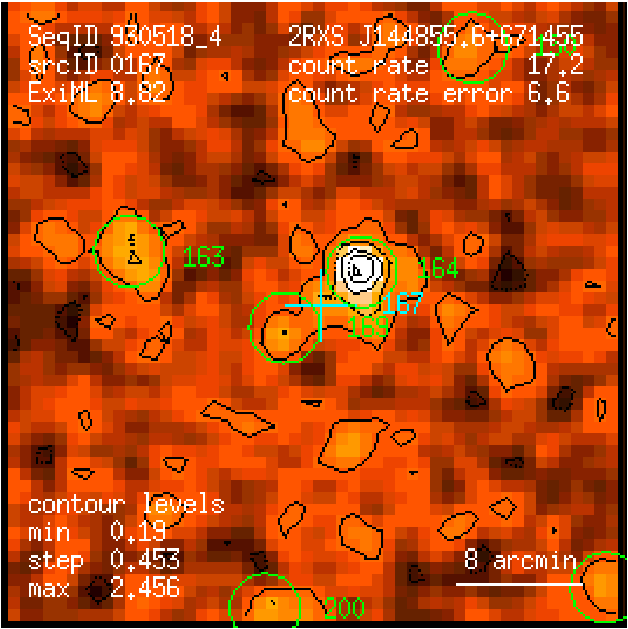}
      \caption{Top: Example of a source that is considered as a correct detection, i.e.\ 
  the source flag is set to zero. The broad-band image is shown.
  The object is identified as white dwarf WD 1501+663. 
  Bottom: Example of an uncertain detection (number 167), located to the lower left of the
          bright central source (number 164). It is flagged and marked in 
  the image with a cyan cross. Again the broad-band image is shown.
              }
             \label{Fig_correct_flagged_detections}
   \end{figure}

The number of detections in the 2RXS catalogue with a detection likelihood
greater than or equal to 6.5 is \TotalNumberDetections. 
We visually inspected all the 2RXS sources to confirm their existence 
and  to identify false detections, particularly in the vicinity of bright 
sources (see, e.g., detection 167 in lower panel of Fig.\ref{Fig_correct_flagged_detections}).
A simple graphical user interface (GUI) based on a MIDAS/EXSAS script was used 
to run through all sources.
The script creates an image with the source to be validated located at the centre.  
The image size is $1\degr$ by $1\degr$ , which allows a zoomed view 
to the central source and the neighbouring objects. 
The image shows the X-ray photon surface density with overlaid source detections. 
By comparing the photon density distribution with the source positions derived 
from the 2RXS detection algorithm, 
correct and uncertain detections can be identified. 
A correct detection is defined by having the source 
position on top of the maximum value of the photon density distribution. 

Using the GUI, we set flags for questionable detections, 
whereas correct source identifications were marked with the flag {\tt SFLAG=0}. 
The top panel of Fig.~\ref{Fig_correct_flagged_detections} shows an example
of a secure detection.
For highly uncertain detections we attributed the flag {\tt SFLAG=1}.
The lower panel of  Fig.~\ref{Fig_correct_flagged_detections} 
shows an example of such a case.  
We flagged \TotalNumberScreened\ 2RXS detections in
total.

\subsection{Screening statistics}\label{sec:ScreeningStatistics}

\begin{figure}
  \centering
  \includegraphics[angle=-90,width=104mm,clip=]{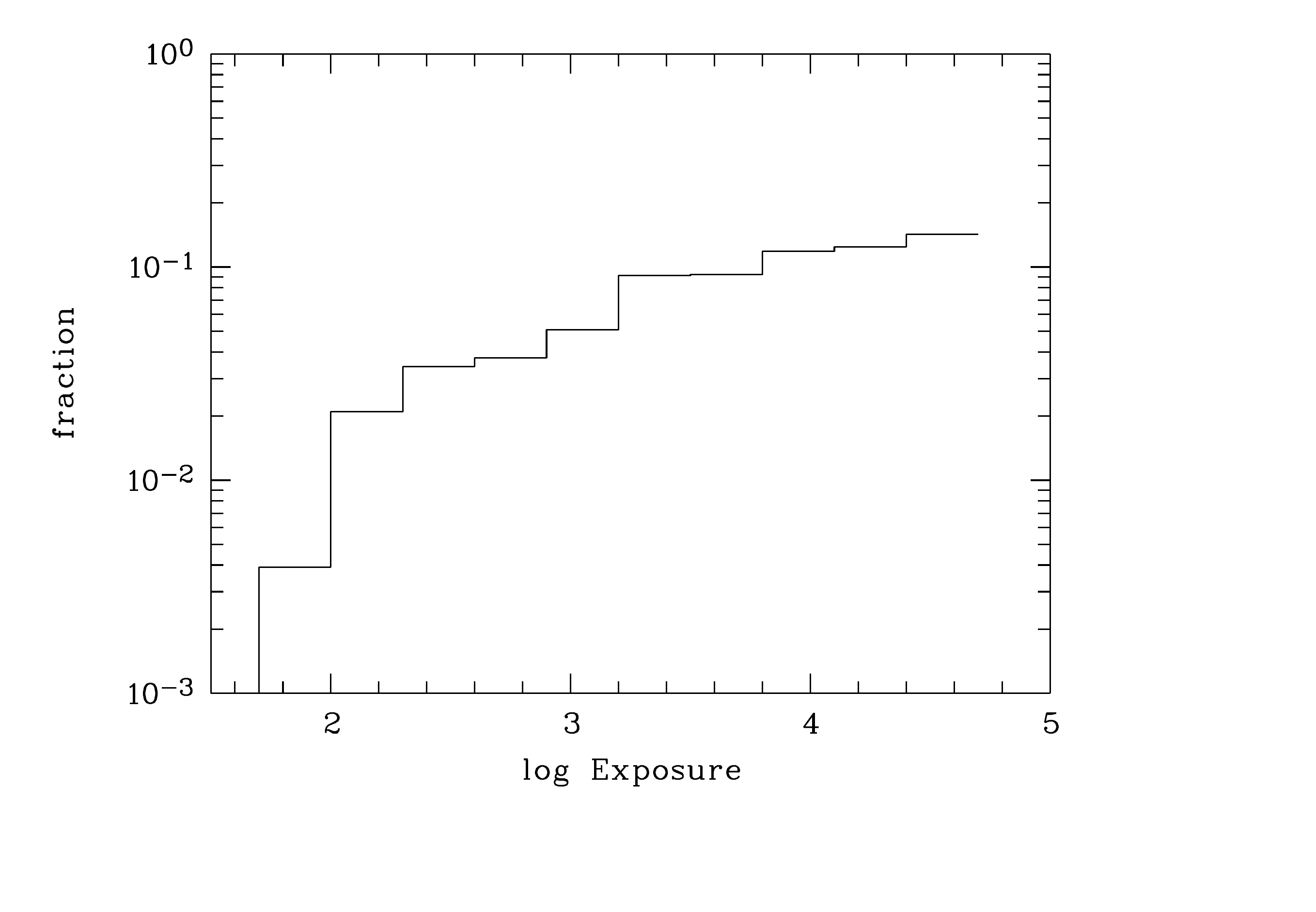}
  \caption{Relative distribution of the \TotalNumberScreened\ flagged detections as a function of the exposure time.}
             \label{screening_statistics}
   \end{figure}

In Fig.~\ref{screening_statistics} we show the distribution of flagged
detections normalised to the total number as a function of the exposure time.

With respect to the exposure time distribution, the number of flagged detections
increases 
with exposure time from 1 to 10 per cent above 100 seconds of exposure time.
The reason for this is the very high exposure around the 
North and South Ecliptic Poles. 
This leads to an increasing number of uncertain detections when approaching
the confusion limit, as the detection software is not optimised for very high 
source densities. 
The number of flagged detections with exposure times greater than 10000 
seconds is 54.

\section{2RXS - 1RXS comparison}\label{sec:2RXS_1RXS}

The 1RXS catalogue consists of the bright source catalogue (BSC), containing 
\TotalNumberBscVoges\ sources \citep{Voges1999}, and the faint source catalogue
(FSC), containing
\TotalNumberFscVoges\ sources \citep{Voges2000}. These are \TotalNumberVoges\ sources in total. 
The number of sources in the 2RXS catalogue without any manual screening flag set is
\TotalNumberSources.

\subsection{Detection statistics and detection validation}\label{sec:detection}

\begin{figure}[htp]
  \centering
  \includegraphics[angle=-90,width=105mm,clip=]{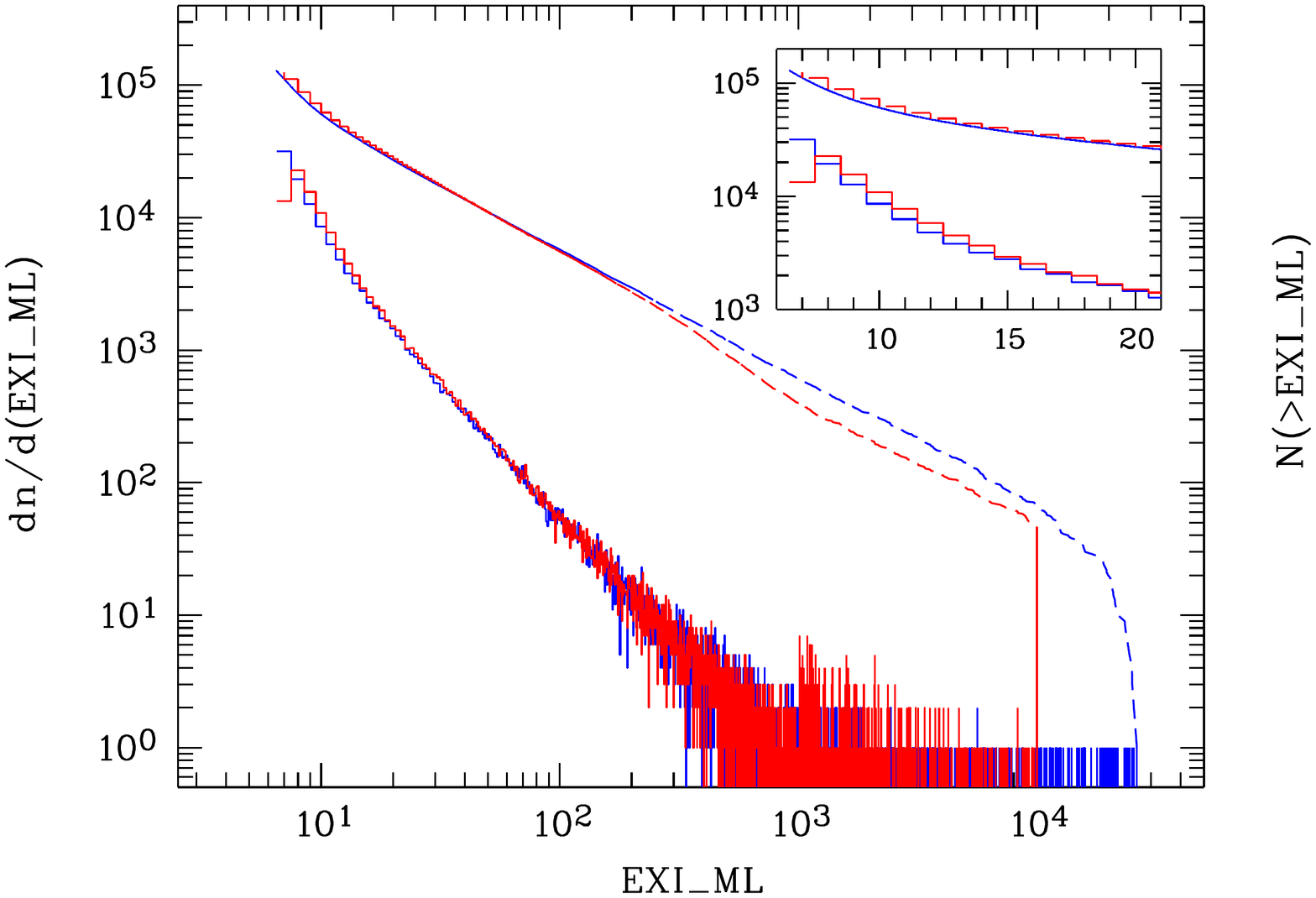}
  \includegraphics[angle=-90,width=105mm,clip=]{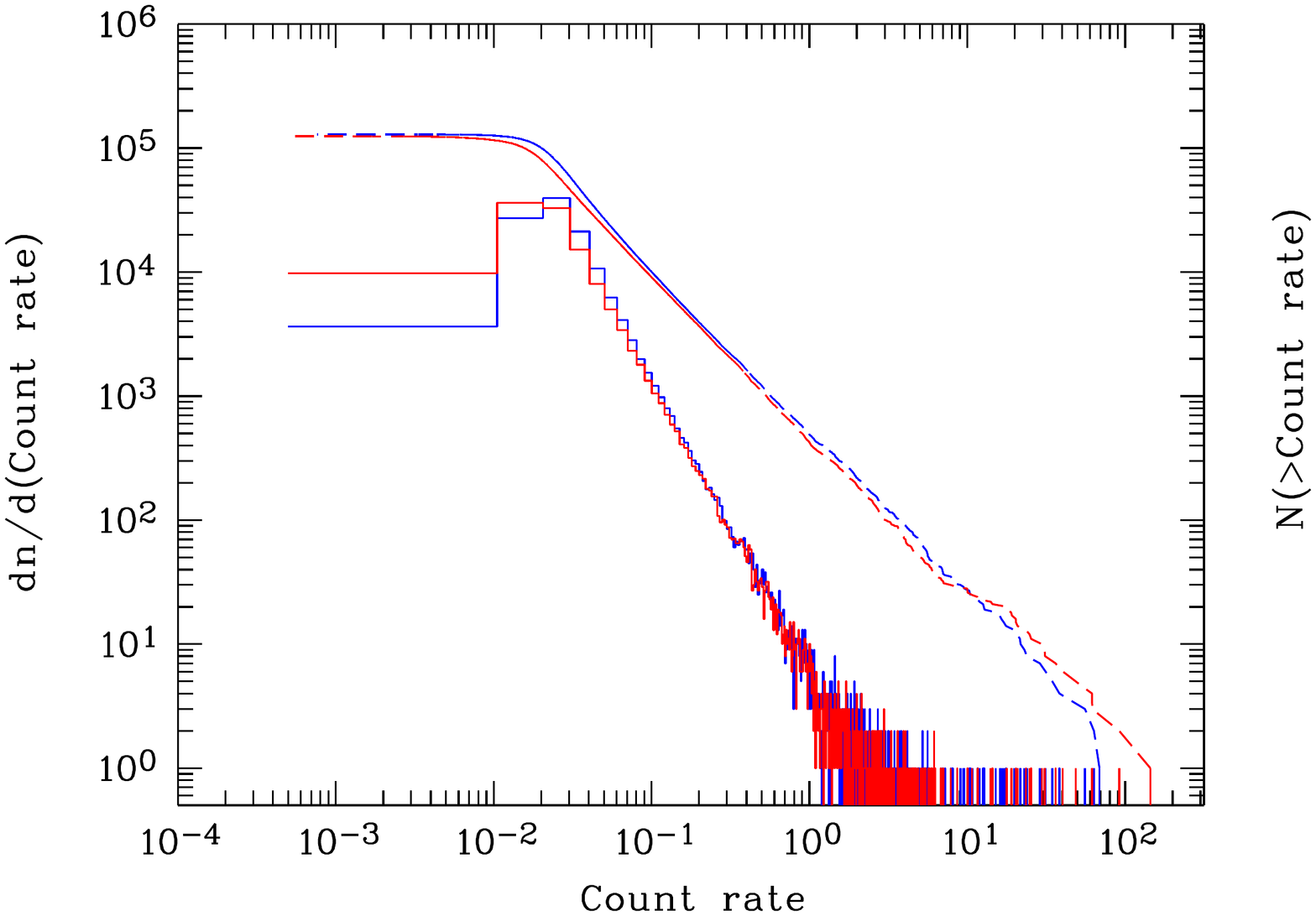}
  \caption{
           Upper panel: Distribution of the detection likelihood values for the 2RXS catalogue 
           (blue) and the 1RXS catalogue (red).
           The inset shows a zoom to the low-likelihood regime.
           Lower panel: Distribution of count rates for the 2RXS and 1RXS catalogue.
           Solid and dashed lines denote differential and integral distributions, respectively.
           The scale for left and right y-axes is identical for all figures with differential 
           and integral distributions.
          }
   \label{Fig_EXIML}
   \end{figure}

In Fig.~\ref{Fig_EXIML} (upper panel) we show the differential and cumulative 
distributions for the 
existence likelihood for the 2RXS (in blue) and the 1RXS (red) catalogue entries.  
In the 2RXS catalogue the lower limit for the existence likelihood is 6.5, 
which is equivalent to the rounded (nearest integer) value of 7 in the 1RXS catalogue.
We note that in the 1RXS catalogue an additional constraint of at least six counts is
present, and therefore especially the lowest likelihood bin is affected.
From the differential distribution, it is evident that there are more 1RXS
detections at the low likelihood end (see inset).
Two aspects might account for this: 
it could be produced by a higher fraction of spurious detections in 1RXS, 
but also by differing detection
likelihood values (cf.\ Fig.\ref{plot_eximl_1rxs_2rxs} in Sect.\ref{sec:eximl_redistribution}).
The second panel of Fig.~\ref{Fig_EXIML} gives the distributions in count 
rates.
In the range between 0.1 and 1 $\rm counts\ s^{-1}$ , the differential
and integral distributions can be fit with a slope of -2.36 and -1.33, 
respectively. Both slopes are consistent within the errors ($\pm 0.04$).
The slopes agree with the log N - log S distribution of AGN \citep{Hasinger2001}. 
However, in the BSC catalogue, the fraction of stars and extragalactic sources
are similar; this may reflect what we see from the 2RXS as a whole. 

\begin{figure}[htp]
  \centering
  \includegraphics[angle=0000,width=90mm,clip=]{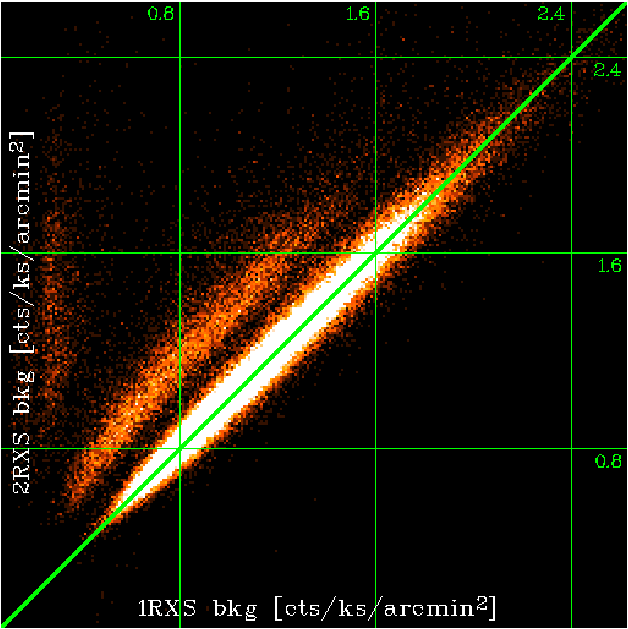}
  \caption{Intensity image of the 2RXS background values on the y-axis versus the 1RXS background values 
           on the x-axis. While most of the objects are located close to the one-to-one line, two other regions
           are identified (one parallel to the one-to-one line, and a third vertical distribution where most of the
           objects are located in the ecliptic poles). 
          }
   \label{sky_fields_green_yellow}
   \end{figure}

To further investigate the differences in the 1RXS and 2RXS detection
algorithms, we have compared the background counts for 1RXS and 2RXS 
sources (Fig.~\ref{sky_fields_green_yellow}). 
While for a number of sources the background values from 
the two catalogues are consistent and follow the one-to-one relation, there 
are sources with significantly higher background values in the 2RXS catalogue.
The sources with the largest background offset are located around the poles, 
where extreme intensity peaks exist, which in turn leads to a large underestimation of 
the background in 1RXS. 
The objects on the one-to-one line and the objects located on the middle strip 
seen in Fig.~\ref{sky_fields_green_yellow} are randomly 
distributed over the entire sky, with the exception of the polar regions. 
Underestimating the background in high-intensity areas leads to an overestimation
of the detection likelihood in the 1RXS catalogue. 
Therefore, the probability that the source is spurious becomes significantly 
higher for detections with low likelihood values in the 1RXS catalogue. 
The total number of 2RXS sources that have 1RXS counterparts and
defined background ratios is 
89,648. 
The number of sources located at the one-to-one line is 80,753.
The object numbers on the strip parallel to the one-to-one line is 
7,614
(background ratios between greater than 1.2 and lower than or equal to 3.0).
The number of objects in the ecliptic pole regions (vertical distribution with
respect to the 1RXS background) is 1,281 (background ratios greater than three).

\begin{figure}[htp]
  \centering
  \includegraphics[width=90mm,clip=]{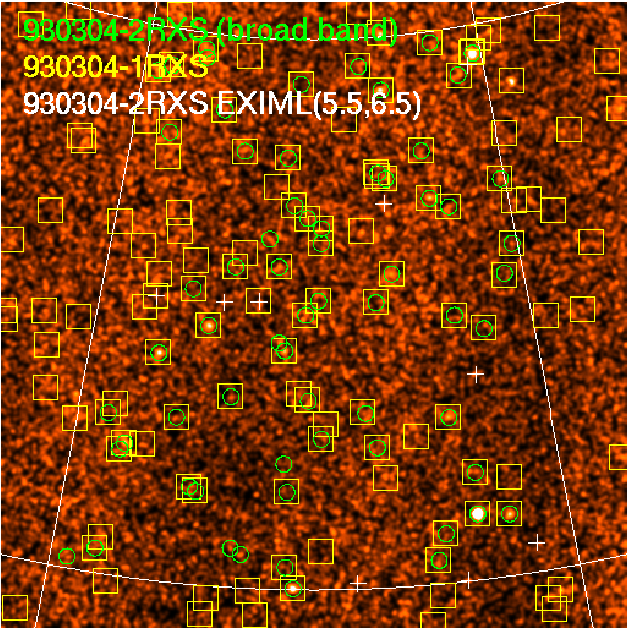}
  \caption{Illustrative comparison of 1RXS (yellow boxes) and 2RXS detections (green circles)
           for the field 930304. 
           The image shows the intensity distribution in counts. 
           Perpendicular to the scan directions the count rate shows intensity variations. 
           At the intensity peaks many 1RXS detections do not have an 2RXS counterpart. 
           As the detection algorithm for the 1RXS sources is run on a much larger
           field size (6.4\degr\ by\ 6.4\degr),  the background at the intensity peaks is
           often underestimated and the detection likelihood of the sources is overestimated.
           For the 2RXS sources the detection is run on nine subfields. With this the 
           detection algorithm can much better follow the count rate variations and the 
           background is determined more precisely. At the intensity peaks the background 
           is higher for 2RXS sources and fewer
           2RXS sources are detected than at 1RXS. With the white crosses
           we show 2RXS detections down to likelihood values of 5.5. Some of the 
           1RXS detections do have an 2RXS counterpart when the likelihood is decreased to 5.5. 
           }
   \label{sky_fields_green_yellow_ima}
   \end{figure}

We show the 1RXS and the 2RXS detections as an example for the different 
detection algorithms applied for the 1RXS and the 2RXS
catalogue in Fig.~\ref{sky_fields_green_yellow_ima} for field 930304.
Here we included those 1RXS detections that do not have a 2RXS counterpart 
down to the 2RXS existence likelihood limit of 6.5, but have 
a 2RXS counterpart down to a likelihood limit of 5.5 
(marked with white crosses in Fig.~\ref{sky_fields_green_yellow_ima}).

\subsection{Comparison of 1RXS and 2RXS detection likelihood distributions}\label{sec:eximl_redistribution}

\begin{figure}[htp]
  \centering
  \includegraphics[angle=-90,width=90mm,clip=]{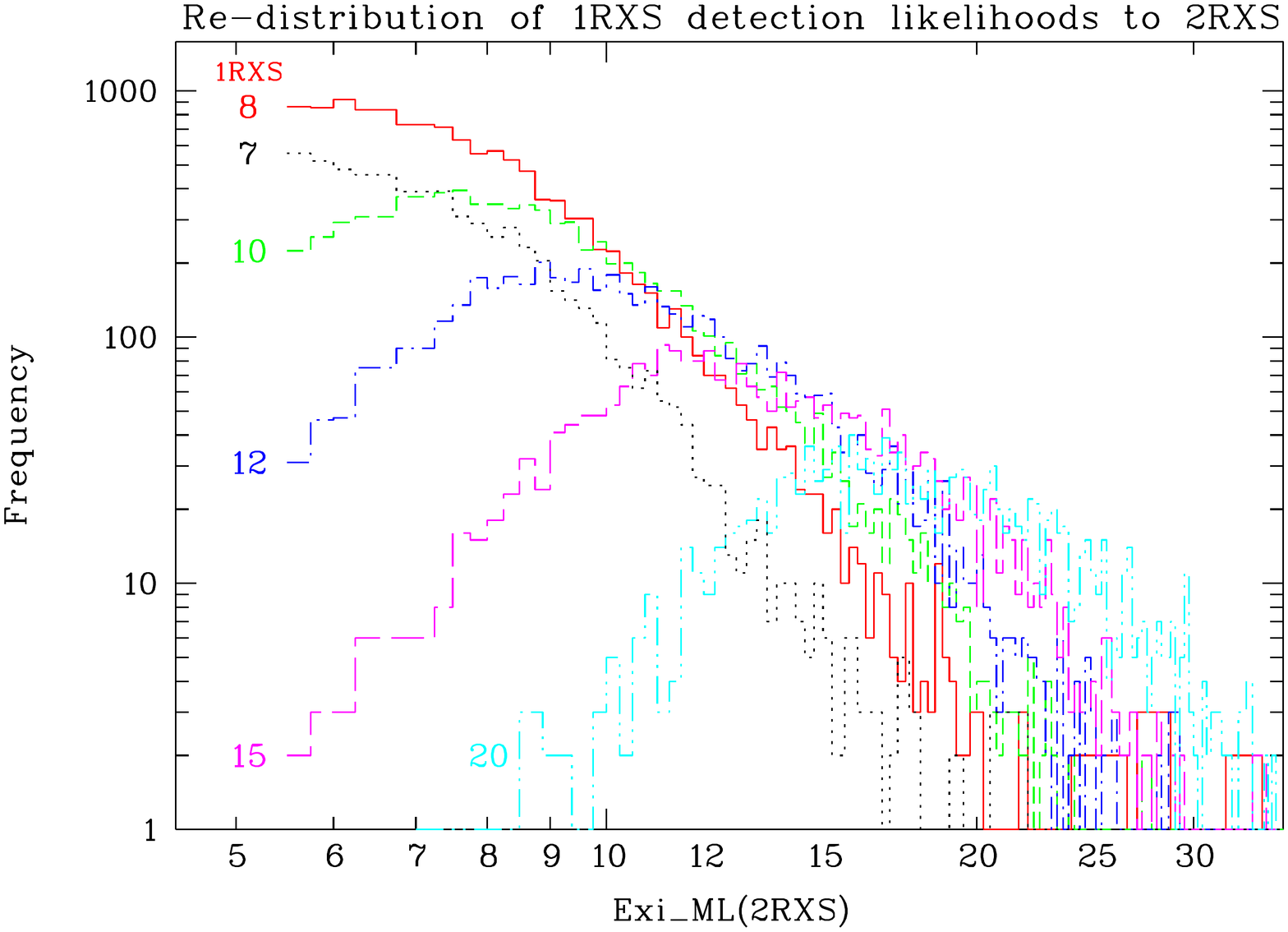}
  \caption{%
  Distribution of 2RXS detection likelihoods
  for catalogue entries common in both 1RXS and 2RXS
  for selected 1RXS detection likelihood values of 
  7 (black dotted), 8 (red solid), 10 (green short-dashed), 
  12 (blue dash-dotted), 15 (magenta long-dash), 
  and 20 (cyan dash-dotted-dotted).
  For details see text.
          }
   \label{plot_eximl_1rxs_2rxs}
\end{figure}

In Fig.~\ref{plot_eximl_1rxs_2rxs} we show the 
distribution of 2RXS detection likelihoods for catalogue entries common
in 1RXS and 2RXS for selected 1RXS detection likelihood values of 
7, 8, 10, 12, 15, and 20.
It is important to note that the distributions are not delta functions but
are rather broad with an extended tail to higher likelihood values.
Furthermore, the peak of the 2RXS distribution is below the 1RXS reference value,
but the mean is quite similar (e.g., for EXI\_ML\_1RXS = 12 the mean EXI\_ML\_2RXS 
is 11.6).
Therefore different threshold effects in 1RXS and 2RXS are expected
that for instance lead to missing and new sources, respectively.
This plot shows that detection likelihood values
obtained with different algorithms cannot be directly compared.

10096 entries in our detection runs with likelihood values
between 5.5 and 6.5 are within 1 arcmin of an 1RXS source
and 47832 detections in this likelihood range do not have a 
(close-by) 1RXS counterpart.
These common detections will also be made available, with a reduced set of
information compared to the main catalogue.
We regard these common detections as more reliable than other low-likelihood detections 
in either 1RXS or 2RXS. 
\subsection{Bright sources in 1RXS and 2RXS}
While the 1RXS catalogue contained \TotalNumberBscVoges\ sources according to the selection
criteria, the number of 2RXS bright sources determined by applying the same criteria is 22228. 
The main reason is that some sources in 1RXS are overestimated in their count rates because the background subtraction is underestimated. 
If we change the BSC count rate criterion from 0.05 to 0.058 counts $\rm s^{-1}$ 
and leave the detection likelihood threshold and number of counts (15 for both catalogues) unchanged, the number of 2RXS bright sources is 18912, 
which is similar to the 1RXS BSC.
Another effect is the re-distribution of 1RXS detection likelihoods
with respect to 2RXS, as shown  in Fig.~\ref{plot_eximl_1rxs_2rxs}. 
The magenta dashed curve shows the 2RXS detection likelihoods for 1RXS sources with 
a likelihood of 15. 
This distribution is very broad, meaning that\ the detection likelihoods in 1RXS range from
2RXS detection likelihoods from about 5.5 up to 30, with a significant fraction below 15.
On the other hand, the blue curve for 1RXS likelihoods of 12 also extends beyond 15,
which means that ``new'' bright sources formally not part of the BSC would now match the criterion.

\subsection{Reliability of 2RXS sources with low existence likelihood}\label{sec:Reliability}

\subsubsection{Simulations}\label{sec:Simulations}
\begin{figure}[htp]
  \centering
  \includegraphics[angle=-90,width=90mm,clip=]{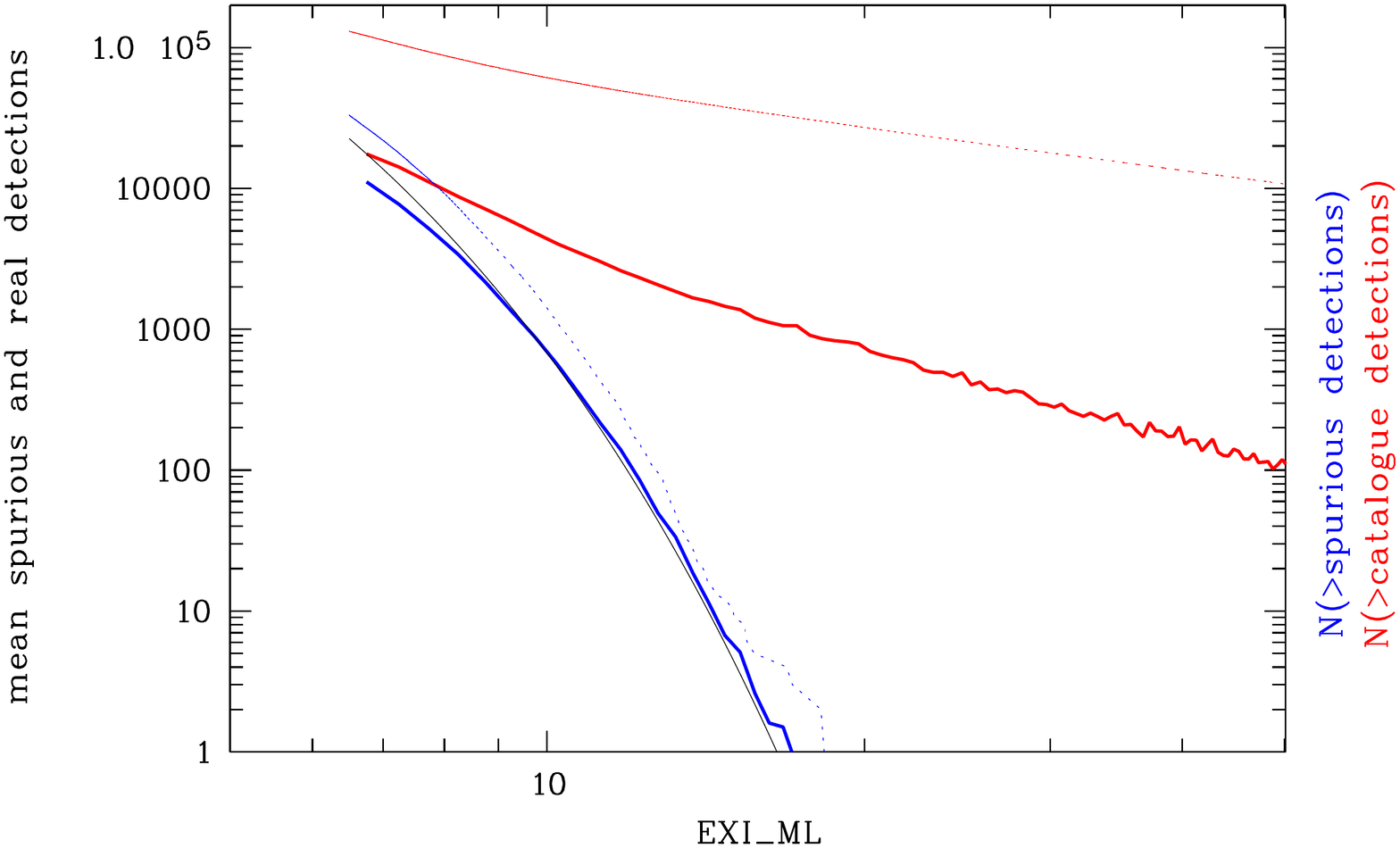}
  \includegraphics[angle=-90,width=90mm,clip=]{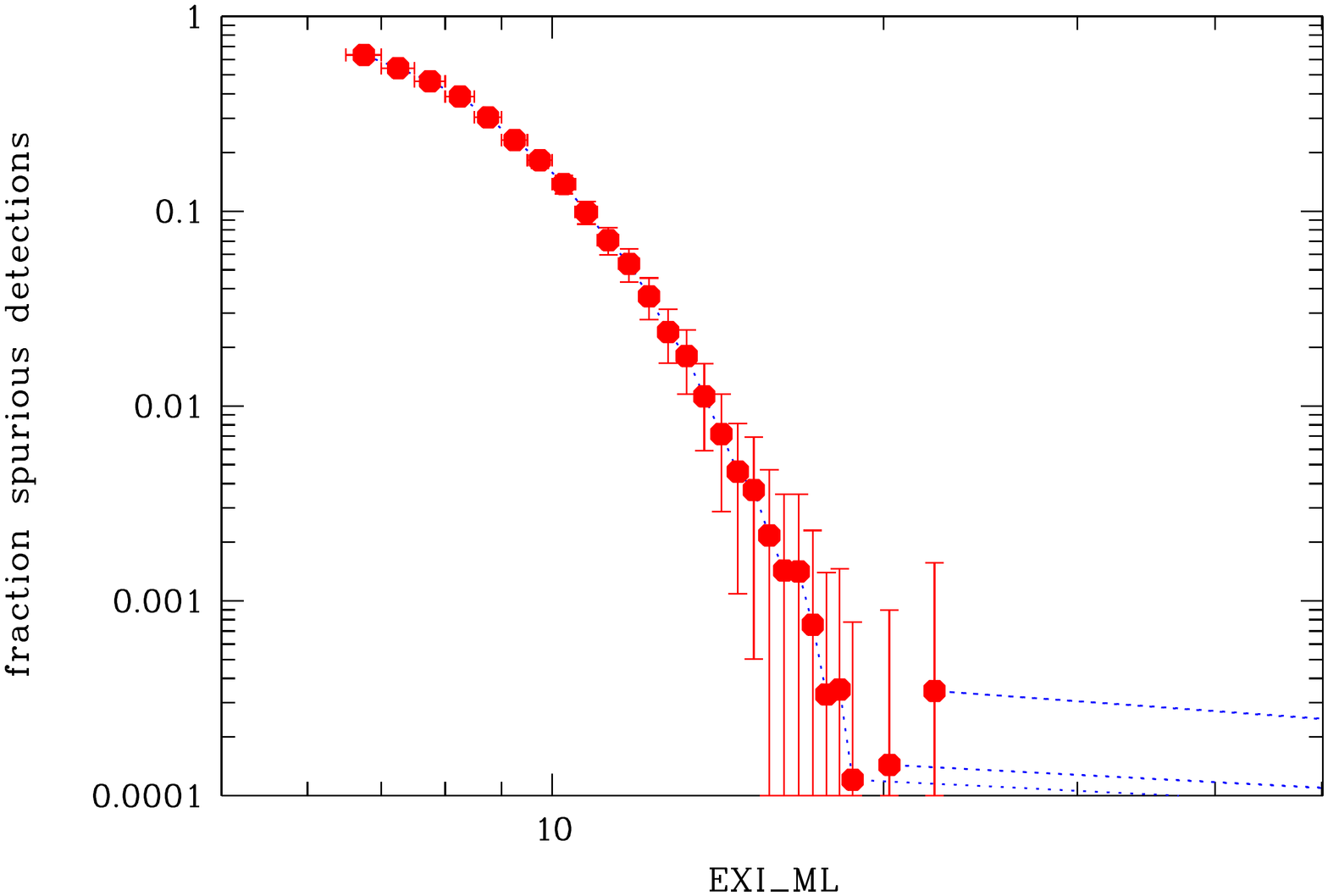}
  \caption{Upper panel: Differential (solid blue) distribution of the number of 
    spurious detections and differential (solid red) distribution of catalogue sources. 
    The integral distributions of spurious detections
    (thin blue, dashed) and catalogue sources (thin red, dashed) are shown in addition 
    with the labelling given on the right y-axis.
    The solid black thin line gives the theoretical differential relation between the 
    detection likelihood and the probability that a source is spurious.                            
    Lower panel: Fraction of spurious detections as a function of
    the existence likelihood.
    Details of the simulations are described in Sect.~\ref{sec:Simulations}.
          }
   \label{Fig_spurious}
   \end{figure}

\begin{figure}[htp]
  \centering
  \includegraphics[angle=-90,width=90mm,clip=]{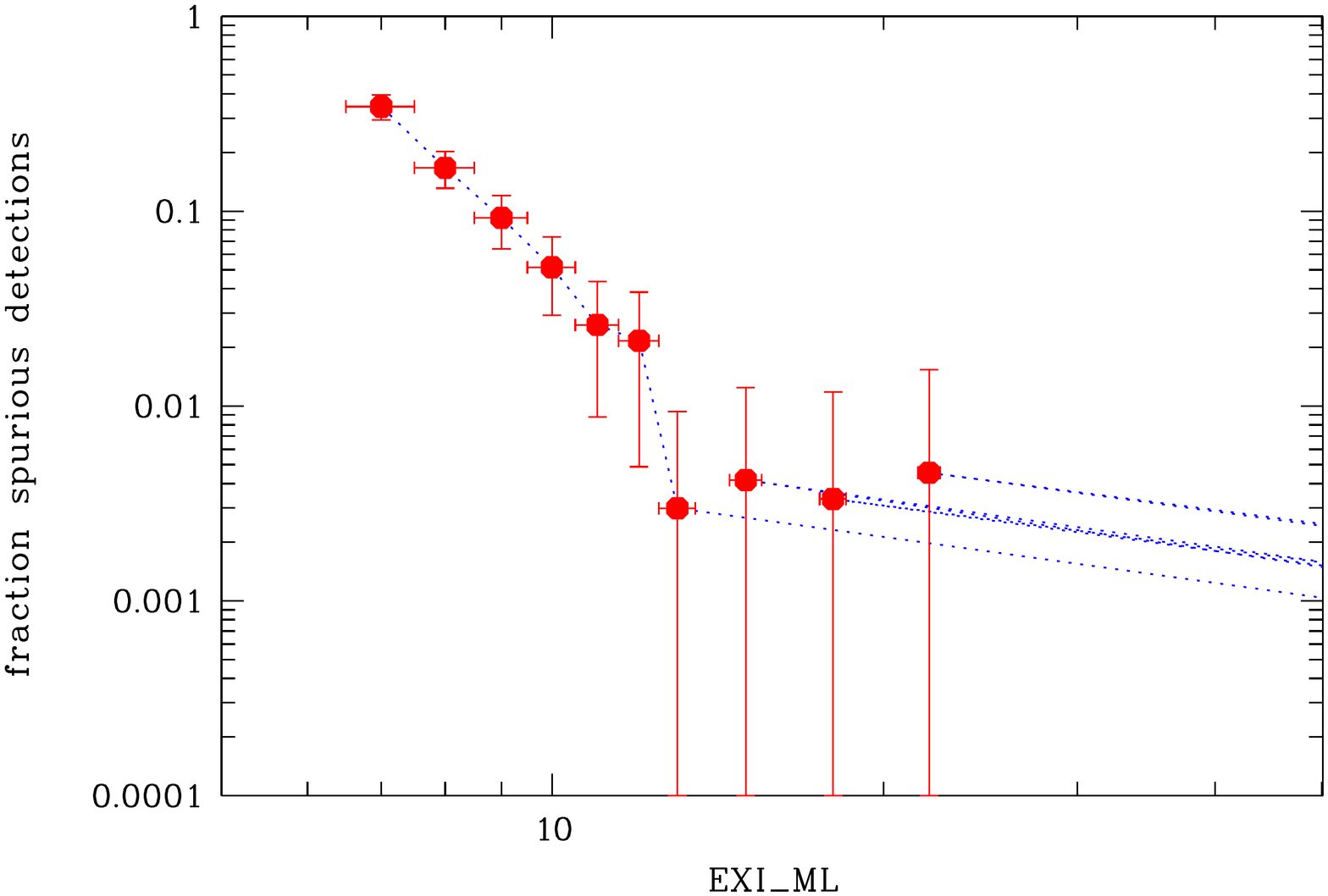}
  \caption{Fraction of spurious X-ray detections as a function of 
   detection likelihood for exposure times greater than 4000 seconds. 
   The differential fraction of spurious detections in the lowest bin 
   decreases to about 30 per cent.
This plot has to be compared with the lower panel of Fig.~\ref{Fig_spurious} , which
shows the fraction of spurious sources for the
whole sky (excluding fields that have been masked).
}
   \label{Fig_spurious_NEP}
   \end{figure}

\begin{figure}[htp]
  \centering
  \includegraphics[angle=-90,width=90mm,clip=]{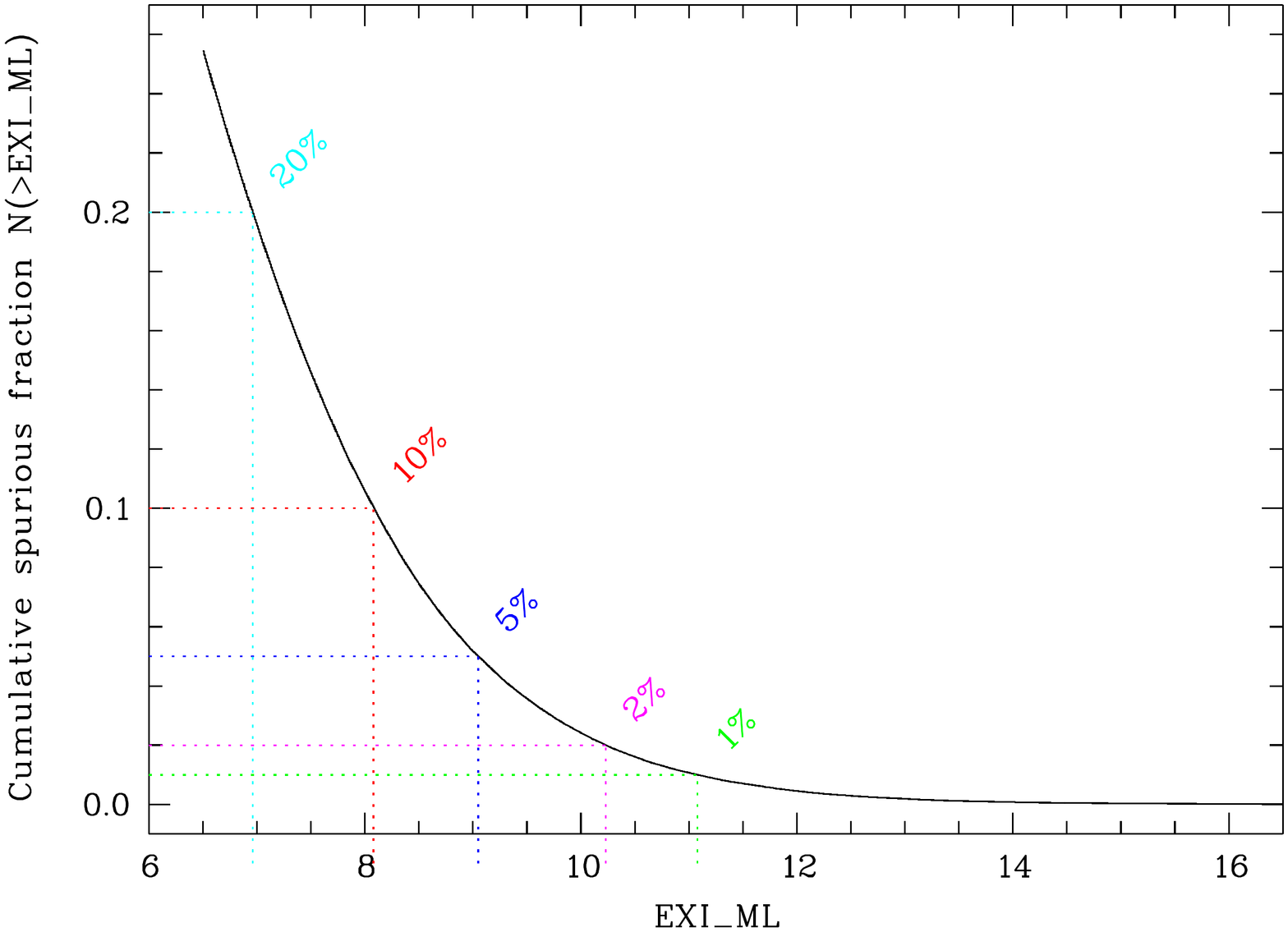}
  \caption{
Cumulative fraction of spurious detections as a function of detection likelihood.
The dashed lines indicate the corresponding values for 20, 10, 5, 2, and 1 per cent 
(from top left to bottom right), see text for details.}
   \label{Fig_spurious_cumulative}
   \end{figure}

\begin{table}
 \caption{Fraction of spurious detections.
The first columns lists the EXI\_ML bin.
The numbers given in brackets refer to the number of spurious and catalogue 
objects in that EXI\_ML bin.
The error to the spurious detection per bin is given in the third column. 
Columns 4 and 5 give the fraction of spurious detections and the error for 
exposure times greater than 4000 sec (fields close to NEP).
We note that we here only consider fields without masked regions.
}
\label{tab:spurious}
{\tiny
\begin{tabular}{lllll}
\hline\hline\noalign{\smallskip} 
EXI\_ML       & fraction bin all-sky        & error  & fraction NEP & error    \\ 
              & (spurious, objects)         &        &              &          \\
\noalign{\smallskip}\hline\noalign{\smallskip}  
6.5-7.0   &  0.63 (11162, 17709) & 0.03 & 0.34& 0.08 \\                             
7.0-7.5   &  0.55 (7657, 13989)  & 0.03 & 0.29& 0.08 \\
7.5-8.0   &  0.46 (4996, 10826)  & 0.03 & 0.15& 0.06 \\
8.0-8.5   &  0.40 (3475, 8612)   & 0.03 & 0.12& 0.05 \\
8.5-9.0   &  0.30 (2134, 6982)   & 0.02 & 0.12& 0.06 \\
9.0-9.5   &  0.23 (1311, 5692)   & 0.02 & 0.09& 0.05 \\
9.5-10.0  &  0.18 (869, 4685)    & 0.02 & 0.06& 0.04 \\
10.0-10.5 &  0.14 (555, 3917)    & 0.02 & 0.05& 0.04 \\
10.5-11.0 &  0.10 (326, 3373)    & 0.02 & 0.04& 0.04 \\
11.0-11.5 &  0.07 (193, 2908)    & 0.01 & 0.02& 0.02 \\
11.5-12.0 &  0.06 (148, 2548)    & 0.01 & 0.02& 0.02 \\
12.0-12.5 &  0.04 (84, 2245)     & 0.01 & 0.02& 0.02 \\
12.5-13.0 &  0.02 (45, 1993)     & 0.008& 0.01& 0.02 \\
13.0-13.5 &  0.02 (36, 1805)     & 0.008& 0.00& 0.00 \\
13.5-14.0 &  0.008 (13, 1637)    & 0.006& 0.00& 0.00 \\
14.0-14.5 &  0.006 (9, 1553)     & 0.005& 0.00& 0.00 \\
14.5-15.0 &  0.001 (2, 1437)     & 0.001& 0.00& 0.00 \\
\noalign{\smallskip}\hline
\end{tabular}
}
\end{table}

We performed simulations for the ROSAT all-sky survey to 
estimate the reliability of the 2RXS catalogue at its faint end.
\citet{Watson2009}
have pointed out that in 3XMM the relation between the existence likelihood and the
detection probability requires calibration through simulations. 
Similarly, \citet{2014ApJS..210....8E} %
have performed this analysis 
for the Swift-XRT 1SXPS catalogue.

To keep all spatial background structures that are due to cosmic and non-cosmic
emission, we have used a specific approach.
First, we sorted the detection list (down to the limit of ${\tt EXI\_ML} = 5.5$)
for each sky field that does not contain masked regions
in increasing order of count rate. 
For each detection we then determined the fraction of 
background photons and source photons within
the source extraction circle from parameters obtained in the detection run,
and correspondingly assigned a probability that a photon is a background photon.
If a photon is covered by more than one source extraction region, then the 
probability is overwritten by subsequent brighter sources. 
``Source'' photons are then statistically removed within the extraction radius, 
such that the background is flat on scales larger than the source extraction radius 
(i.e. several arcmin), 
while all large-scale structures (on scales of degrees) remain unaffected.
The remaining photons are then randomly redistributed within the actual 
field of view at the photon arrival time.
This preserves the spatial and temporal structure of the background. 
Finally, the last step was performed ten times for each of the RASS sky fields 
without masked emission regions, and the detection algorithm was run identically
to the actual observations (see Sect.~\ref{sec:SourceDetectionGeneral}).
The results are shown in Fig.~\ref{Fig_spurious} and Table~\ref{tab:spurious}.
To our knowledge, this is the first time that such extensive and realistic simulations 
were performed for ROSAT all-sky survey point source detections. 

The solid thin black line in Fig.\ref{Fig_spurious} (upper panel)
represents the theoretical relation 
between the probability P that a source is
spurious as a function of the detection likelihood EXI\_ML, that
is, $\rm P = \exp(-EXI\_ML)$. 
This relation refers to the detection likelihood per detection cell.  
To calculate the differential number of spurious detections, we have to multiply 
this by the number of detection cells in the ROSAT all-sky survey, if known {\em \textup{a priori}}.
Here this number was determined to match the simulated distribution 
of spurious sources for likelihood values exceeding 10 (see Sect.~\ref{sect_psf_cellsize}).
For lower likelihood values the theoretical curve is slightly 
below the differential distribution of spurious detections.
This may be due to incompleteness of the input candidate lists 
to the ML source detection runs
in the low-likelihood regime. 
Our simulation approach differs from the 2XMM \citep{Watson2009} 
and Swift \citep{2014ApJS..210....8E} simulations of spurious detections. 
In this paper we have cut out all detections, while in the
XMM and Swift approaches the simulated images still include simulated sources.
Typical CCD detector features present in pointed observations, 
such as hot pixels or columns, read-out streaks, or single reflection rings,
are not relevant in PSPC scanning mode observations because they either do not exist
or are smeared out to an additional large-scale background.
Therefore we only have to consider purely statistical spurious sources.
As all detected ``real'' sources have been removed before the simulations, we regard
all remaining detections from our simulated data as ``spurious''.
Most of the systematic artefacts that are due to bright sources have been cleaned by our 
visual screening procedure because the visual screening flag setting is a human and subjective process.

With this work we are able to quantify the number of spurious detections and
for a given detection likelihood value estimate the mean fraction of spurious 
detections. 

The acceptable ratio of real sources to spurious detections is strongly dependent 
on the scientific application. 
We choose to release the catalogue down to a likelihood limit of 6.5, 
which corresponds to the same limit as the ROSAT FSC. 
This allows users to search for (real) X-ray emission down to very faint limits, 
but results in a high percentage of detections, 
mostly at low likelihoods, which are likely to be spurious. 
Specifically, our estimates show that around 30~\% of the sources in the entire 
2RXS catalogue could be spurious down to this limit. 
If a lower spurious fraction is considered important by the user, 
a higher likelihood threshold can be chosen to generate a more reliable catalogue.  
To give some examples, 
the likelihood threshold 
and the integral spurious numbers for existence likelihood values
of 20, 10, 5, 2, and 1 per cent are %
EXI\_ML =  6.96,\, 22602  sources (20~\% spurious),
EXI\_ML =  8.08,\, 8547   sources (10~\%),  
EXI\_ML =  9.03,\, 3517  (5\%),  
EXI\_ML = 10.18,\, 1189  (2\%), and
EXI\_ML = 11.01,\,  533  (1\%), respectively.
In Fig.~\ref{Fig_spurious_cumulative} we show the cumulative fraction
of spurious detections as a function of detection likelihood.

The \textup{{\em \textup{fraction}}} of spurious detections depends on the number of ``real'' 
sources per field, which means that fields with high object numbers possess a lower fraction
of spurious detections because the mean number of spurious detections does not
depend on the number of catalogue sources in the field 
(this holds only for our simulation approach with cutting out all sources before the simulation runs).
As an example in Fig.~\ref{Fig_spurious_NEP}, we show the fraction of spurious 
X-ray detections as a function of detection likelihood for exposure times
greater than 4000 seconds (i.e., close to the ecliptic poles):
the differential fraction of spurious detections in the lowest bin 
(${\tt EXI\_ML} = 6.5 - 7.5$) 
decreases to about 30 per cent.

\subsubsection{Sources only present in either 1RXS or 2RXS}

Unfortunately, the 1RXS source detection software is not operational anymore
and a direct comparison of the number of spurious detections cannot be made
using simulations.
To quantify which catalogue is more reliable, we have made correlations with
other X-ray catalogues for those survey sources that are included in either
the 1RXS or the 2RXS catalogue for detection likelihoods between 7.5 and 14.5. 
(in 1RXS the lowest likelihood bin is strongly incomplete 
because of the additional constraint of ${\tt counts} \geq 6$). 
A search radius of 60 arcsec was applied for the 2RXP and the XMMSL1 catalogues.
Because of the better spatial resolution, a search radius of 30 arcsec was used
for the 3XMM and the 1SXPS catalogues.
\\
For sources unique to 2RXS, we find 523 2RXP matches, while for
sources unique to 1RXS the number of matches is lower with 394, 
which is about a factor of 1.4 in the percentages.
The same holds for the correlations with the slew survey XMMSL1,
where 136 1RXS and 211 2RXS counterparts are found, which is about
a factor of 1.6 in percentages.
For the 3XMM catalogue we find similar numbers in the fractions of 
associations, while for the 1SXPS catalogue a higher fraction of 2RXS counterparts with a 
factor of  1.16 is found. We speculate that the 3XMM catalogue might be
also affected by a high number of spurious detections at the
faint end, which appears not be the case of the
other catalogues
(see Table~\ref{tab:2rxs1rxs_cross}). 
In summary, the fraction of unique 2RXS sources
(i.e. not contained in 1RXS) with counterparts in other X-ray catalogues is higher
at the faint end than there are 1RXS sources without
2RXS counterparts (except 3XMM). This indicates that the 2RXS catalogue is more reliable in
terms of spurious source content than 1RXS.

Although at the faint end the 2RXS catalogue contains a substantial number of
spurious  sources (as expected from low likelihood values), 
the contamination in the 2RXS catalogue is a lower limit for the 1RXS catalogue.

\begin{table*}
\caption{Cross-correlations 
for 1RXS and 
(clean) 2RXS catalogue sources with EXI\_ML in the range $7.5-14.5$.}
\label{tab:2rxs1rxs_cross}
\begin{tabular}{ll|llll}
\hline\hline\noalign{\smallskip} 
EXI\_ML $7.5-14.5$& Sources       & 2RXP            & XMMSL1\_dr6    & 3XMM\_dr4       & 1SXPS          \\ 
\noalign{\smallskip}\hline\noalign{\smallskip}   %
1RXS              &  70909        & 2.91\,\% (2070)   & 1.52 \,\% (1077) & 1.31 \,\% (935)  & 1.55\,\%  (1101) \\
2RXS              &  58766        & 3.36\,\% (1975)   & 1.94 \,\% (1138) & 1.46 \,\% (859)  & 1.85\,\%  (1089) \\
\noalign{\smallskip}\hline\noalign{\smallskip}
\noalign{\smallskip}\hline\noalign{\smallskip}
\noalign{\smallskip}\hline\noalign{\smallskip}  %
1RXS no 2RXS     & 21241         & 1.85\,\% (394)   & 0.64 \,\% (136)  & 1.32 \,\% (281)  & 1.09 \,\% (232)  \\
2RXS no 1RXS     & 20766         & 2.51\,\% (523)   & 1.01 \,\% (211)  & 1.32 \,\% (273)  & 1.27 \,\% (264)  \\
\noalign{\smallskip}\hline
\end{tabular}
\end{table*}

\subsection{Positional offset in scan direction}\label{sec:plot_offset}

\subsubsection{RASS-2 versus RASS-3 processing}\label{sec:rel_astro}

We found a systematic positional offset in ecliptic coordinates 
between the 1RXS catalogue (RASS-2 processing) and detections based on 
event files from the RASS-3 processing.
This shift is preferentially along ecliptic great circles (i.e. scan direction).
An important parameter influencing the astrometry of the ROSAT sources 
is a time delay between the star tracker time and the photon arrival times.
This time delay was determined for the RASS-2 processing to 2.53\,s.
After the RASS-2 processing, a rounding error was found in the software and 
corrected for the RASS-3 processing, but without re-adjusting the time delay.
While in the RASS-2 processing the rounding error was largely compensated 
for by the time delay, a systematic coordinate shift along
the scan direction was 
introduced in the RASS-3 processing that is present in the public event files.
Because the scan direction reversed several times during the survey, the 
coordinate shift changes sign at scan reversals.

\begin{figure}[htp]
  \centering
  \includegraphics[angle=-90,width=90mm,clip=]{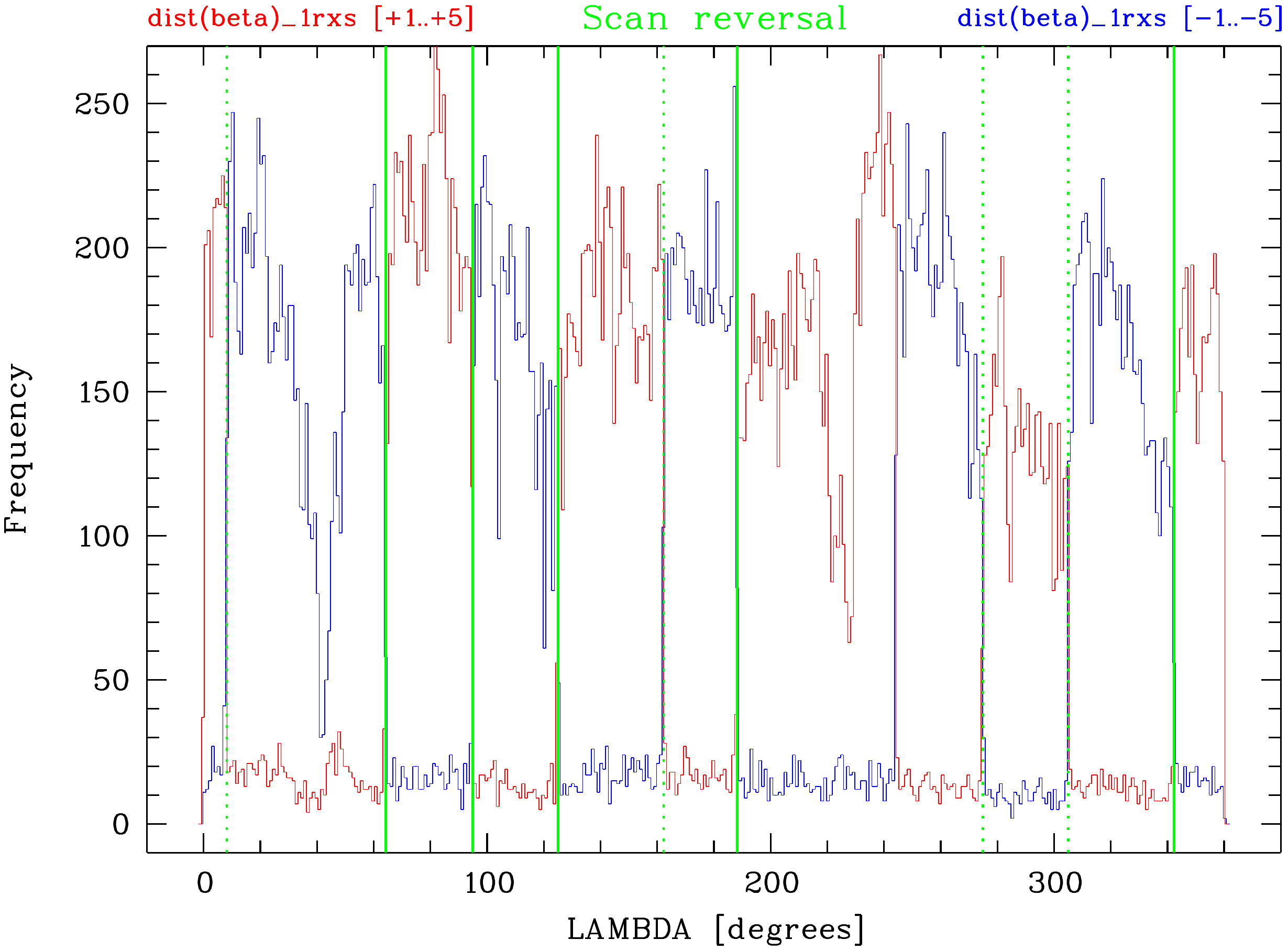}
  \caption{       
   Histogram of the positional offsets in ecliptic latitude between the 1RXS and 2RXS
   catalogues as a function of ecliptic longitude (LAMBDA). 
   The blue lines indicate distance differences between $-1\arcsec$ and 
   $-5\arcsec$, the red lines refer to differences between $+1\arcsec$ and $+5\arcsec$. 
   The green vertical lines indicate the scan reversals.
              }
             \label{Fig_offset1}
   \end{figure}

In Fig.~\ref{Fig_offset1} a histogram of the average position offsets in
ecliptic latitude, that is, in scan direction, as a function of ecliptic 
longitude (progressing survey) is presented.
At the scan reversals, the offsets change sign. 
Offsets between RASS-2 and RASS-3 in the ranges [-1,-5]\arcsec\ and [+1,+5]\arcsec\
are indicated with blue and red, respectively, showing 
the clear split depending on scan direction.
In the following we refer to the longitude ranges
(equivalent to time ranges of the six-month survey phase) 
with offset shifts preferentially negative and positive as 
blue and red periods.

\begin{figure}[htp]
  \centering
  \includegraphics[angle=-90,width=45.654mm,clip=]{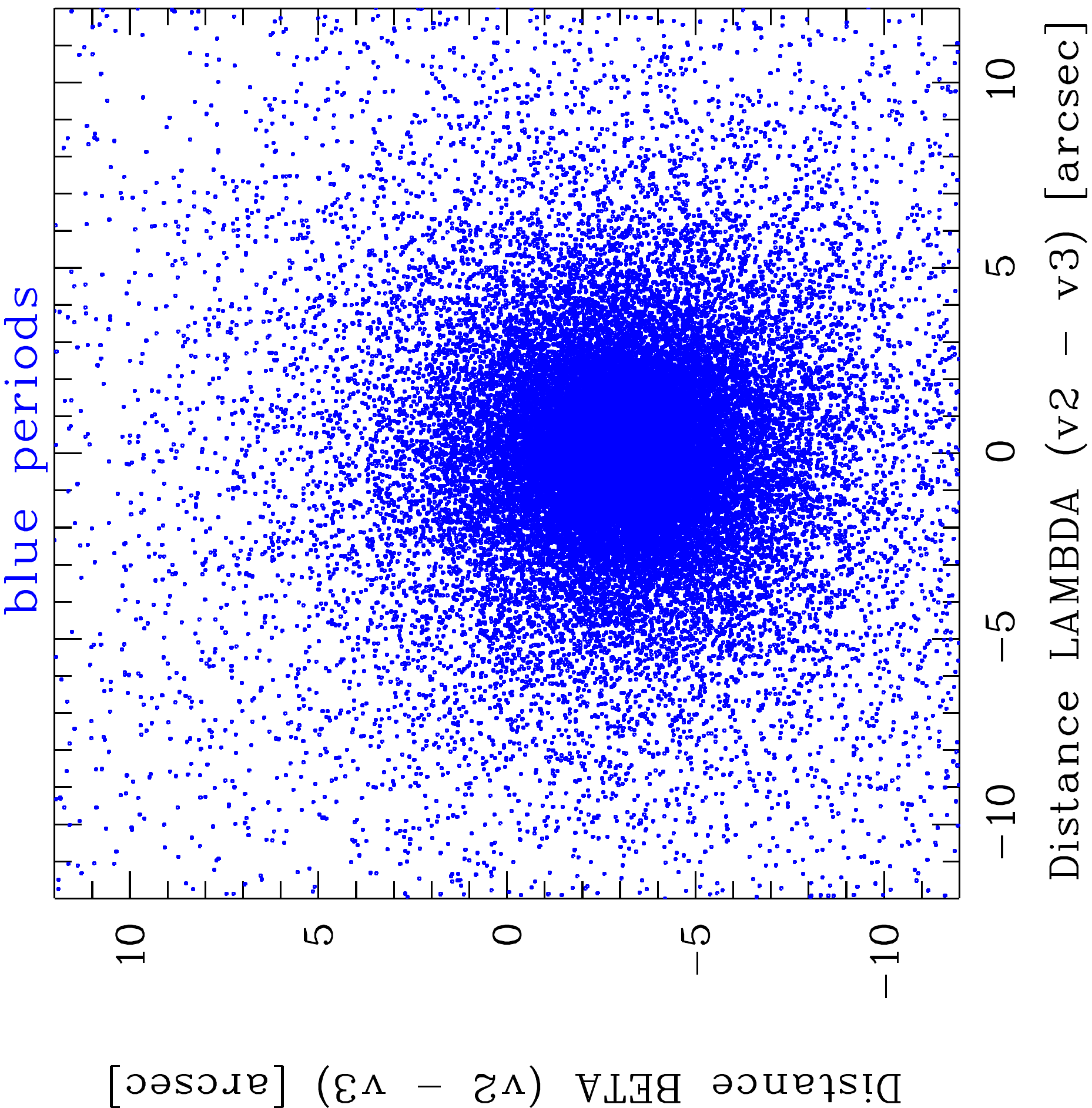}%
  \includegraphics[angle=-90,width=42.291mm,clip=]{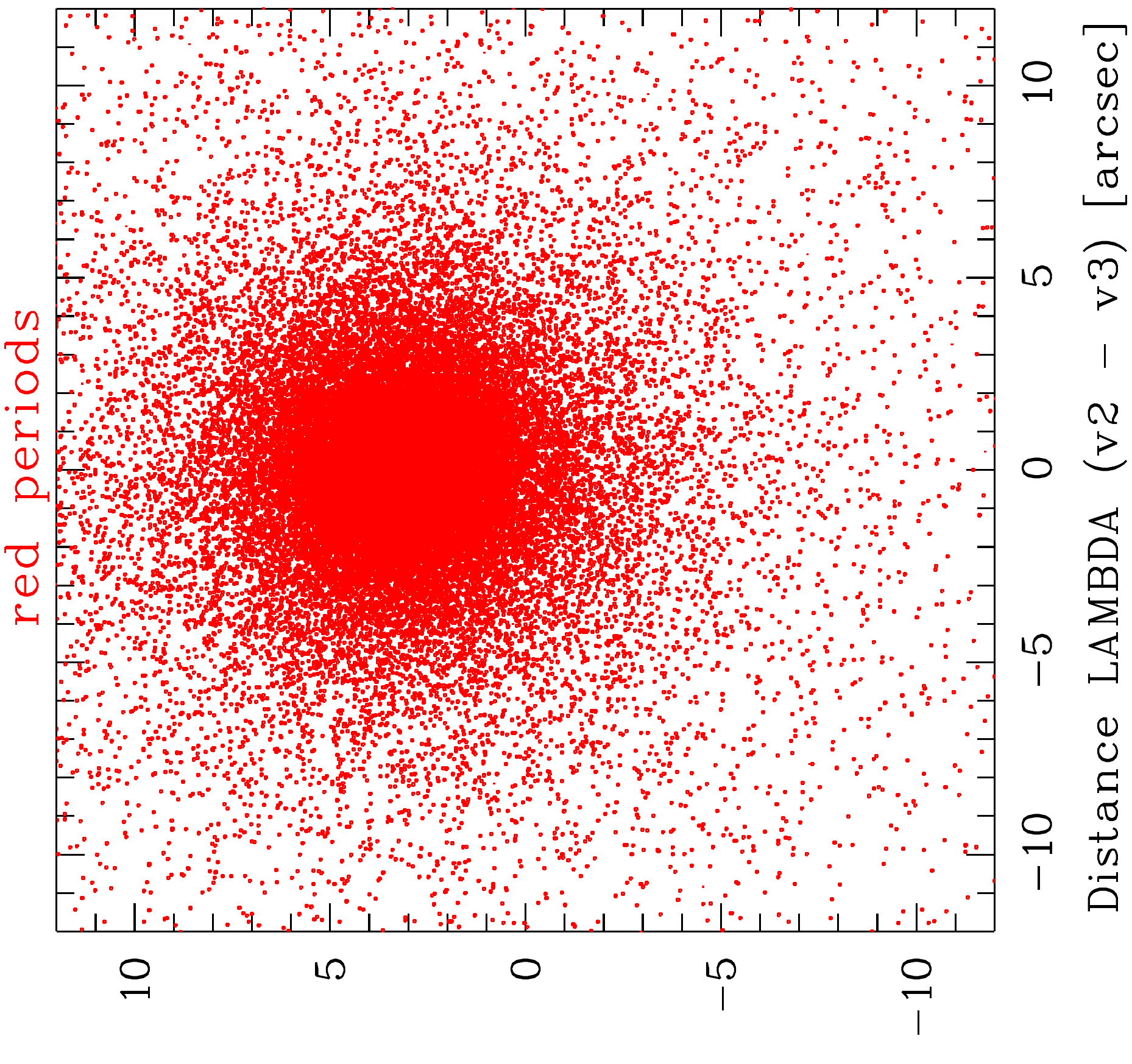}
  \caption{Offsets in ecliptic coordinates between RASS-2 (1RXS) and RASS-3 positions, 
      that is, before corrections applied for 2RXS.
           Depending on scan direction, a shift in scan direction of about $+3.1\arcsec$ or $-3.1\arcsec$ is seen.
            For more details see Sect.~\ref{sec:rel_astro}.
           }
   \label{Fig_offset2}
   \end{figure}

In Fig.~\ref{Fig_offset2} we show the positional offsets in ecliptic 
coordinates for RASS-3 detections that have 1RXS counterparts. 
For the blue periods of Fig.~\ref{Fig_offset1} the 
mean
positional offset in ecliptic latitude $\beta$ is $\rm \Delta\beta=-3.110 \arcsec$ and the 
mean
positional offset in ecliptic longitude $\lambda$ is
$\rm \Delta\lambda=+0.031 \arcsec$. 
For the red periods the corresponding offsets are
$\rm \Delta\beta=+3.134 \arcsec$ and $\rm \Delta\lambda=+0.031\arcsec$.
The positional offset is dominant in scan direction $\beta$. 

This systematic positional offset can be corrected for each photon of which the 
scan direction is known. The values determined above adjust the positions
from RASS-2 and RASS-3 processing. Comparison of the coordinates with optical 
catalogues shows that an additional correction is required 
to minimise positional offsets in ecliptic latitude.
As an intermediate step, we therefore corrected the missing time delay adjustment by applying a 
$\pm 3.14$\arcsec\ positional shift in ecliptic latitude 
depending on the scan direction (and scan reversals) in our RASS-3 processing as a first 
iteration and used the new positions for comparison with optical catalogues (see next subsection).

\begin{figure}[htp]
  \centering
  \includegraphics[angle=-90,width=90mm,clip=]{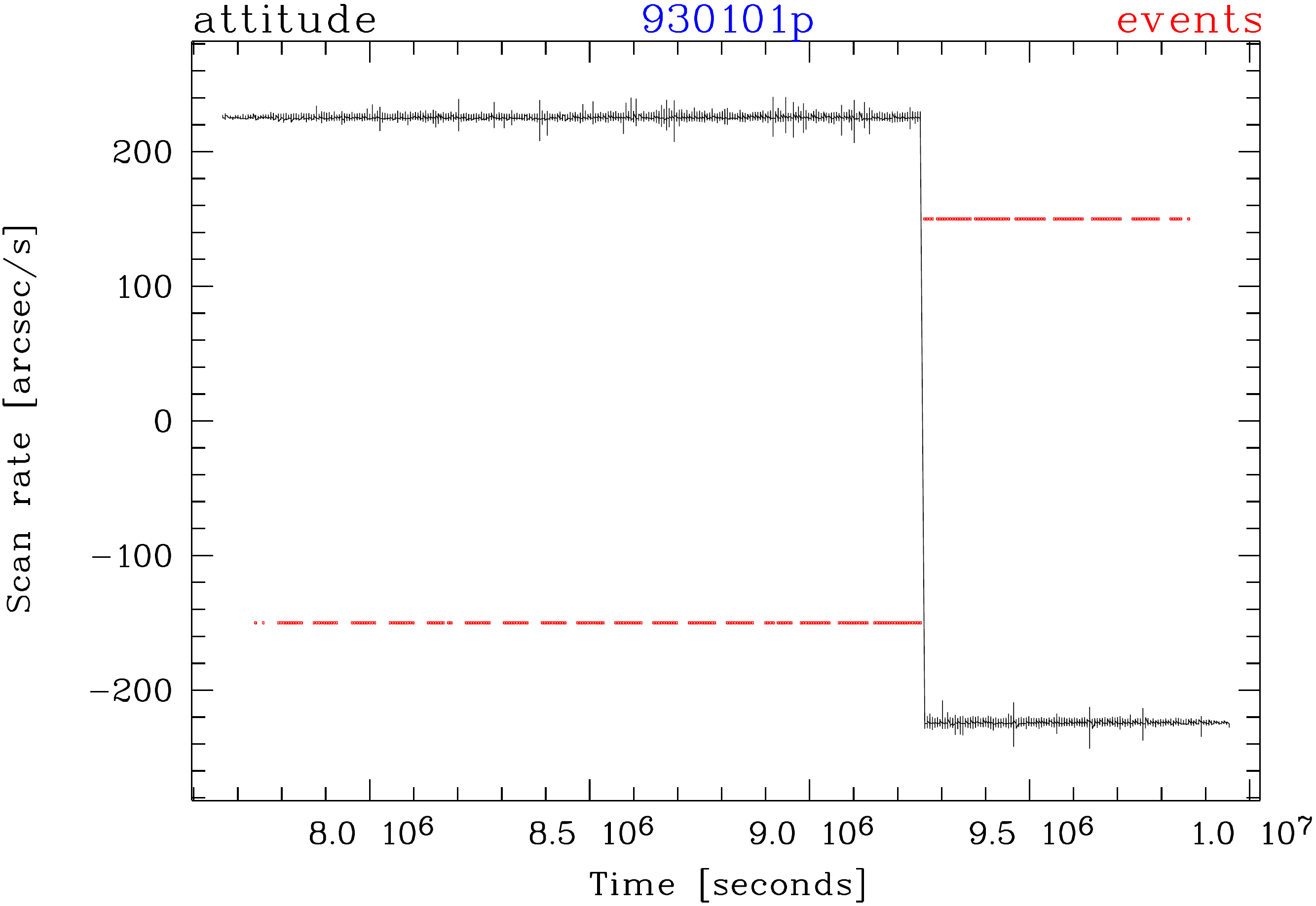}   
\caption{Example for the speed of the ROSAT all-sky survey scanning as a 
function of time before and after a scan reversal. 
The alternating spreads correspond to a positional offset in scan direction 
of $+3.14\arcsec$ and $-3.14\arcsec$ or a time delay between
the photons in the RASS-2 and RASS-3 processing of 14.3 ms. 
The dashed red lines indicate time intervals when photons have been
collected.}                                                                                                                                
             \label{Fig_offset3}
   \end{figure}

In Fig.~\ref{Fig_offset3} we show an example for the scan rate (black line) determined from  
the attitude data that is available in 1\,s steps, in units of arcsec\,s$^{-1}$,
before and after a scan reversal. 
The red dashed lines indicate at which times the ROSAT PSPC detector was 
switched on and good time intervals have been identified. 
For these periods photons are available for correction.

\begin{figure}[htp]
  \centering
  \includegraphics[angle=-90,width=45.654mm,clip=]{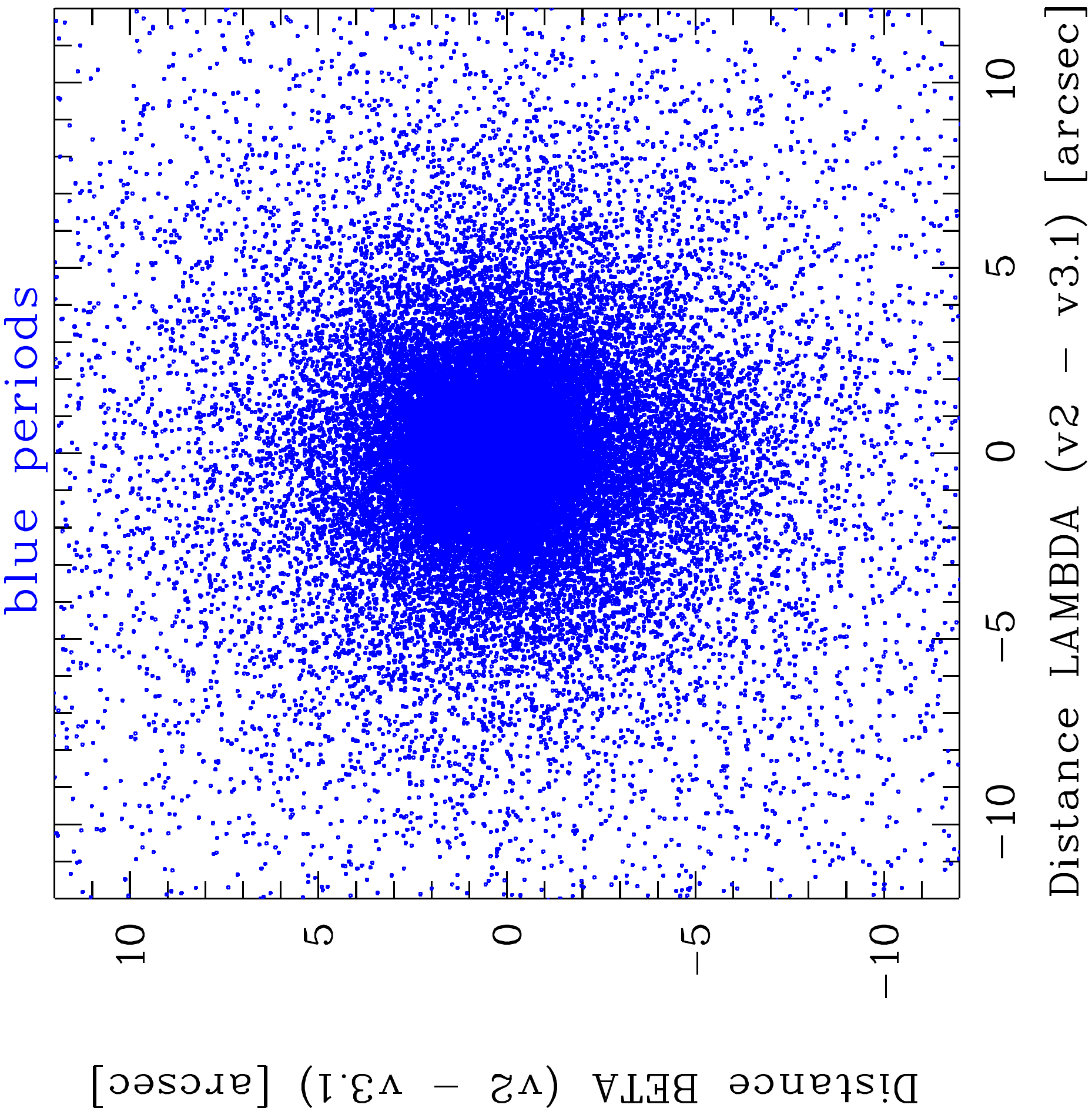}
  \includegraphics[angle=-90,width=42.291mm,clip=]{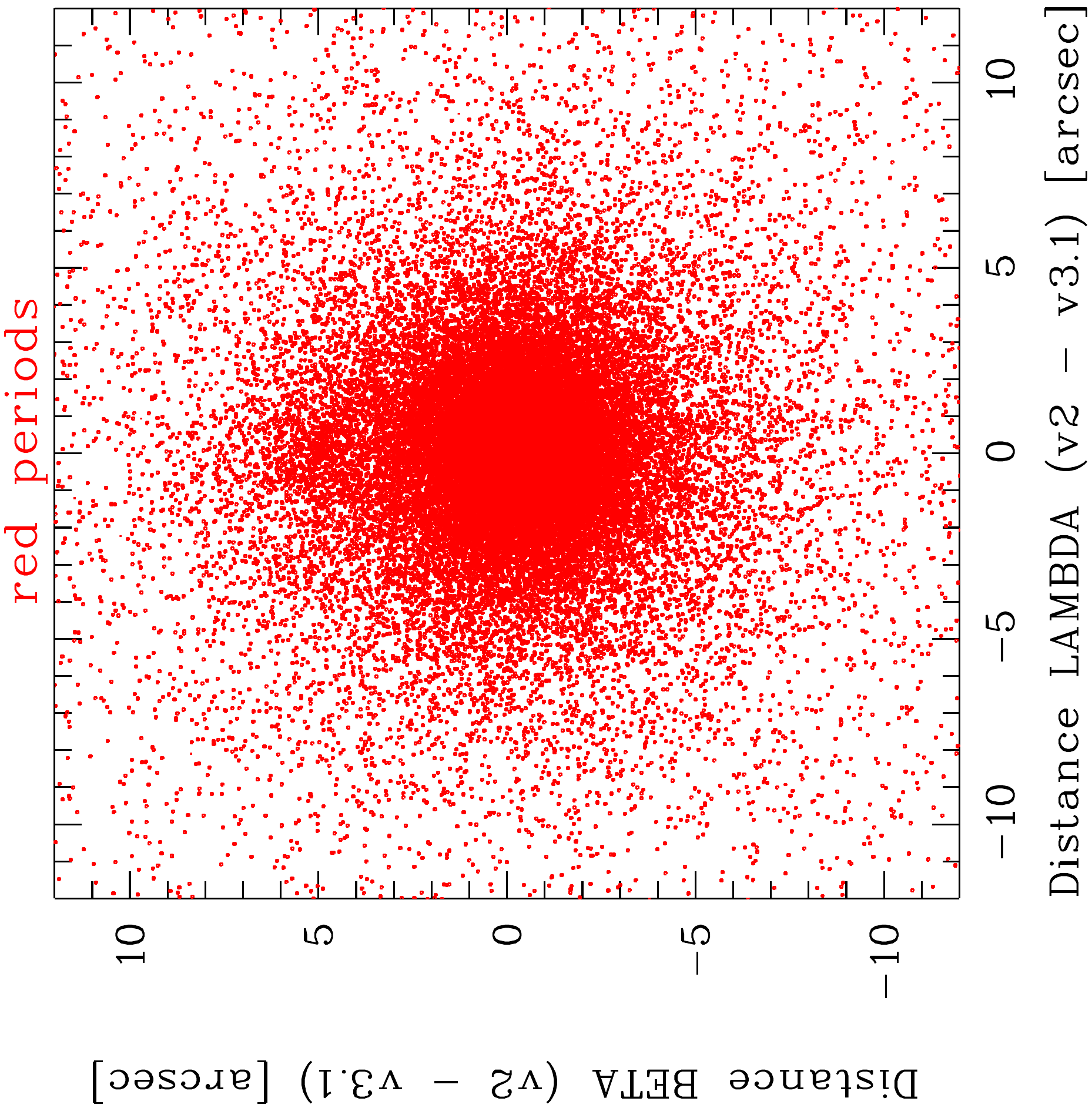}
  \caption{
           Same as Fig.~\ref{Fig_offset2} after applying a $\pm 3.14\arcsec$ shift 
           in ecliptic latitude to the events of RASS-3 processing. 
           The systematic offset in scan direction has been reduced by a factor of about 10, and the positional offsets 
           in ecliptic coordinates are in the sub-arcsec range.
           }
   \label{Fig_offset4}
   \end{figure}

In Fig.~\ref{Fig_offset4} we present the positional offset in ecliptic coordinates 
after applying a shift of $\pm 3.14$\arcsec\ in ecliptic latitude to each RASS-3 event 
depending on scan direction.
The systematic offset in ecliptic latitude present in Fig.~\ref{Fig_offset2} is now 
reduced to sub-arcsec level.
For the blue periods of Fig.~\ref{Fig_offset4} the remaining mean positional offset in 
ecliptic latitude is $\rm \Delta\beta=-0.308$\arcsec\ and in ecliptic longitude 
$\rm \Delta\lambda=0.039$\arcsec.
For the red periods the positional offsets are
$\rm \Delta\beta=+0.314$\arcsec\ and $\rm \Delta\lambda=+0.052$\arcsec.

The absolute astrometry is described in the following section. 
Section~\ref{sec:tycho_correlation} describes the correlation between the Tycho-2 catalogue and 2RXS, which 
gives an additional indication that 2RXS is more reliable.
\subsubsection{Absolute astrometry}\label{sec:abs_astro_new}

The absolute astrometry of the RASS-3 (after the 3.14\arcsec\ correction) 
and RASS-2 positions can best be tested with catalogues of point-like 
ROSAT sources that have been identified optically and whose optical positions have been accurately 
determined.
Such catalogues are the ROSAT Bright Survey (RBS) catalogue from 
\citet{Schwope2000}, and the SDSS catalogue of stars 
detected in the RASS \citep{SDSS_stars_rass}. 
We have selected entries from the two catalogues that possess optical identifications
for ROSAT sources with optical positional errors smaller than 1\arcsec. 
For RBS we restricted ourselves to point-like entries of the classes
AGN, star, cataclysmic variable, and X-ray binary, 
and did not consider extended emitters such as clusters 
or groups of galaxies, for which the extent increases the intrinsic X-ray positional 
uncertainty.

\begin{figure}[htp]
  \centering
  \includegraphics[angle=-90,width=45.30mm,clip=]{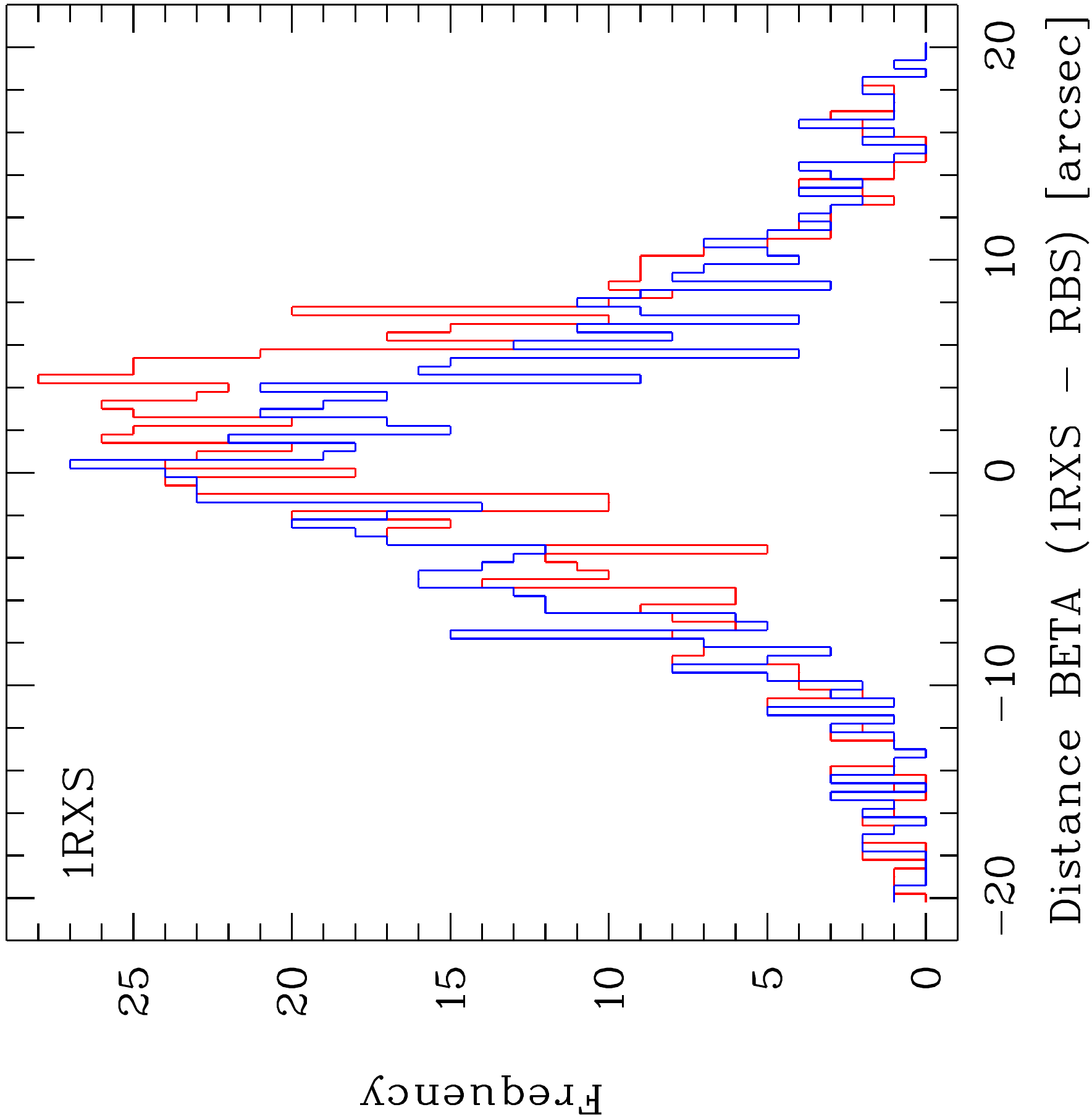}
  \includegraphics[angle=-90,width=42.69mm,clip=]{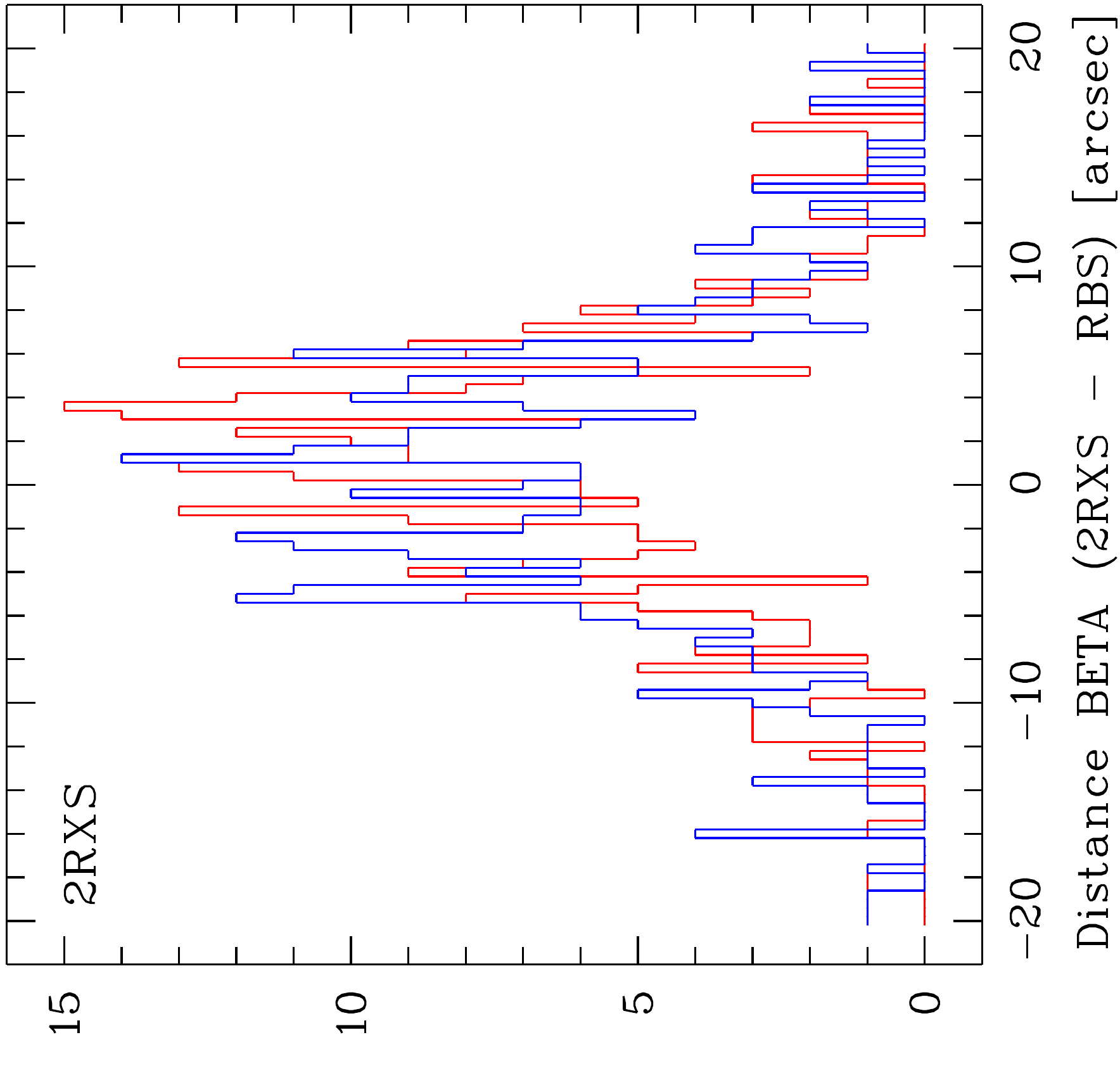}
  \caption{%
  Positional offsets in ecliptic latitude between the 1RXS catalogue and the 
  optical positions from the \citet{Schwope2000} catalogue for red and blue 
  periods (left panel),
  and similarly with respect to 
  the 2RXS catalogue (right panel). 
              }
             \label{Fig_offset5}
   \end{figure}

\begin{figure}[htp]
  \centering
  \includegraphics[angle=-90,width=45.30mm,clip=]{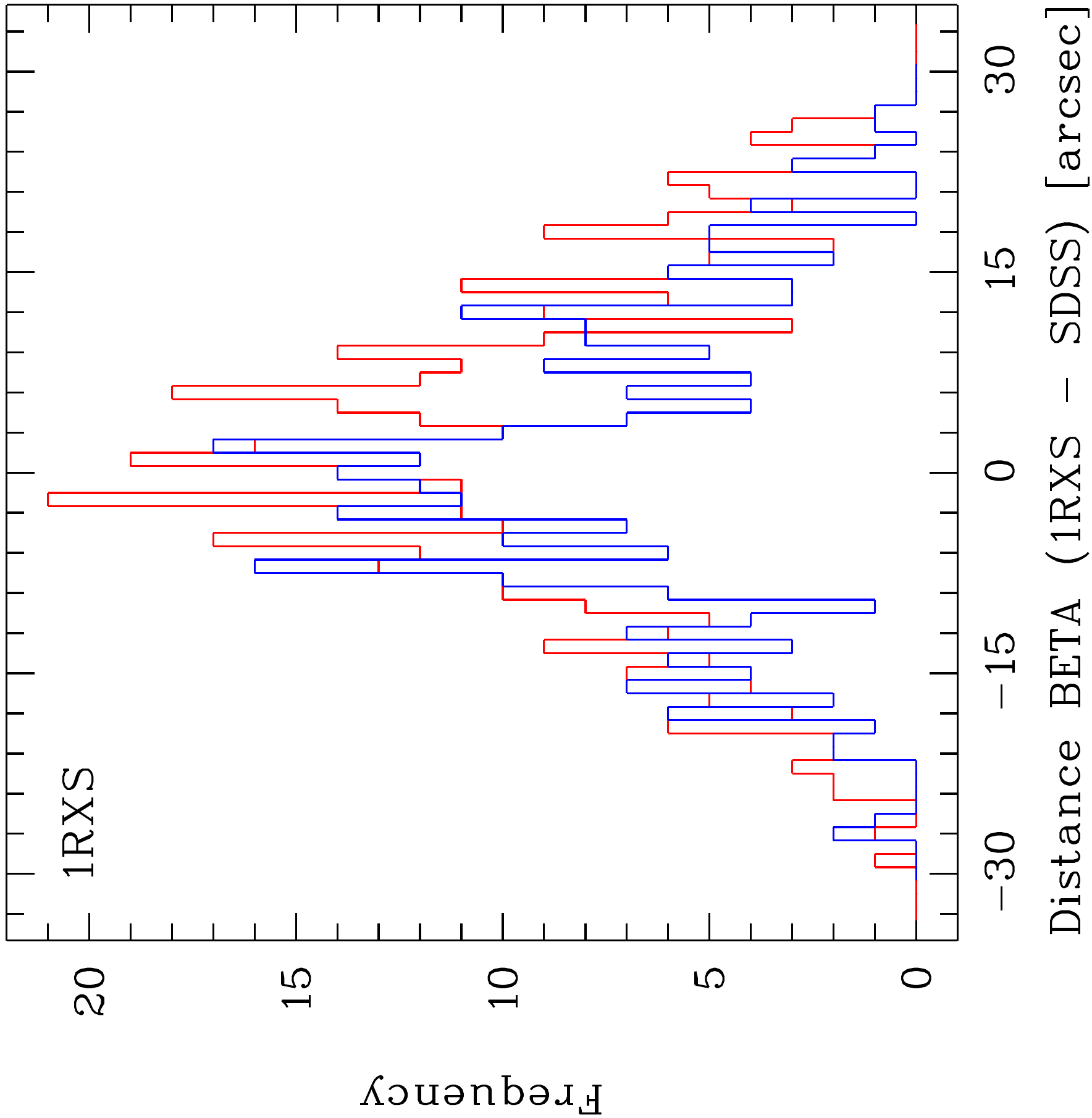}
  \includegraphics[angle=-90,width=42.69mm,clip=]{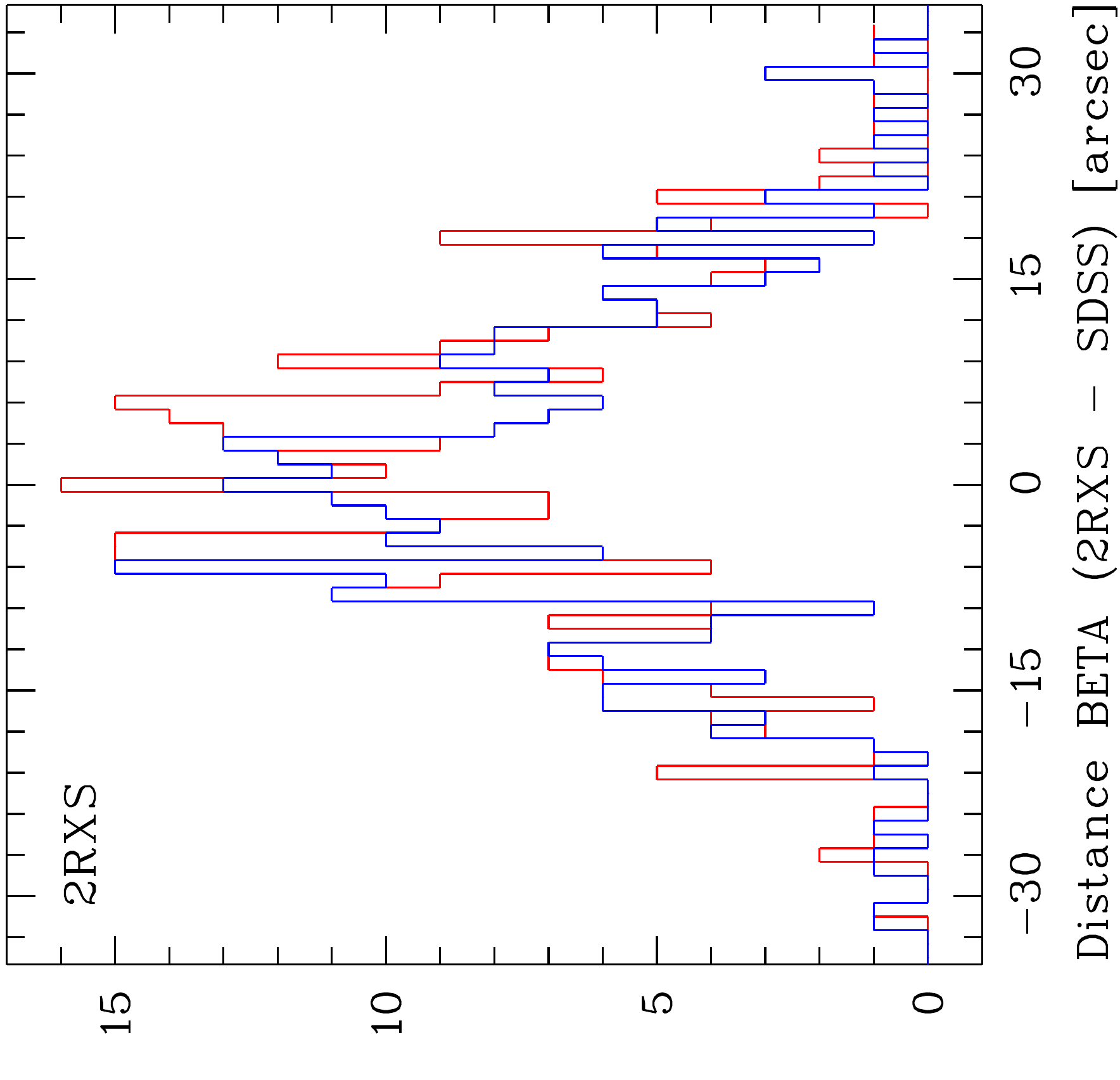}
  \caption{%
  Positional offsets in ecliptic latitude between the 1RXS catalogue and the 
  optical positions from SDSS data \citep{SDSS_stars_rass} for red and blue periods 
  (left panel),
  and similarly with respect to 
  the 2RXS catalogue (right panel). 
              }
             \label{Fig_offset6}
   \end{figure}

The correlation results with the RBS and SDSS catalogues are 
summarised in Table~\ref{tab:astrom:rbs}. 
The positional offsets for RASS-2 (1RXS) and RASS-3 still show a split depending 
on scan direction, indicating that there is a remaining component related to an 
uncorrected time delay shift.
As this time delay was only given with two decimals for 1RXS, this corresponds to an
uncertainty of $\pm 5$\,ms ($\sim \pm1$\arcsec).
Therefore, a further shift in $\beta$ beyond the pure rounding error adjustment 
is justified, and we applied a final $\pm 3.70$\arcsec\ shift
depending on scan direction.
A 3.70\arcsec\ positional offset in scan direction corresponds to a star tracker 
time delay of 2.532\,s 
(using the satellite rotation period of 96 minutes to scan a full great circle)
compared to the previously implemented 2.53\,s.
The final offset histograms for point-like RBS sources and X-ray stars in SDSS 
are shown in Figs.~\ref{Fig_offset5} and \ref{Fig_offset6}, respectively.

For the 2RXS catalogue production we have shifted the sky positions for each 
individual event in ecliptic latitude (scan direction) and created new photon 
event tables. We repeated the source detection procedure with the same 
parameters as for the original RASS-3 processing. We made the new event tables 
publicly available and refer to the final corrected files as RASS-3.1 photon 
event tables. 

\begin{table*}
\caption{
Shift and separation of RASS X-ray and optical positions 
(from RBS and SDSS cross-correlation catalogues)
in ecliptic latitude $\beta$ 
in arcsec depending on ecliptic longitude (i.e. scan direction).
The table is described in Sect.~\ref{sec:abs_astro_new}.
}\label{tab:astrom:rbs}\label{tab:astrom:sdss}
\begin{tabular}{c|cccc|c}
\hline\hline%
   & RASS-2          & RASS-3          & rounding error        & RASS-3.1      & comment \\
   & (1RXS)          &                 & re-correction         & (2RXS)        & (reference)  \\
\hline
RBS &\quad $0.894 \pm0.418$\quad  &\quad $0.861 \pm 3.401$\quad &\quad $0.630 \pm 0.984$\quad &\quad $0.573 \pm 0.539$\quad &
 only point-like \,\citep{Schwope2000}  \\
SDSS&\quad $0.615 \pm0.551$\quad  &\quad $0.663 \pm 3.136$\quad &\quad $0.430 \pm 0.734$\quad &\quad $0.630 \pm 0.327$\quad &
 only northern sky \,\citep{SDSS_stars_rass} \\
\hline
\end{tabular}
\tablefoot{
\tablefoottext{a}{For the RASS-2 column, for instance, the value
of 0.894 is the mean 1RXS-RBS distance in arcsec. 
The value of 0.418 refers to the
systematic separation between the 1RXS and RBS optical positions 
in arcsec depending on scan direction. 
The same holds for
columns RASS-3 and RASS-3.1.
The large separation of more than 3 arcsec 
for RASS-3 arises because only one of two compensating errors
present in RASS-2 had been corrected.
Applying an {\em \textup{a posteriori}} rounding error re-correction (attitude time shift)
reduces the separation significantly.
We have applied an additional correction to minimise the the X-ray -- optical offsets
(column RASS-3.1, as in 2RXS, and new public event files).
}
                 }
\end{table*}

\begin{figure}[htp]
  \centering
  \includegraphics[angle=-90,width=89mm,clip=]{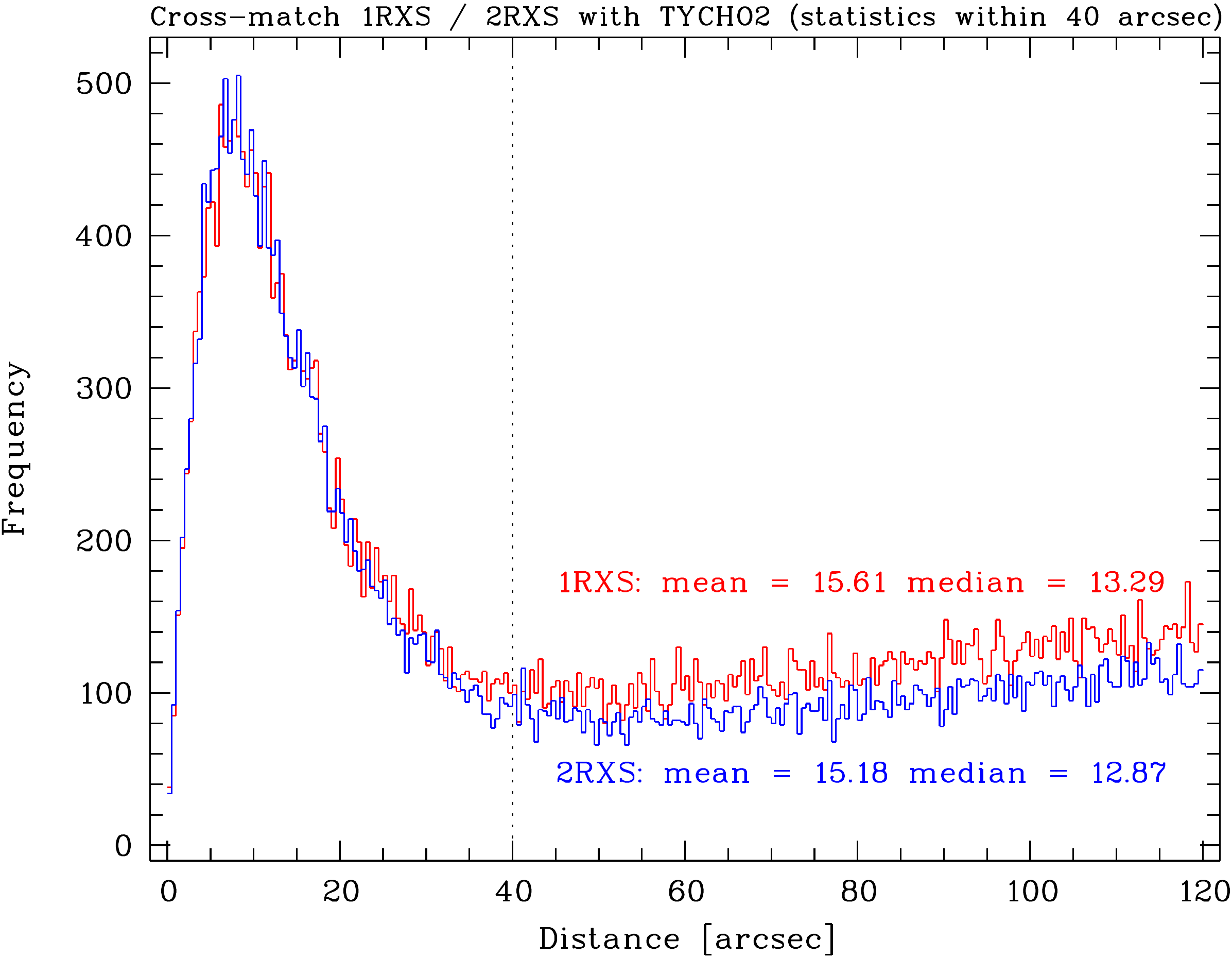}
  \caption{
  Distribution of the angular separation of 
  1RXS (red)
  and 2RXS (blue)
  sources from nearest Tycho-2 catalogue entries.
  Statistical values for mean and median are computed below 40 arcsec
  (chance coincidences start to dominate above 40 arcsec, 
  but have not been subtracted here). 
  No selection on Tycho-2 positional errors has been performed in this plot.}
   \label{xmatch_tycho2_1rxs_2rxs}
\end{figure}

\begin{figure}[htp]
  \centering
  \includegraphics[bb=98 312 515 730,angle=0,width=0.408\textwidth,clip=]{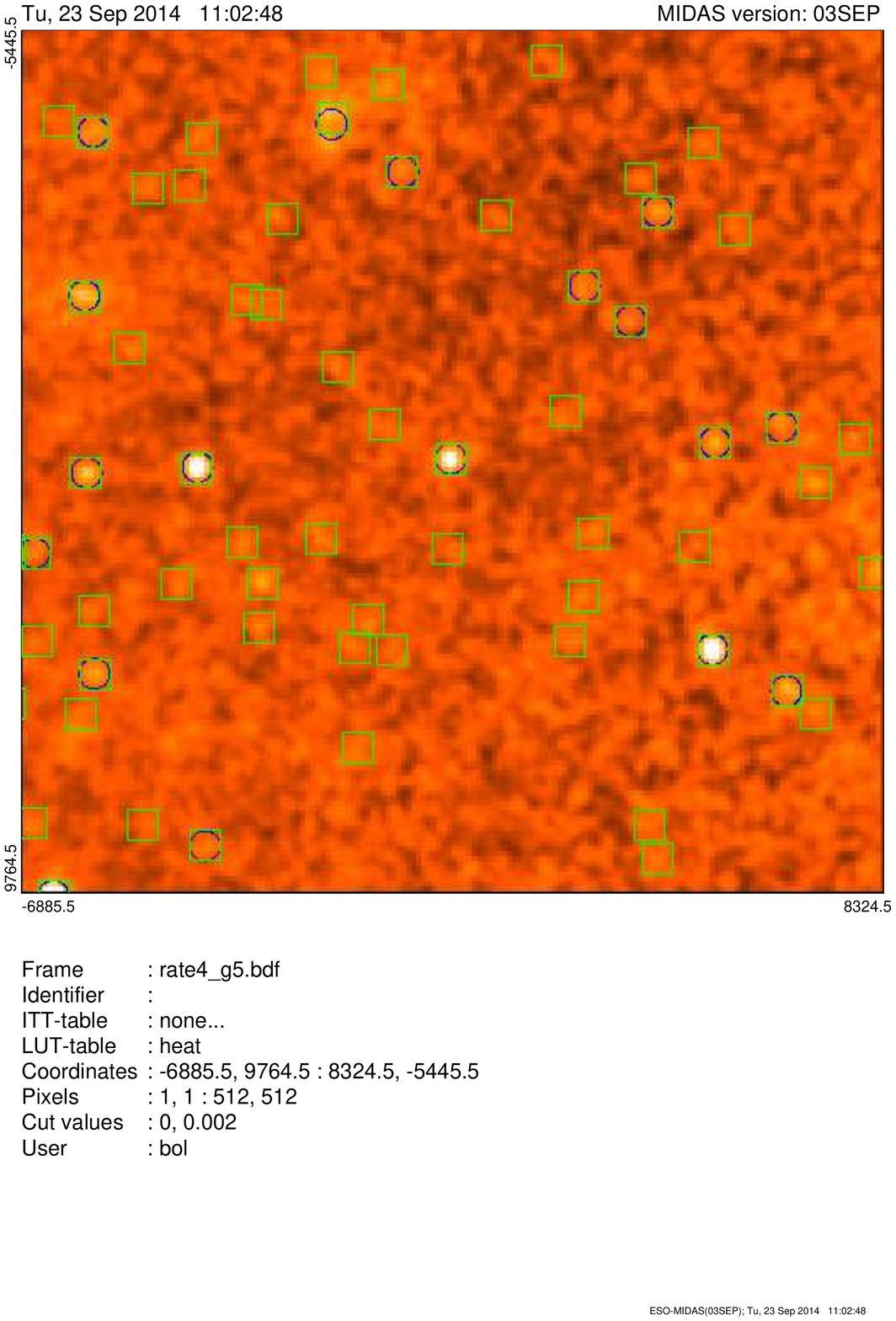}\vspace*{0.62mm} 

  \includegraphics[bb=98 312 515 730,angle=0,width=0.408\textwidth,clip=]{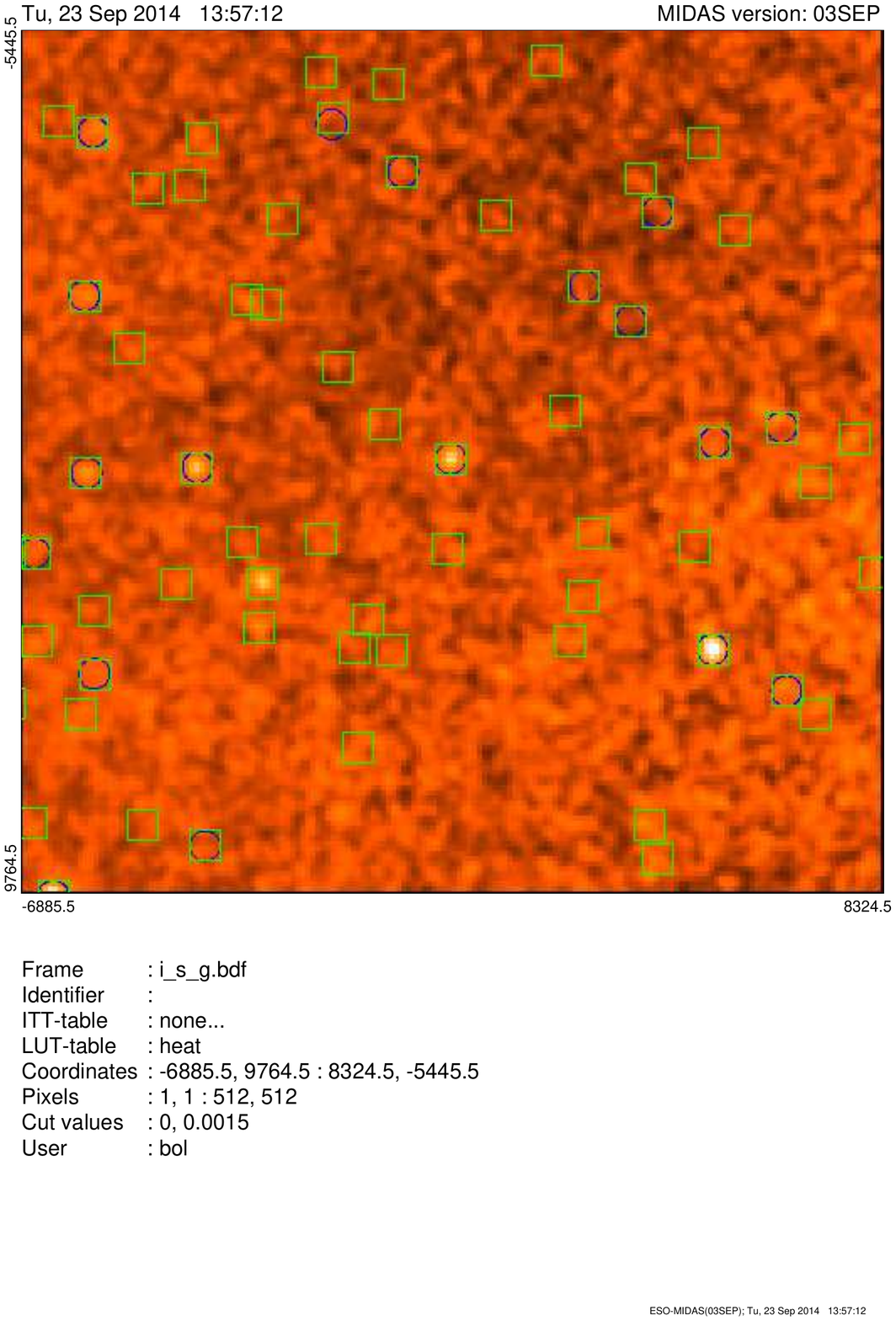}\vspace*{0.62mm} 

  \includegraphics[bb=98 312 515 730,angle=0,width=0.408\textwidth,clip=]{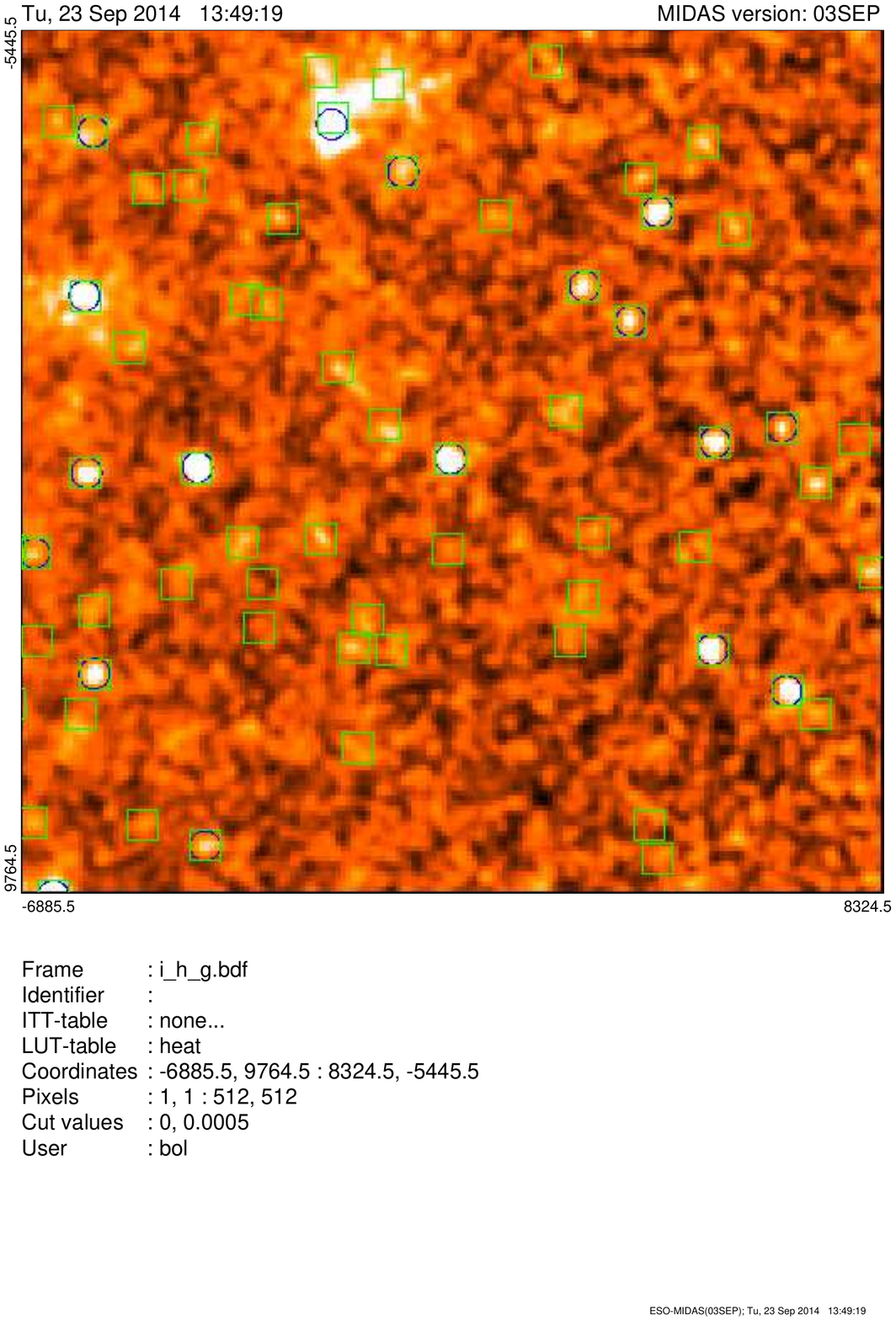}\vspace*{0.62mm} 

     \caption{NEP analysis by \citet{Henry} compared to the 2RXS sources for
              field 930521. 
               Shown are the broad- (channels 11-235), soft- (11-41), and hard- (52-201) band images from top to bottom.
               All sources from the catalogue of \citet{Henry} are marked with blue circles and are also detected
               in our analysis (green squares). 
              }
             \label{Fig_Henry}
   \end{figure}

\subsubsection{Correlation with Tycho-2}\label{sec:tycho_correlation}

The ROSAT survey sources were correlated with the stellar Tycho-2 catalogue
\citep{Hog1998} to compare absolute
positional accuracies of 1RXS and 2RXS detections.
In Fig.~\ref{xmatch_tycho2_1rxs_2rxs} we show the distributions of
angular separations for
1RXS (red) and 2RXS (blue).
The significantly higher tail of 1RXS with respect to 2RXS might be interpreted
as more spurious sources and/or greater positional errors for
sources with larger separations for Tycho-2 counterparts. 
For distances $\leq 40$\,arcsec, the mean and median values for 1RXS and 2RXS 
are comparable (mean 15.6 and 15.2, median 13.3 and 12.9, respectively),
with 2RXS values being slightly more accurate.

\section{Comparison with other %
        source detections in deep ROSAT exposure regions}\label{sec:Henry_detections}

The comparison of different source detection algorithms is of importance 
to evaluate the quality of the 2RXS source detections, especially in the 
Ecliptic Pole regions, which have the deepest exposures of the entire ROSAT all-sky survey. 
\citet{Henry} performed such an analysis and produced the deepest large 
solid-angle contiguous sample of 442 X-ray sources in a 80.6 deg$^2$ region around
the North Ecliptic Pole (NEP).
In Fig.~\ref{Fig_Henry} we compare the sources detected in the NEP region by 
\citet{Henry} with our 2RXS sources. 
The figure shows, from top to bottom, the images of broad, soft, and hard energy bands 
(see Sect.~\ref{sec:SourceDetectionBands} for the definition of the energy bands). 
We use these three energies in the selected images because  this
lets us evaluate better the detection of soft and hard 
2RXS sources.  
All sources from the analysis of \citet{Henry} are also detected
as 2RXS sources. As the 2RXS detection limit extends down to an existence likelihood 
value of 6.5 \citep[in contrast to the limit of 10 used by][]{Henry}, we find
additional, weaker sources, which are shown as green squares.

\section{Cross-matches of the 2RXS sources}\label{sec:CrossMatches}

We performed spatial cross-correlations with various other catalogues. 
Using simply the nearest neighbour to the X-ray position within 1\arcmin\,, 
we point out that these cross-correlations do not always provide the most likely 
identification of the X-ray source, but reveal only potential counterparts.
For the cross-correlations we included the following X-ray source catalogues: 
1RXS \citep{Voges1999},
2RXP \citep{2RXP}, 
3XMM \citep{2015arXiv150407051R},
XMMSL1 \citep{XMMSL1},
1SXPS \citep{2014ApJS..210....8E},
and object lists 
for active galactic nuclei \citep{VV10},
for stars \citep[Tycho2,][]{Tycho2},
the bright star catalogue \citep[BSC,][]{BrightStar}, 
a catalogue of low- and high-mass X-ray binaries \citep{LHMXRB}, 
a catalogue of high-mass X-ray binaries \citep{HMXRB},
a pulsar catalogue \citep{pulsar}, and
the catalogue of variable 1RXS sources \citep{FUH}.
The nearest counterpart in the cross-matching catalogues to a 2RXS source is listed in 
our catalogue.
A more detailed and sophisticated source identification work is beyond the scope of 
this paper and will be presented in a paper by Salvato et al.\ (in prep). 

\begin{figure}[htp]
  \centering
  \includegraphics[angle=-90,width=89mm,clip=]{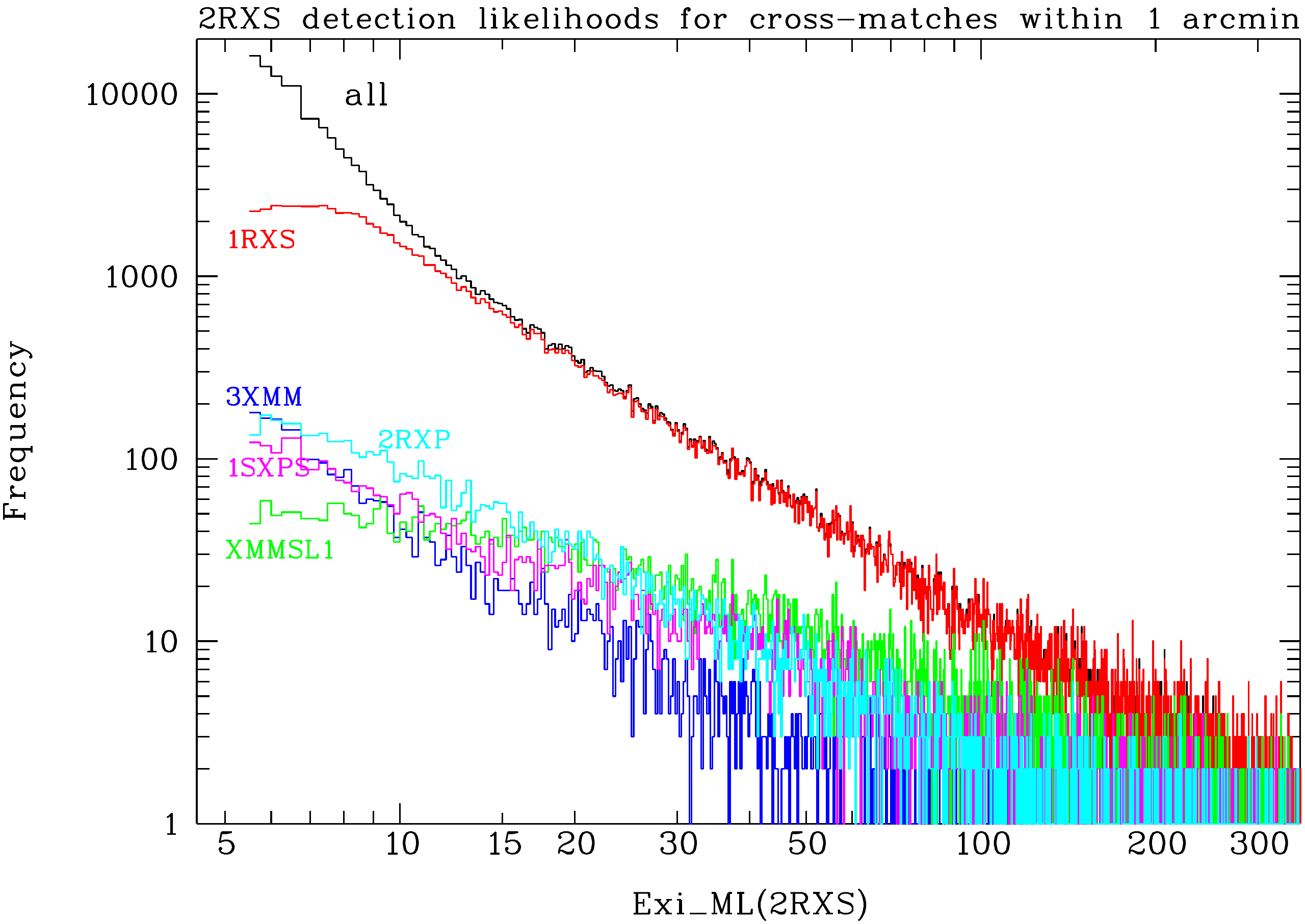}
  \caption{%
     2RXS detection likelihood distributions 
     for all detections (black) and for purely spatial
     cross-matches within 60 arcsec
     with the X-ray catalogues 1RXS (red), 2RXP (cyan), 
            3XMM\_dr4 (blue), XMMSL1\_dr6 (green), and 1SXPS (magenta), 
            with a bin size of 0.25 of {\tt EXI\_ML}.
    }
   \label{plot_eximl_2rxs_match}
\end{figure}

In Fig.~\ref{plot_eximl_2rxs_match} we illustrate the performed cross-matches 
with 
the catalogues listed above
as a function of the 2RXS detection likelihood.
At high 2RXS detection likelihoods these (usually bright) sources should almost always be
detected in other X-ray catalogues, if they have been spatially covered (unless they show
strong variability or an extremely different spectrum).
The XMM slew survey catalogue XMMSL1 has the highest matching fraction 
at high EXI\_ML values because it has the highest spatial coverage.
At low EXI\_ML values the sensitivity of XMMSL1 is generally too low to detect
many of the 2RXS sources, and the matching fraction curve flattens.
The 3XMM catalogue has the smallest spatial coverage but is also deepest,
and therefore the curve is steepest.
The comparison with the 1RXS catalogue shows 
that it becomes incomplete toward the faint end with respect to 2RXS.
This is explained in Sect.~\ref{sec:eximl_redistribution} and is related to the
width of the EXI\_ML distribution.

\section{2RXS catalogue properties}\label{sec:2RXS_properties}

\subsection{Sky distribution in count rate and hardness ratio}\label{sec:skydistribution}

\begin{figure*}
  \centering
  \includegraphics[width=18cm,clip=]{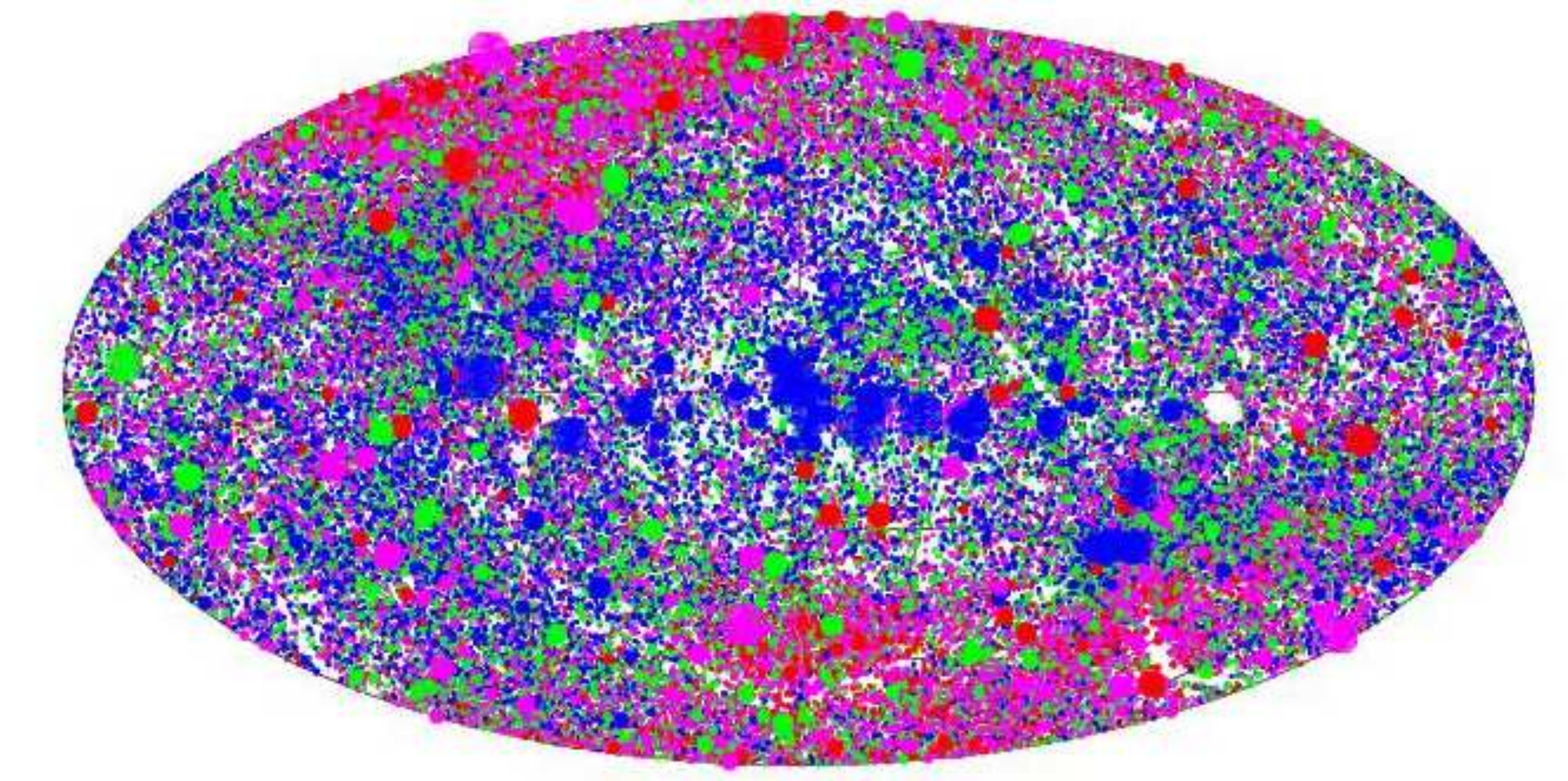}
  \caption{Aitoff projection in Galactic coordinates of the sky distribution of 2RXS sources 
  with HR1 errors smaller than 0.5. 
  The size of the data 
  points scales with the source count rate, and the colour represents the
  HR1 value. Sources in the
  hardness ratio intervals
  $\rm HR1=[-1.0,-0.5]$ are indicated in red,
  $\rm HR1=[-0.5,0.0]$ in magenta,
  $\rm HR1=[0.0,+0.5]$ in green, and
  $\rm HR1=[+0.5,+1.0]$ in blue.
  }
              \label{Fig_sky_distribution}
\end{figure*}

The sky distribution of the 2RXS sources is shown in Fig.~\ref{Fig_sky_distribution}.
The size of the symbols scales with count rate\footnote{using the hyperbolic tangent 
function $\tanh(x) = (e^x - e^{-x})/(e^x + e^{-x})$}, covering values
between 0.001 and 68 counts s$^{-1}$, while 
the colours represent different hardness ratio ranges
(for the definition of hardness ratios see Appendix~\ref{sec:SourceDetectionBands}).
Red sources indicate soft and super-soft sources, often characterised with 
steep X-ray photon indices, while blue sources mark
hard sources, typically with flat X-ray photon indices.
We only show the 88,586 objects whose errors on the hardness ratio 1 (HR1) are smaller than 0.5.
The faintest sources are detected in the North and South Ecliptic Pole regions, 
where the exposure time is longest.

\subsection{General properties}\label{sec:GeneralProperties}

In this section we discuss the distributions of the source count rates, 
the source counts, the existence
likelihood, and the exposure time of the
\TotalNumberDetections\   
2RXS sources. 
The distributions of existence likelihoods, count rates, and the source counts
are given in Fig.~\ref{Fig_EXIML}. 
The distribution of source counts 
ranges between 3 and 35,033 counts.
The distribution of the existence likelihood ranges between 6.5 and 26,198. 
Exposure times range between 7\,s and 39,214\,s.

\subsection{Timing properties}\label{sec:Timing}

In this section we compare the count rates of the 2RXS sources with those from 
ROSAT pointed observations, the XMM-Newton slew survey, and the 3XMM source catalogue. 
Additionally, we discuss the 2RXS source variability during the survey scans. 
We list the sources with most interesting timing properties based on our 
light curve analysis. A complete and detailed analysis is beyond the scope of this paper.

\subsubsection{2RXS versus ROSAT pointed observations}\label{sec:2RXS_ROSAT_pointings}

In Fig.~\ref{Fig_2RXS_2RXP_XMMSLEW_3XMM} we compare the mean count rates of 
sources both detected in 2RXS and in ROSAT PSPC pointed observations (2RXP).
This plot illustrates the degree of variability between the ROSAT survey and 
pointed observations. 
Count rate variations by more than a factor of 100 are found. 
We list the sources whose count rate 
variations between survey and pointed observations exceed a factor of 50 in Table~\ref{tab:2rxs_2rxp_50}.
This is expected because of the longer exposures and therefore higher sensitivity 
of the pointed observations. 
This is also seen in Fig.~\ref{Fig_2RXS_2RXP_XMMSLEW_3XMM}, where lower count 
rates are reached in the pointed observations.

\subsubsection{2RXS versus XMM-Newton slew survey}\label{sec:2RXS_XMMSLEW}

Here we compare the 2RXS count rates with count rates of XMM-Newton slew survey counterparts (XMMSL1). 
Although the XMM-Newton slew survey and the ROSAT survey observations
have similar overall sensitivities in the 0.5$-$2\,keV energy range \citep{XMMSL1}, 
the energy dependence of the effective areas is different between ROSAT and the EPIC-pn 
instrument of XMM-Newton. 
Assuming a power-law spectral model with photon index 1.7 and observed X-ray absorption of 
3$\times$10$^{20}$ cm$^{-2}$ \citep{Watson2009}, %
a factor of 8.38 higher count rate is expected for EPIC-pn. 
We have therefore used the XMM-Newton slew conversion factor of
8.38 to convert 2RXS count rates into 2RXS scaled count rates.
We compare the 2RXS scaled count rates with count rates 
for the XMM-Newton slew survey objects (EPIC-pn, medium filter, 0.2$-$2\,keV band) in Fig.~\ref{Fig_2RXS_2RXP_XMMSLEW_3XMM}.
Outliers in Fig.~\ref{Fig_2RXS_2RXP_XMMSLEW_3XMM}, far away from this relation, are 
candidates for high variability.
The deviation from the one-to-one line arises because XMMSL1 has a lower sensitivity than 2RXS, which is of the order of about one $\rm counts\ s^{-1}$,
while the 2RXS count rates decrease to about a few $\rm 10^{-3}\ counts\ s^{-1}$.
This is also obvious from Fig.~\ref{plot_eximl_2rxs_match}, where the XMMSL1 is shown with the green line, 
becoming incomplete at low detection likelihood values. 

\subsubsection{2RXS count rates versus 3XMM fluxes}\label{sec:2RXS_3XMM}

\begin{figure}
  \centering
  {\includegraphics[angle=-90,width=85mm,clip=]{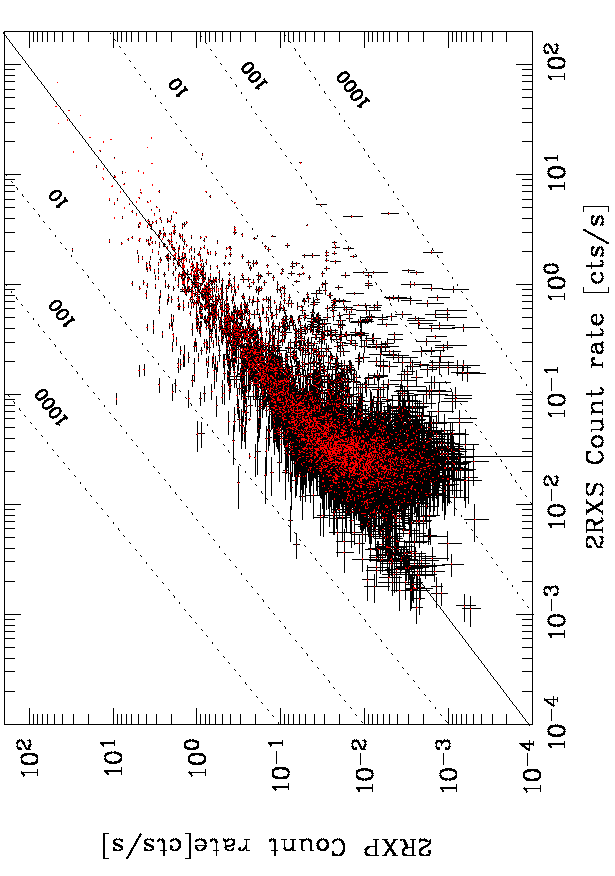}}
  {\includegraphics[angle=-90,width=104mm,clip=]{pl_2RXS_xmmsl1_08122015.png}}
  {\includegraphics[angle=-90,width=104mm,clip=]{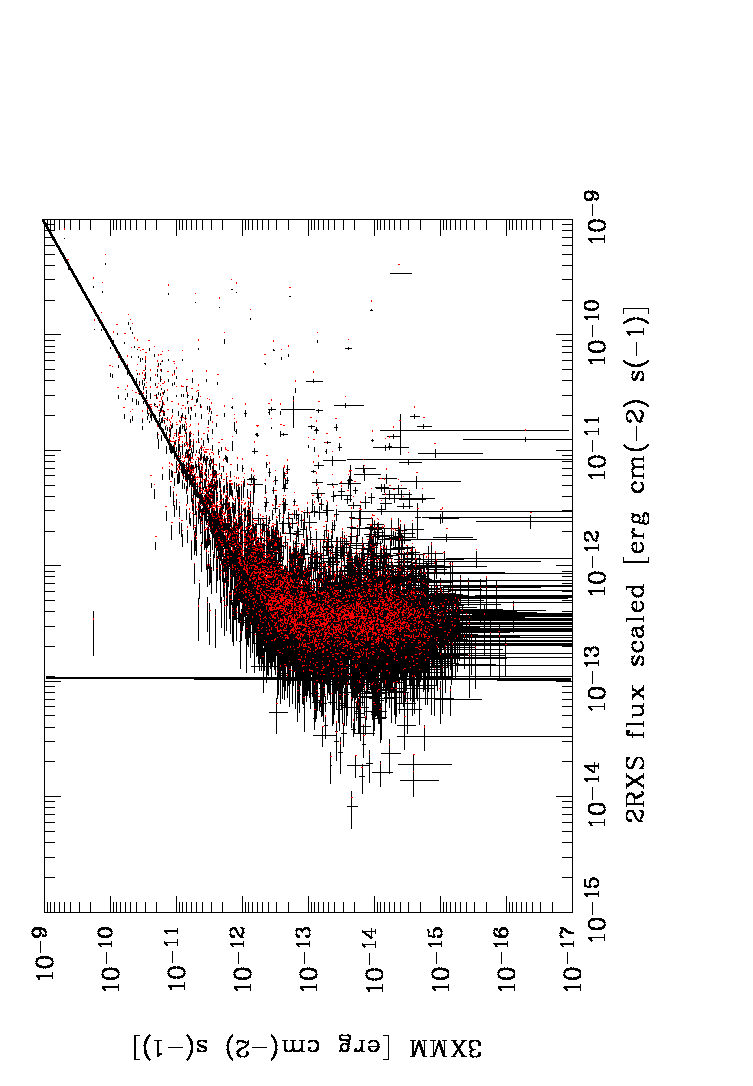}}
  
  \caption{Count rates of ROSAT 2RXS sources versus count rates derived from 
           2RXP ROSAT pointed observations (top), 
           XMM-Newton slew survey count rates (middle), and
           3XMM fluxes (bottom). 
           For the assumed power-law model and derived conversion factors see text.
            }
   \label{Fig_2RXS_2RXP_XMMSLEW_3XMM}
   \end{figure}

To convert 2RXS count rates into fluxes (assuming 
the power-law model), a factor of 1.08$\times$10$^{-11}$ erg cm$^{-2}$ needs to be applied.
In addition to the different effective areas, the flux limits are very different
for 2RXS sources and the deeper 3XMM pointed observations. 
The 2RXS flux limit is about $10^{-13}$ erg cm$^{-2}$ s$^{-1}$ 
, indicated 
by the vertical line in Fig.~\ref{Fig_2RXS_2RXP_XMMSLEW_3XMM},
while the 
3XMM flux limit is much deeper with about $10^{-16}$ erg cm$^{-2}$ s$^{-1}$. 
Most of the 
faint 3XMM sources near
the ROSAT flux limit have no real 
counterparts in the 2RXS catalogue and are spurious chance associations. 
This is shown in Fig.~\ref{Fig_2RXS_2RXP_XMMSLEW_3XMM}, where 
the correlation is clearly no longer linear near the ROSAT flux limit.
Most of the 3XMM counterparts have fainter fluxes (below the one-to-one line) because the
3XMM count rate limit is lower than that of 2RXS.

\subsubsection{Source variability during ROSAT survey scans}\label{sec:variab}

To characterise the temporal behaviour of the 2RXS sources, we have calculated 
the normalised excess variance with its uncertainty and the maximum amplitude 
variability during the survey scans.
\noindent
The light curves were background subtracted as described in 
Sect.\ \ref{sec:lightcurves}.
We automatically searched for cases where the net count rate decreases for up to
three data points to low values, mostly below $\sim 1\,\hbox{\rm counts s}^{-1}$, 
caused by an increase in the background count rate.
Most likely, events of strong solar activity or an increased particle background 
are responsible for this.
Because the extraction for source and background events is not simultaneous
(Sect.\ \ref{sec:lightcurves}), short background flares are not always subtracted properly.

The normalised excess variance is a powerful and commonly used method to determine the
probability that a time series shows variability above a certain threshold 
\citep[e.g.][]{Nandra1997,Vaughan2003,Ponti2004}.
For the 2RXS sources we
have calculated the normalised excess variance with the formula
$$
\sigma^2_{rms} = \frac{1}{N \mu^2} \sum_{i=1}^N[(X_i - \mu)^2 - \sigma^2_i] 
$$
and the uncertainty
$$
\left(\Delta \sigma^2_\mathrm{rms}\right)_\mathrm{meas} = 
\sqrt{ \left(\sqrt{\frac{2}{N}} \frac{ <\sigma^2_i> }{\mu^2}\right)^2 + \left( \sqrt{\frac{<\sigma^2_i>}{N} \frac{2F_\mathrm{var}}{\mu}} \right)^2 }
$$
where $\rm F_{var} = \sigma_{rms} / \mu$ is the fractional variability,
$N$ is the number of time bins, $\mu$ is the 
weighted (by exposure time)
mean of 
the count rates, $\rm X_i$ and $\sigma_i$ are the count rates and the 
corresponding uncertainties. 
The weighted mean was used because during the survey scans some time intervals exhibit
low exposure times (of about 10 seconds) 
that are due to periods of high background, resulting in large 
errors bars in the data points in the survey light curves. With the unweighted mean
the calculation of the mean count rate would result in incorrect values for the
mean count rate. 
In combination with the uncertainty of the normalised excess variance, 
the ratio of both quantities gives the probability that
a 2RXS source is time variable in units of Gaussian $\sigma$.
For 0.9 per cent of the objects we find extremely short exposure times
(shorter than or equal to 6 seconds) for the data points in the survey light curve.  
We have flagged these light curves and objects. 
\citet{FUH}  have performed a systematic study of X-ray variability in the 
ROSAT all-sky survey for 1RXS sources in the BSC and FSC with an existence likelihood greater than or equal to 15. 
2RXS sources with X-ray variability significance values above $10\,\sigma$, but
not listed \citet{FUH}, are given in Table~\ref{tab:2rxs_exc_10}. 
2RXS sources with significance values above $20\,\sigma$ are listed in Table~\ref{tab:2rxs_exc_20}. 
These sources are listed in \citet{FUH}, but they are not shown 
as graphical representations. 
An example is shown in Fig.~\ref{excess_example}.

  \begin{figure}
  \centering
  \includegraphics[angle=-90,width=105mm,clip=]{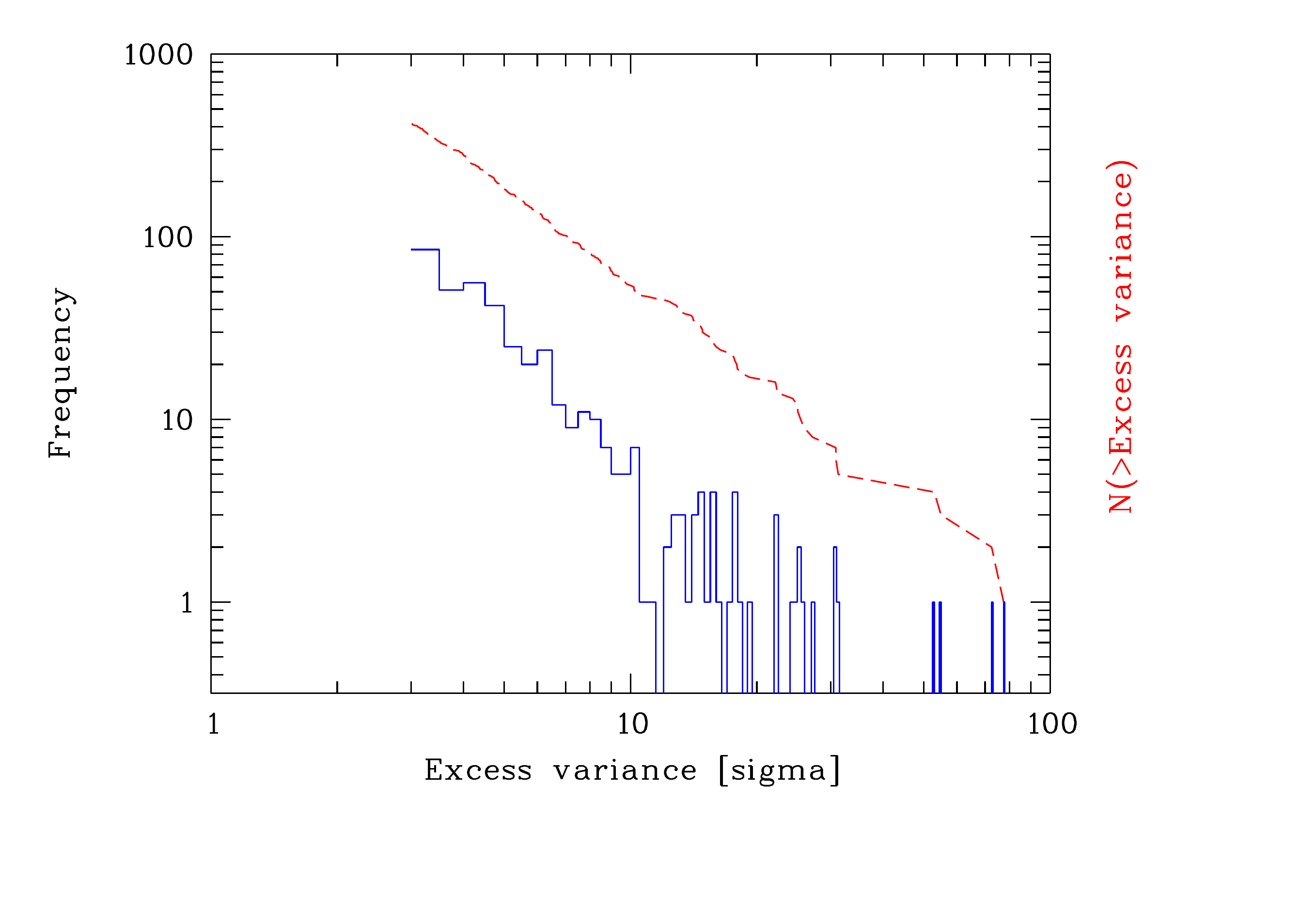}
  \caption{Histogram of sources with excess variance values above $3\,\sigma$.
           The differential distribution is shown as the blue histogram. 
           The solid red-dashed lines delineate the integral distribution.}
   \label{Fig_excess_variance}
   \end{figure}

  \begin{figure}
  \centering
  \includegraphics[angle=-90,width=90mm,clip=]{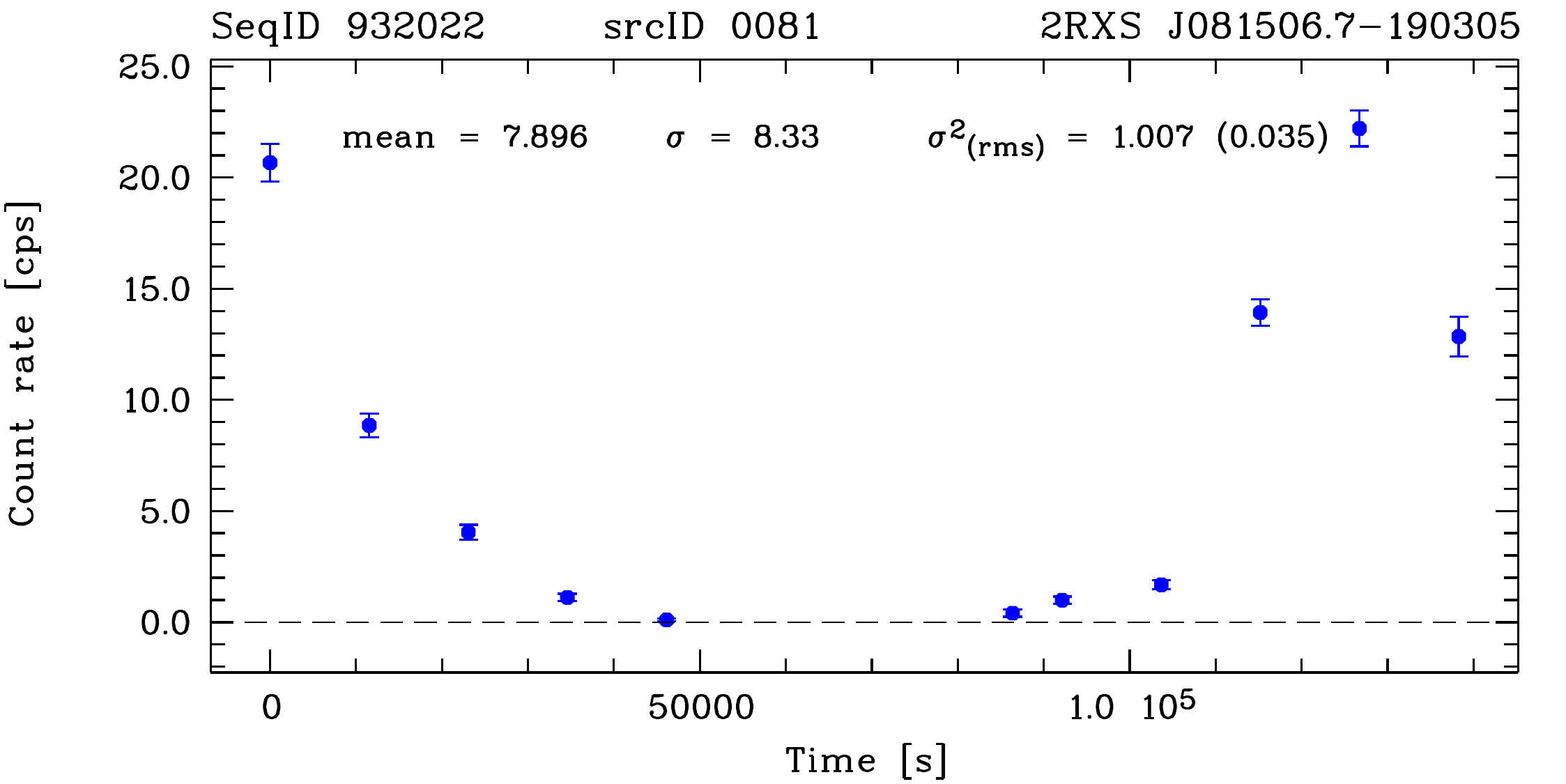}
  \caption{Example for a source with excess variance above the $3\,\sigma$ limit.
           The source is known as VV Pup, an AM CV with an orbital period of about 100 minutes.
           As the ROSAT light curve samples close to the intrinsic period, 
           we see an aliasing effect caused by the convolution of the true variation 
           and the window function.
}
   \label{excess_example}
   \end{figure}

\begin{figure}
  \centering
  \includegraphics[angle=-90,width=105mm,clip=]{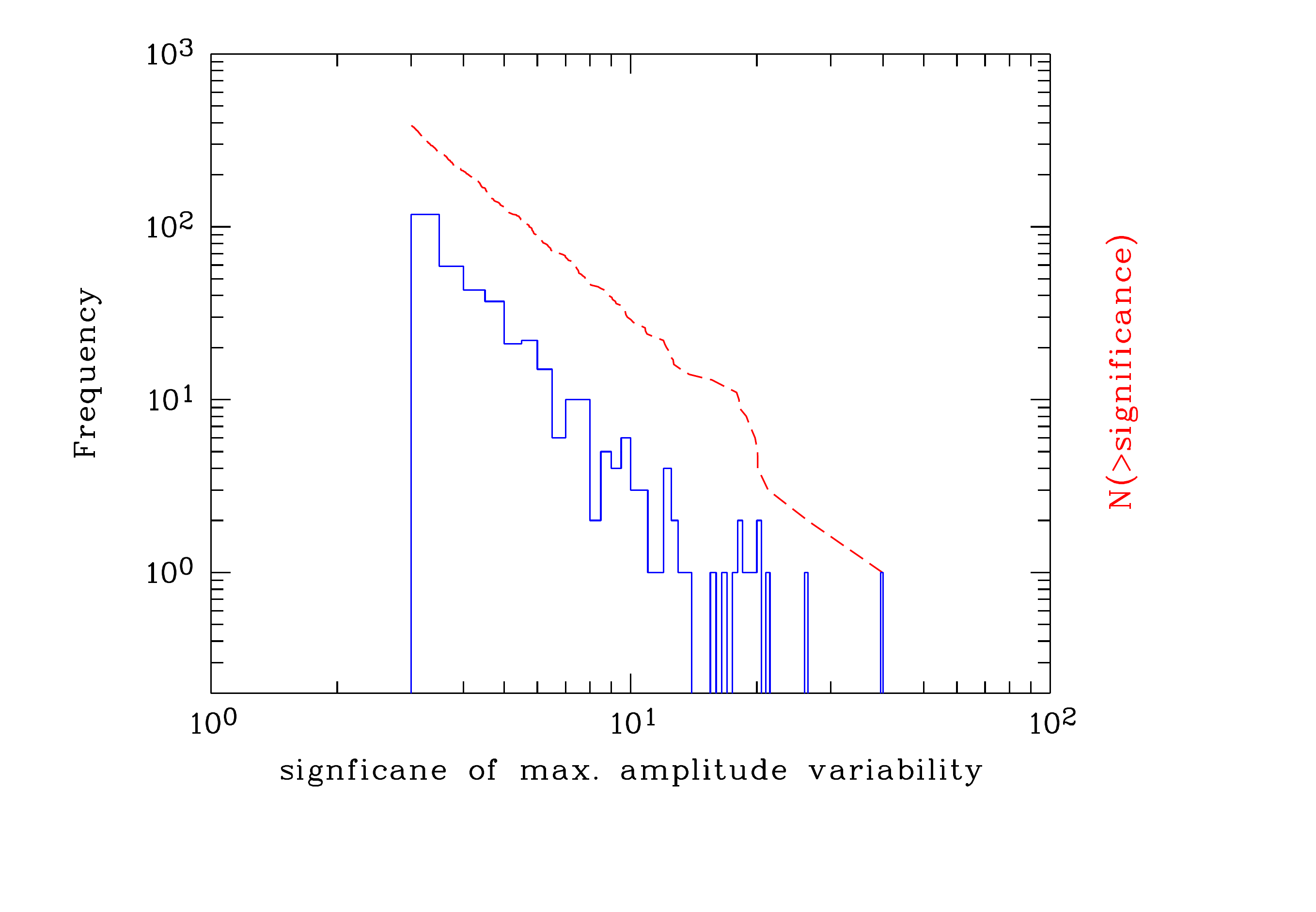}
 \caption{Distribution of the significance values of the maximum amplitude variability for 2RXS above the  3\,$\sigma$ limit.
           The differential distribution is shown as the blue histogram. 
           The solid red-dashed line delineates the integral distribution.}
   \label{Fig_ampl_max}
   \end{figure}

\begin{figure}
  \centering
  \includegraphics[angle=-90,width=90mm,clip=]{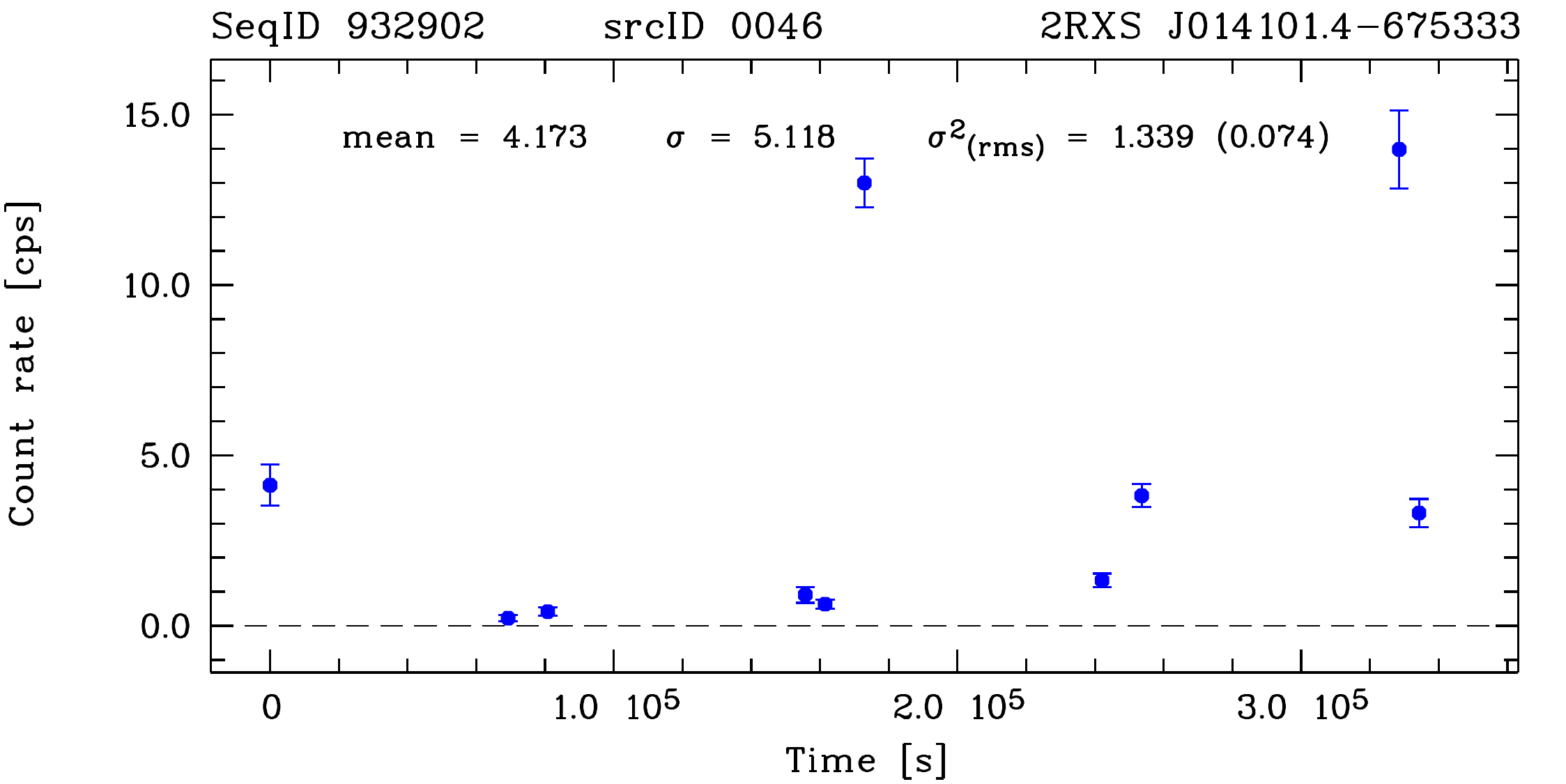}
  \caption{Example for a source passing the maximum amplitude variability test.}
   \label{max_example}
   \end{figure}

In addition to the normalised excess variance, which describes the variability of a survey light curve
as a whole, we calculated the maximum amplitude variability to search for flaring events during
the survey observations. 
To calculate the maximum amplitude variability $ampl\_max$ and its significance $ampl\_sig$ ,
we used the maximum count rate
$cmax$, the corresponding error $cmax\_{err}$, the minimum count rate $cmin,$ and its 
corresponding error $cmin\_{err}$ for each survey light curve. 
The maximum amplitude variation and its significance is then
\begin{eqnarray}
ampl\_max &\!=\!& (cmax-cmax\_{err})-(cmin+cmin\_{err}) \nonumber \\
ampl\_sig &\!=\!& ampl\_max / \sqrt{cmin\_{err}^2 + cmax\_{err}^2} \nonumber
\end{eqnarray}
In Table~\ref{tab:2rxs_amp_10} we list objects with significance for the maximum amplitude variability above  
10\,$\sigma$.  Figure~\ref{Fig_ampl_max} shows the distribution of sources with
maximum amplitude variability above the 3\,$\sigma$ limit.
An example is shown in Fig.~\ref{max_example}.

A variability test was applied to the 3XMM catalogue by the authors  using a $\rm \chi^2$ test. 
Sources with a probability lower than $\rm 10^{-5}$ of being constant were flagged as variable 
sources.
%
%
%
About 30,000  of the 151,524 1SXPS
sources are classified as variable sources.

\subsection{Spectral analysis}\label{sec:SpectralProperties}

\begin{figure}
\centering
\includegraphics[angle=-90,width=90mm,clip=]{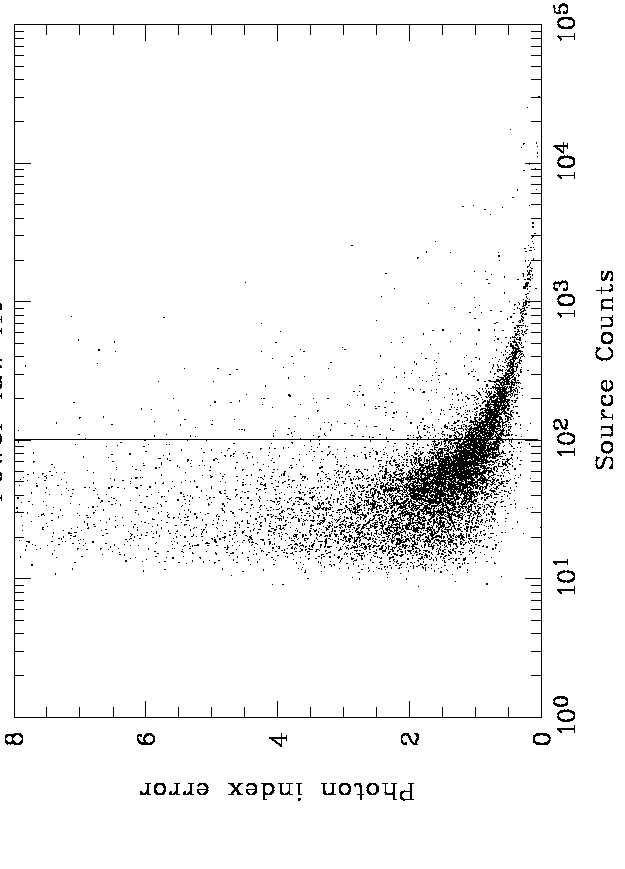}
\includegraphics[angle=-90,width=90mm,clip=]{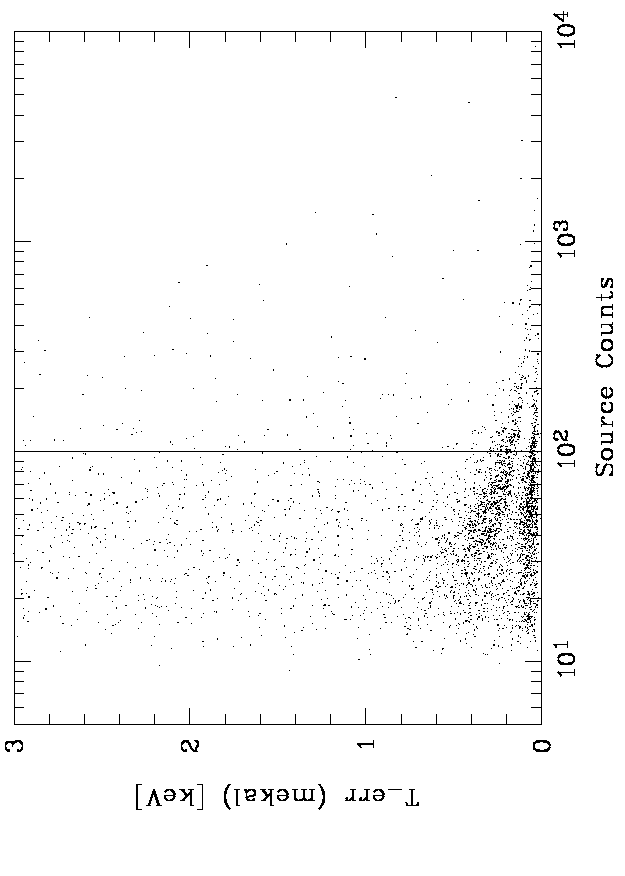}
\includegraphics[angle=-90,width=90mm,clip=]{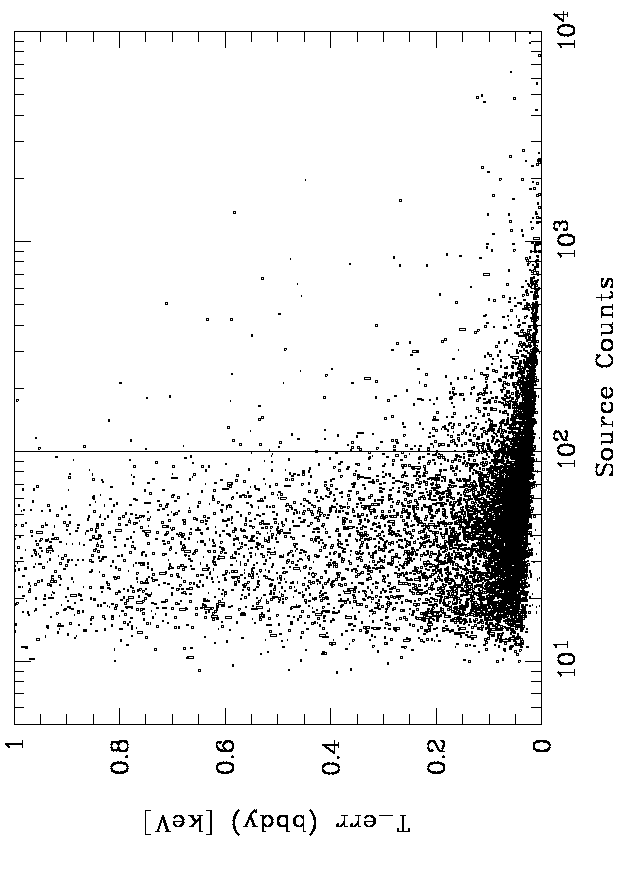}
 \caption{
 Distribution of the source counts in relation to the error of the
 photon index, the temperature error derived from the mekal, and the temperature error from the black-body fits (from top to bottom). 
 For all three spectral models the errors for fewer than 100 source counts are broadly distributed. Therefore, we applied a limit of at least 100 source counts for the spectral fit results presented 
 here.  
 }
             \label{Counts_100_all_fits}
\end{figure}

We have performed spectral fits using three different models:
(i) a power law ({\it powl}), 
(ii) an optically thin plasma emission model ({\it mekal}),
and (iii) an optically thick black-body model ({\it bbdy}). 
Spectral fitting was performed in EXSAS \citep{Zimmermann1998}.
Absorption by neutral gas with solar elemental abundances 
\citep{1983ApJ...270..119M}
was included for all models.
From the fits, we stored the absorbing hydrogen column 
density ($N_{\rm H}$), {\it powl} photon index, {\it mekal,} and {\it bbdy} temperatures and
model normalisations together with the corresponding errors, reduced $\chi^2_{\rm red}$ value, 
$\chi^2$ value, the number of data points used in the fit, and the 
number of degrees of freedom. The absorption-corrected flux was 
calculated for all spectral models. We note that the (absorption-corrected) fluxes for a given source for the 
different models can be very different when the $N_{\rm H}$ values differ strongly.
For comparison with the $N_{\rm H}$ values derived from the spectral fits, 
we determined the Galactic
absorption in the direction of each source $N_{\rm H,\,gal}$ following \citet{DickeyLockman}.

To include only spectral parameters in the catalogue that result from 
acceptable fits, we only accepted spectral fits with a reduced $\chi^2$ 
lower than 1.5 and with at least four degrees of freedom.

In Fig.~\ref{Counts_100_all_fits} we show the errors of the principal model parameters - photon index, 
plasma temperature and black-body temperature - as a function of the number of source counts.
In all three plots we find that the parameter errors strongly increase for spectra with fewer than 100 source 
counts. 
Therefore, we applied an additional cut to
require at least 100 source counts in the spectrum.
We note that in all three plots in Fig.~\ref{Counts_100_all_fits}  there are still sources with
large parameter errors, even for large numbers of source counts.
We inspected these spectra and found that they are mainly from highly absorbed sources,
with $N_{\rm H}$ values close to or even above $\rm 10^{22}\ cm^{-2}$. 
In such cases the photon indices and the temperatures can only be poorly constrained in the available narrow 
energy band. However, the information on the $N_{\rm H}$ value is important, which is the reason for providing these fit parameters.  
The number of sources that fulfil the criteria on reduced $\chi^2$, number of degrees of freedom 
(which is practically fulfilled for sources with more than 100 counts), and the minimum number of source counts 
are 2722, 455, and 1769 for the power law, the mekal, and the black-body fits, respectively. 
For the mekal and black-body fits we furthermore required that the error of the temperatures is smaller 
than one-third of the temperature values.
For the power-law fit the limit on source counts is sufficient to constrain the 
photon indices with adequate precision.

\subsubsection{Power-law model}\label{sec:powerlaw}

\begin{figure}
  \centering
  \includegraphics[angle=-90,width=90.5mm,clip=]{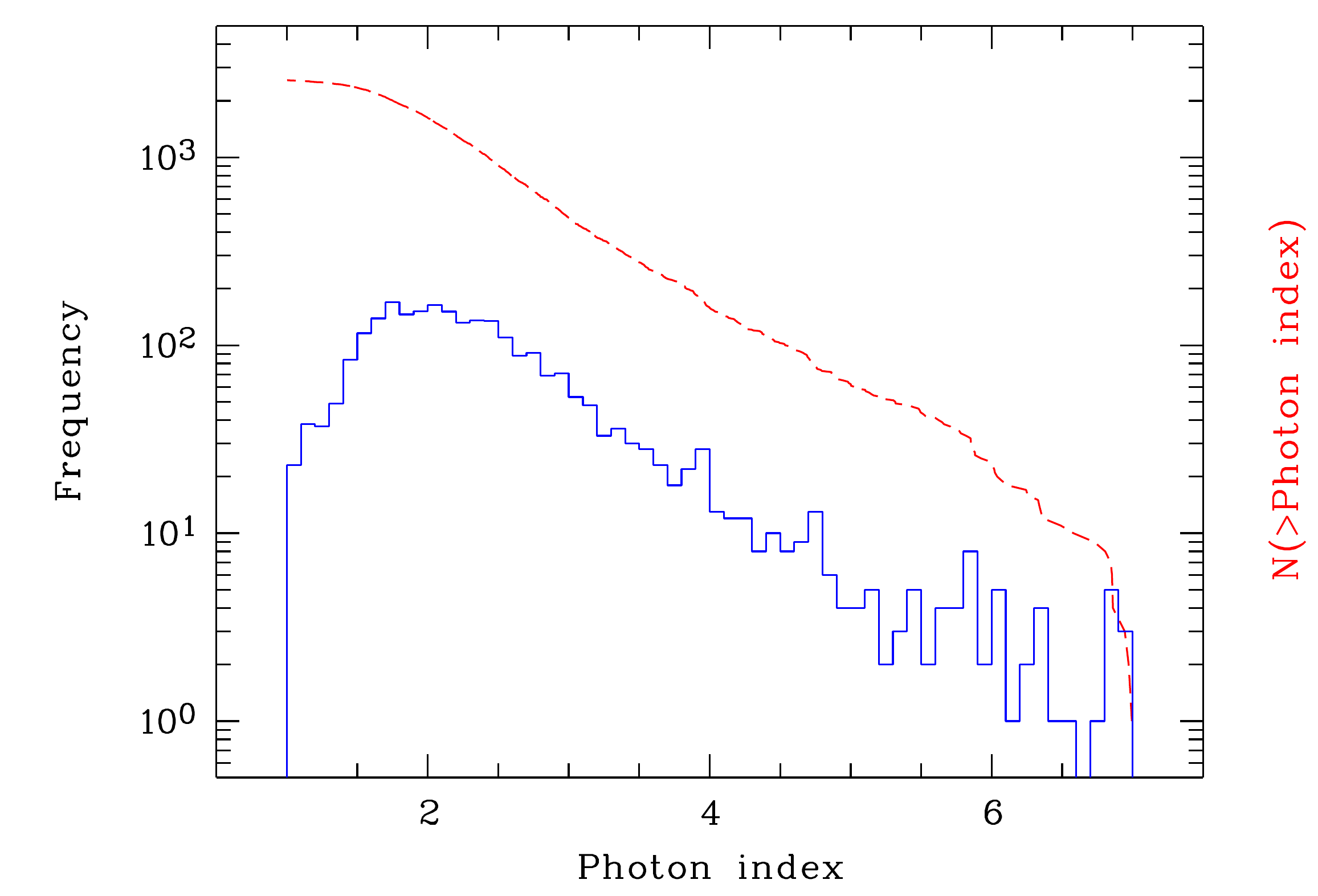}
  \caption{Histogram (blue) of the distribution of photon indices between 1 and 7
   for sources with at least 100 source counts, at least four degrees of freedom 
 and reduced $\chi^2$ values lower than 1.5. 
The red dashed line gives the integral distribution.
     }
   \label{Fig_histo_Gamma}
   \end{figure}

The parameters obtained from a simple power-law fit are the photon index $\Gamma$, the
normalisation parameter, the $N_{\rm H}$ , and their corresponding errors. 
From the spectral fits we calculated the absorption-corrected flux. 
Figure~\ref{Fig_powl_unique} (upper panel) shows an example for a spectral fit with a simple
power-law model for the 2RXS source 
931037\_0107 (Mrk 501).
In Fig.~\ref{Fig_histo_Gamma} we show the distribution of the photon indices.
The histogram peaks at around 2, 
decreasing towards lower and higher photon indices. 

\subsubsection{Plasma-emission model}\label{sec:mekal}

In addition to the power-law fit, we have performed spectral fits for 
optically thin plasma emission using the {\it mekal} model. 
The parameters obtained from these fits are the plasma temperature, 
the  normalisation, and the $N_{\rm H}$ value.
Figure~\ref{Fig_powl_unique} (middle panel) gives an example of a spectral fit with the mekal model.

\begin{figure}
  \centering
  \includegraphics[angle=-90,width=90mm,clip=]{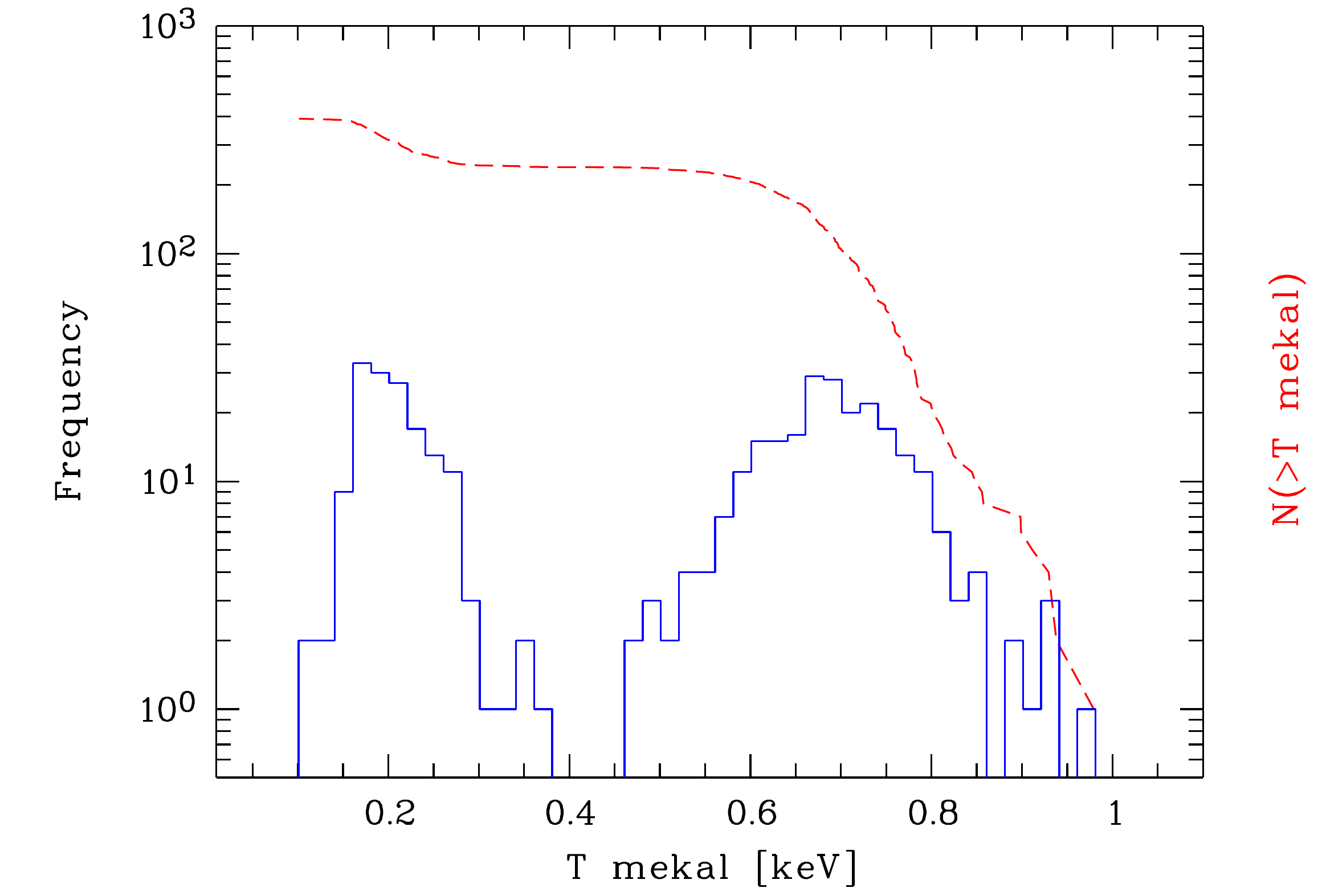}
 \caption{Distribution of the temperatures obtained from the fits using the mekal model
for sources with at least 100 source counts, reduced $\chi^2$ values lower than 1.5, 
at least four degrees of freedom and temperature error smaller than one-third of the best-fit value.
The differential distribution is shown as the blue histogram. The red dashed line gives
the integral distribution.
     }
   \label{Fig_histo_mekal}
   \end{figure}

In Fig.~\ref{Fig_histo_mekal} we show the distribution of the temperatures obtained from 
the mekal fit of 
\citet{Mewe1985}
and a temperature error smaller than one-third of the best-fit temperature.
A bimodal distribution is clearly visible in the temperature, with one peak at around 0.2 keV, and the second peak 
centred
on around 0.7 keV. 
We note that the 0.2 keV peak might be slightly affected by the C-K$\alpha$ absorption edge, which
may not be adequately accounted for by the instrumental corrections, therefore not all of these
objects may have the correct mekal temperature. This was pointed out by \citet{Prieto1996}. 
The second peak at around 0.7 keV is also found in XMM-Newton data for optically thin diffuse emission in 
 NGC 6240, for instance \citep{Boller2003}. 
Similar plasma temperatures have been found in nearby galaxies. 

\subsubsection{Black-body model}\label{sec:blackbody}

As a third spectral model we have applied black-body fits with neutral foreground absorption to the
2RXS sources. The parameters obtained from these fits are the black-body temperature,
the normalisation, and the $N_{\rm H}$ value.
Figure~\ref{Fig_powl_unique} (lower panel) gives an example for a spectral fit with a black-body model.

\begin{figure}
\centering
\includegraphics[angle=-90,width=89.5mm,clip=]{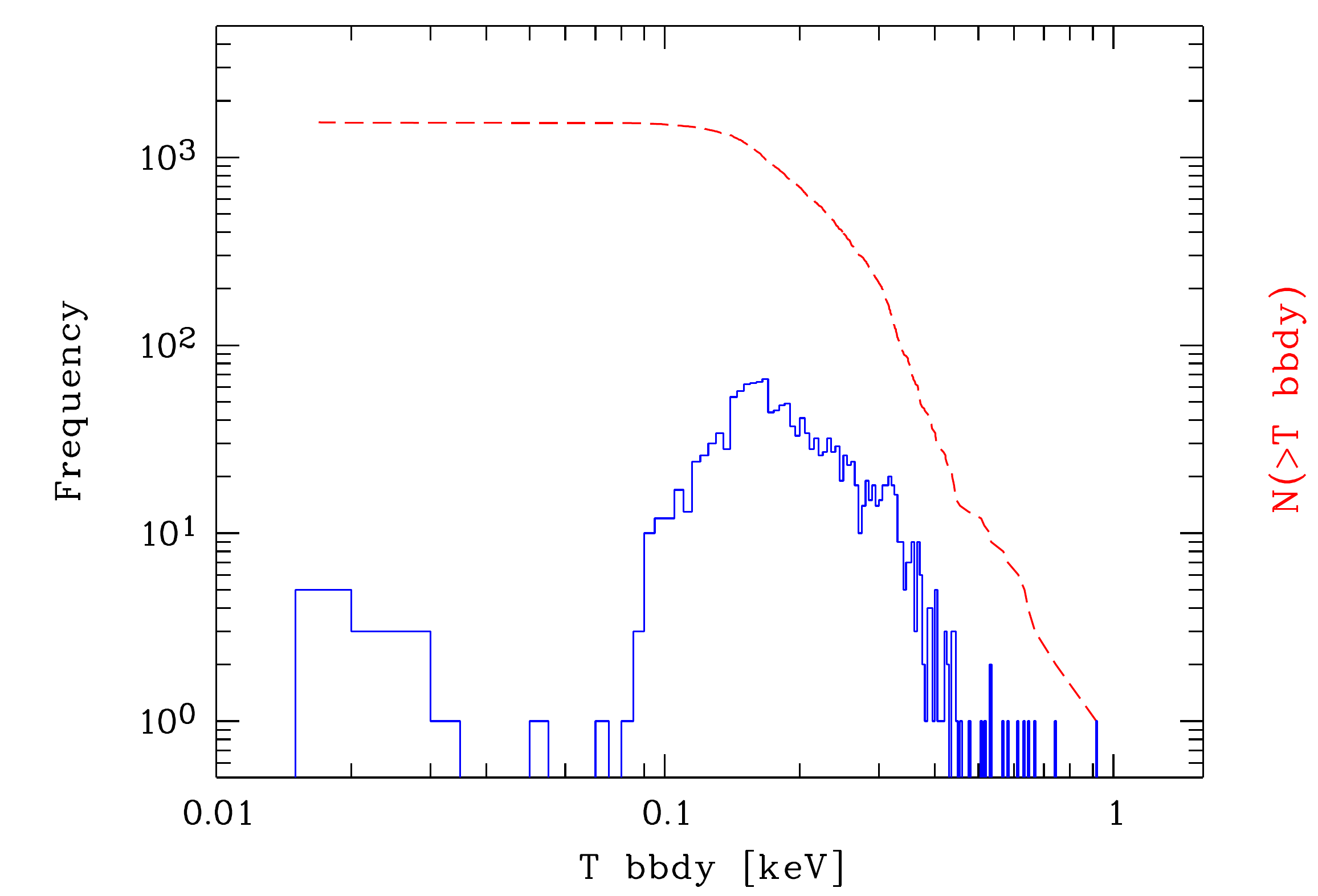}
\caption{Same as Fig.~\ref{Fig_histo_mekal} for the black-body model.
        The differential distribution is shown as the blue histogram. 
        The solid red dashed lines delineates the integral distribution.
       }
\label{Fig_histo_bbdy}
\end{figure}

In Fig.~\ref{Fig_histo_bbdy} we show the distribution of the temperatures obtained from 
the black-body fit, again for sources fulfilling the criteria described by the mekal model.
As for the mekal model, a bimodal distribution in the temperature is found with one peak at around 0.2 keV, 
similar to the peak found for the mekal fits, and the second peak centred
on around 20--30 eV (containing 11 sources).  Nine of these latter sources are optically
identified white dwarfs, and the other two are AM-Her type cataclysmic variables. These nine are 
also listed in \citet{Fleming1996} as White Dwarfs.  
All 11 objects are detected with the ROSAT Wide Field Camera \citep{Pye1995}, 
which is sensitive in the extreme soft band from
60 to 200 eV.

The lower peak is a factor of 10 lower than the 0.2 keV peak in the mekal fits, indicating
that black-body fits yield better results for objects with lower temperatures. This is most probably due to the fitting of different
physical emission mechanism, for instance,\, optically thin gas with the 
{\em mekal} fits, and optically thick black-body emission
from accreting objects.

\subsubsection{Constraints from spectral fits}\label{sec:unique_fitting}

As a result of the rather narrow spectral band-path of ROSAT - between the low-energy cutoff by interstellar 
absorption and the high-energy cutoff by the mirror - and the limited spectral resolution (E/$\Delta$E $\sim$ 4),
the significance of the spectral fits depends on the number of counts and on the column density. 
Of the 2722 spectra with an acceptable fit quality 
for the power-law model, 
1769 and 455 spectra also yield acceptable fits for the black-body and mekal model, respectively. 
For 247 sources all three spectral models give an acceptable fit (for an illustration see
the top panel of Fig.~\ref{Fig_Venn}). 
However, the following question is far more interesting than the statistics of overlaps:
How many sources with more than 100 counts 
can be fitted with only one of the spectral models with $\chi^2$ lower than 1.5?
The answers are 1117, 119, and 79 for the power-law, black-body, and mekal fits, respectively.
The lower panel of Fig.~\ref{Fig_Venn} shows that the number distributions for sources with a unique 
acceptable spectral fit are not
significantly different, which indicates that the uniqueness of the acceptable spectral model does 
not depend on the photon statistics.

\begin{figure}[thb]
   \includegraphics[angle=-90,width=89.6mm,clip=]{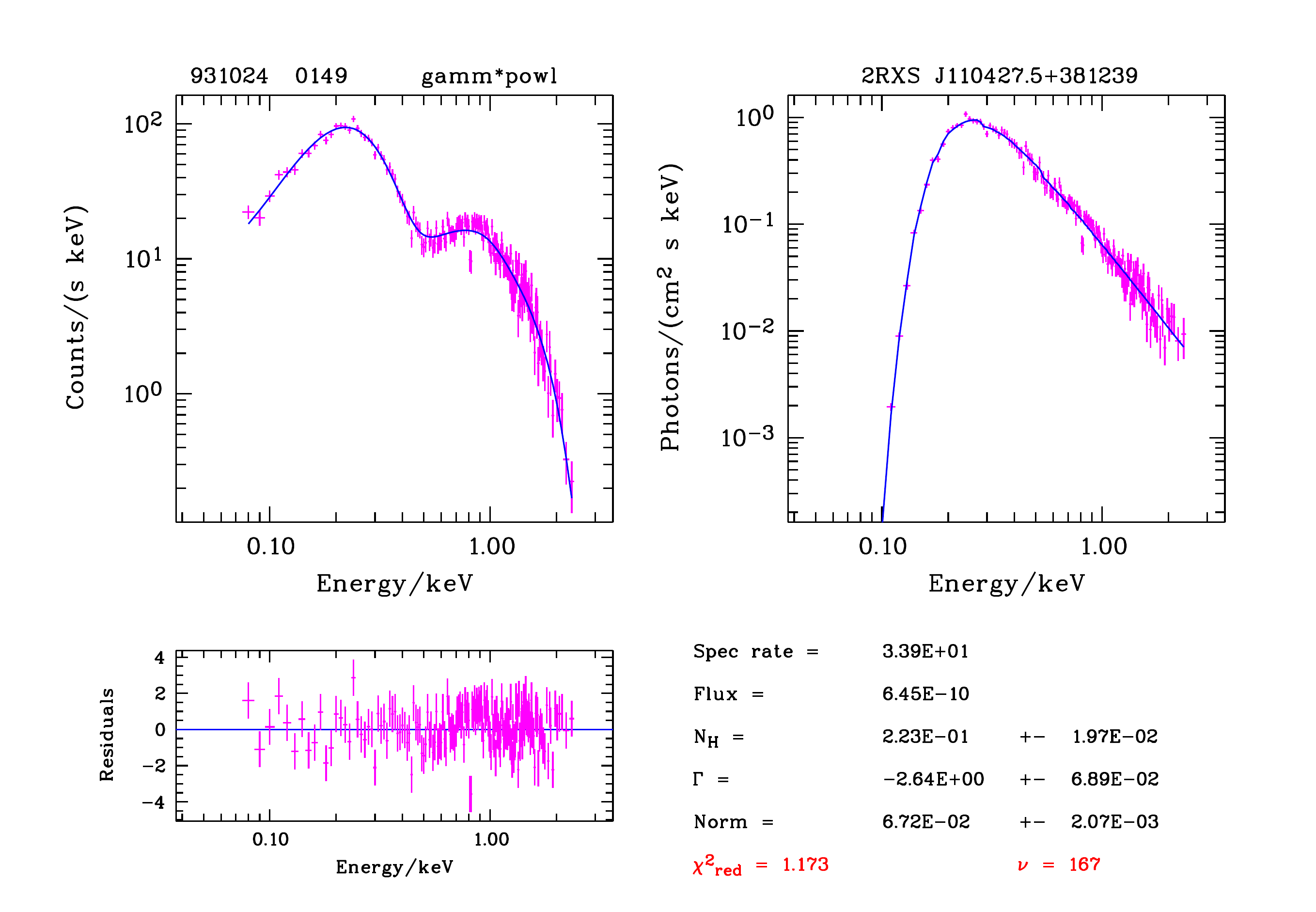}
   \vspace*{2mm}

   \includegraphics[angle=-90,width=89.6mm,clip=]{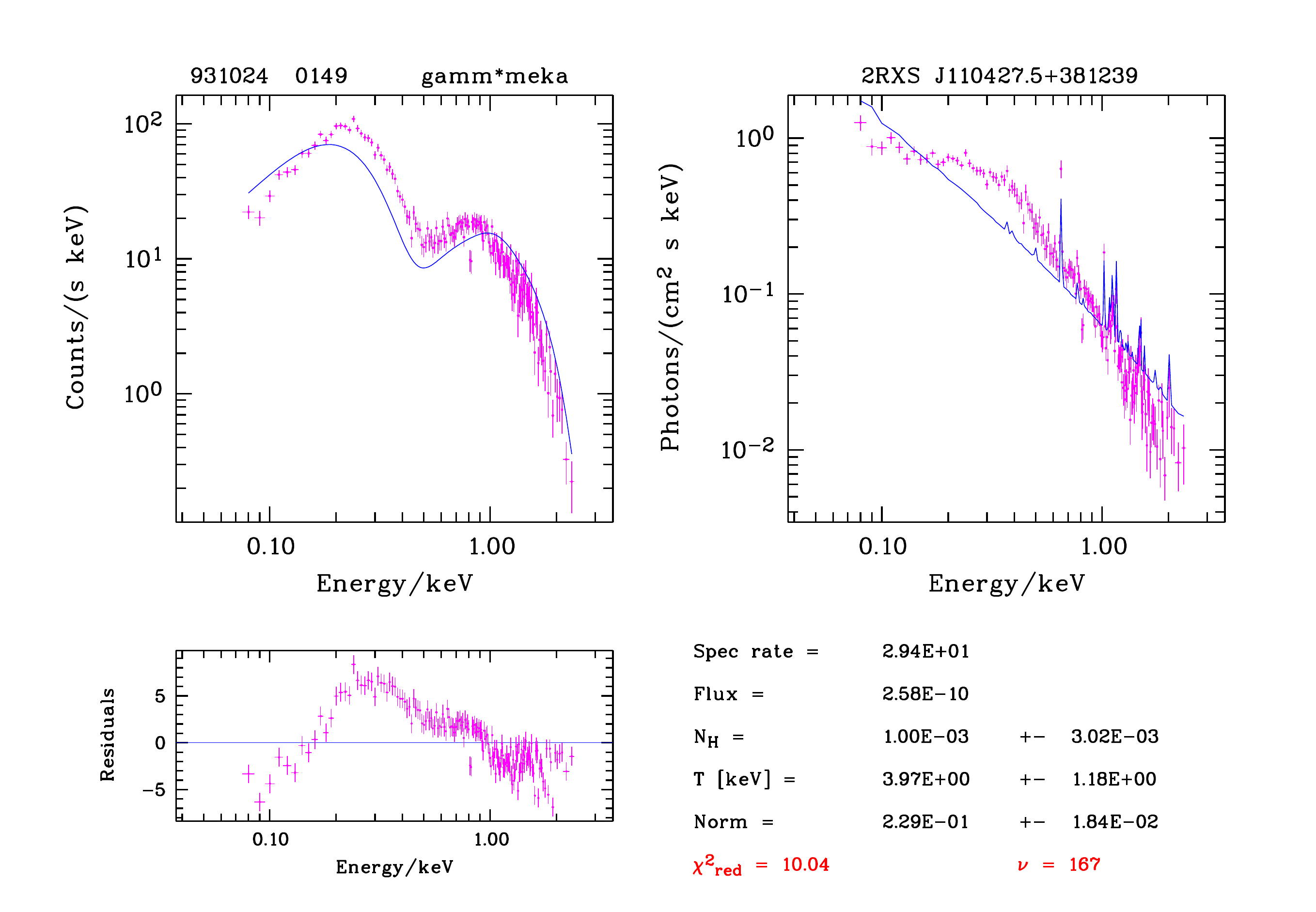}
   \vspace*{2mm}

   \includegraphics[angle=-90,width=89.6mm,clip=]{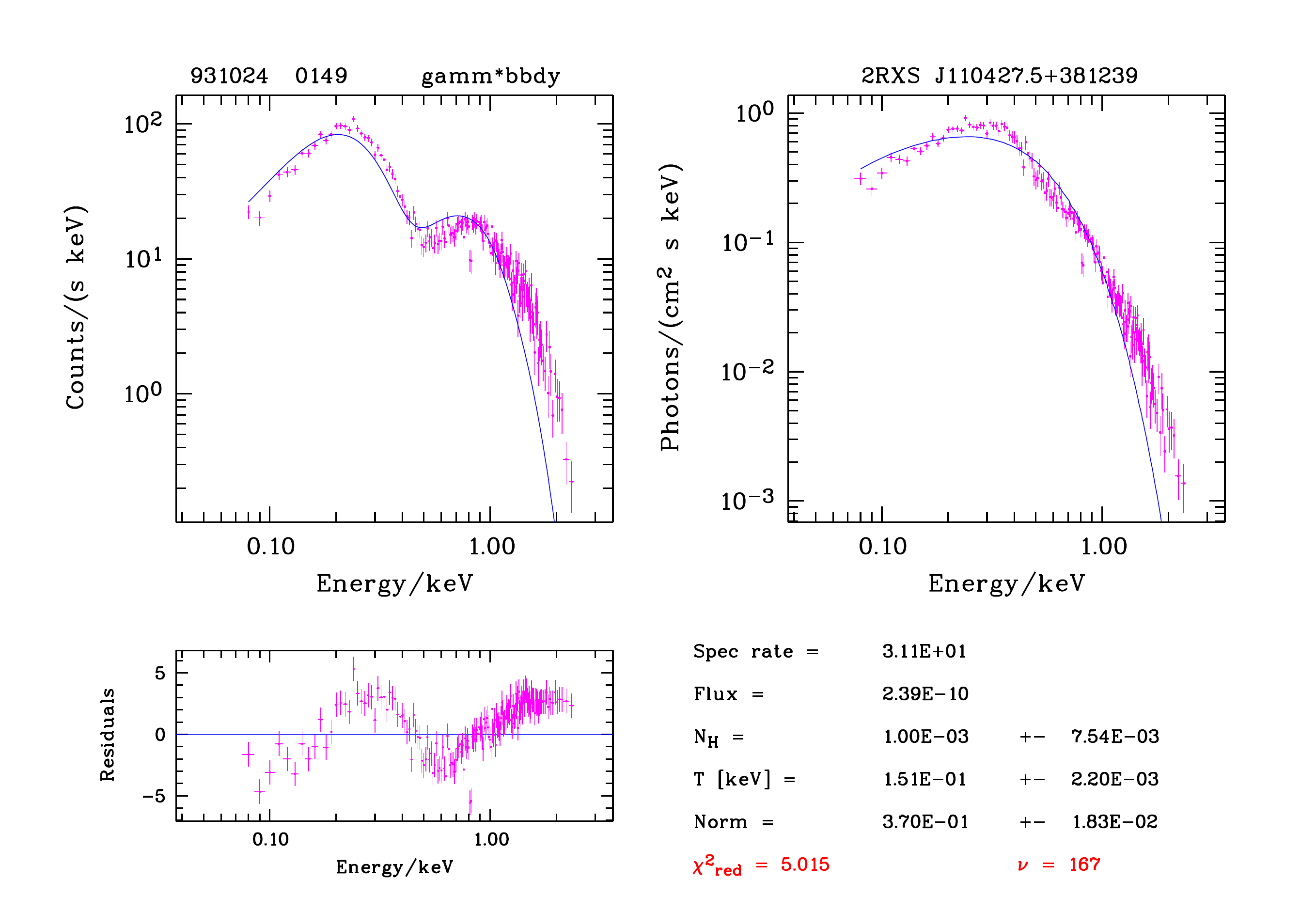}
  \caption{%
   Example of individual 2RXS source-level product:
   Power-law, mekal, and black-body fit to Mrk 421 (from top to bottom).
   The reduced $\chi^2$ values are 1.1, 10.0, and 5.0 for the 
   power-law, mekal, and black body fits, respectively. 
}              \label{Fig_powl_unique}
  \end{figure}

\begin{figure}
\centering
\includegraphics[bb=149 94 524 416,angle=0,width=90mm,clip=]{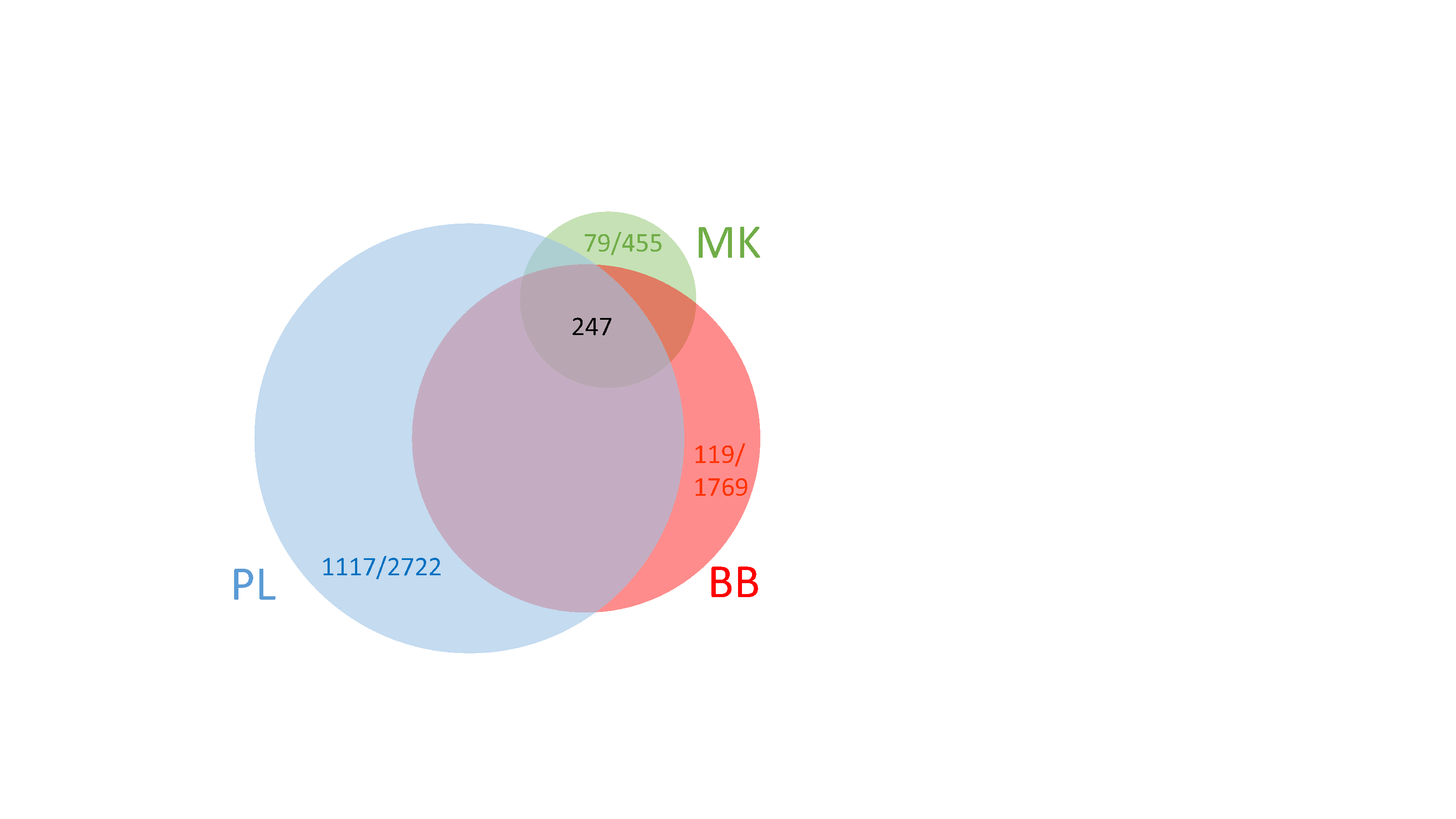}
\includegraphics[angle=-90,width=90mm,clip=]{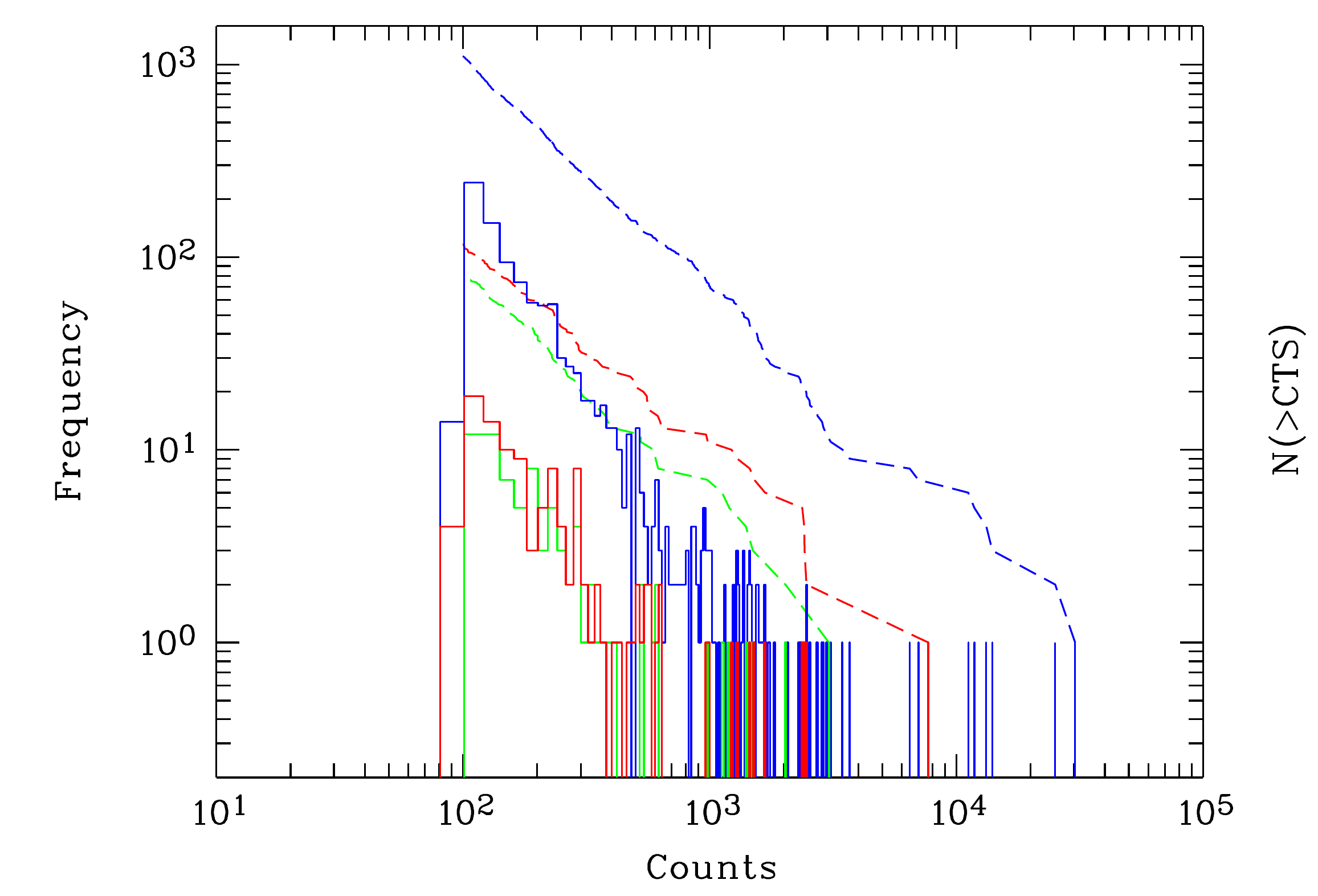}
\caption{Top panel: Graphical illustration of the number of acceptable spectral fits 
        for the three models. 
        The size of the circles scales with the number of spectra with acceptable fits. 
        The number of sources with acceptable power-law, mekal, and black-body fits 
        are 2722, 455, and 1769, respectively, while 1117, 79 and 119 sources have an 
        acceptable (unique) fit with only one of the models.
        For more information see Sect.~\ref{sec:unique_fitting}.
        Bottom panel: Differential and integral distributions for sources with unique power-law, black-body,  
        and mekal fits in blue, red, and green, respectively.
       }
\label{Fig_Venn}
\end{figure}

An example for an acceptable power-law fit to Mrk 421 and for the 
unacceptable
mekal and black-body fits is given in Fig.~\ref{Fig_powl_unique}.
For each fit three panels are presented, showing 
(1) the PSPC spectrum (data points with errors) together with the best-fit model as a solid line, 
(2) the unfolded spectrum with the model, and 
(3) the residuals. Best-fit model parameters are listed.

\subsubsection{Flux distributions}\label{sec:fluxes}

Additionally, we have calculated absorbed fluxes for the sources selected in
the previous subsections for spectral fitting. 
This is important for two reasons: 
(i) some sources are highly absorbed, and only photons above
1.5 keV are detected. 
The extrapolation to an intrinsic unabsorbed flux becomes highly uncertain 
and unphysical. 
(ii) Absorbed fluxes can be related to count rates by an energy-conversion
factor.
For the power-law, the mekal, and the black-body models, the absorbed fluxes
with unique spectral fits range between
$\rm 1.0\times10^{-9}\ and\ 8.8\times10^{-14}$,  
$\rm 1.0\times10^{-10}\ and\ 1.7\times10^{-13}$ \ and  
$\rm 9.8\times10^{-10}\ and\ 1.4\times10^{-13}\,erg\,cm^{-2}\,s^{-1}$, respectively.
For all three spectral models we find a similar lower flux limit of
the 2RXS sources of a few times of  $\rm 10^{-13}\,erg\,cm^{-2}\,s^{-1}$. 
\section{2RXS web page, catalogue interface, and help desk}\label{sec:WEB}

A description of the catalogue is available on the catalogue 
web site.
There, we also provide an interface to access the 2RXS catalogue and
related products.
The basic X-ray properties, correlations with sources 
from other X-ray missions and cross-matches as described in 
Sect.~\ref{sec:CrossMatches}, an X-ray image, the X-ray light curve, and the 
spectral properties (for sources fulfilling the criteria described in 
Sect.~\ref{sec:SpectralProperties}) are available for each 2RXS source.
A complete description of the catalogue entries is available at the catalogue 
web site. %
In addition, we present X-ray images where markings are applied to the source
and background extraction regions. 
These were used to produce background-subtracted light curves and spectra.
A cone search has been implemented, available at the catalogue web site,
 to efficiently access the 2RXS source properties and data products.
Individual sources, the complete 2RXS catalogue, and corrected photon
event files (RASS-3.1 processing) in FITS format can be accessed in the
download area.
We provide a 2RXS help desk for the community at the
E-mail adress {\tt 2rxs@mpe.mpg.de}.

\section{Summary}

We have re-analysed the photon event files from the ROSAT all-sky survey.
The main goal was to create a catalogue of point-like sources, which is
referred to as the 2RXS source catalogue.
We improved the reliability of detections by an advanced detection
algorithm and a complete screening process. 
New data products were created to allow timing and spectral analysis. 
Photon event files with corrected astrometry and Moon rejection 
(RASS-3.1 processing) were made available in FITS format.
The 2RXS catalogue will serve as the basic X-ray all-sky survey catalogue
until eROSITA data become available.

In this paper we list the most interesting objects in terms of their
timing and spectral properties. A discussion of the science highlights is
beyond the scope of the paper. With the publication of the 2RXS catalogue
and its data products, the detailed science specific exploration is now available for the astrophysical community. 

The experience gained by the High-Energy Group at MPE in creating the new 
ROSAT all-sky survey X-ray source catalogue is being and will be fed into
the data reduction analysis and scientific exploration of the forthcoming
eROSITA all-sky survey.

\begin{acknowledgements}
We thank the anonymous referee for detailed comments, which improved this
paper significantly.
The authors thank Damien Coffey for critical reading of the manuscript,
especially for improving the language.
J\"urgen Schmitt provided us unpublished light curves from their 1RXS
source variability analysis.
Joachim Paul performed detective work in our old archives for hints on
attitude errors and has set up the 2RXS web page and catalogue interface.
We appreciate discussions with Damien Coffey and Mara Salvato on positional
offsets and source identification procedures.
This research has made extensive use of the SIMBAD database and of the 
VizieR catalogue access tool, both operated at CDS, Strasbourg, France 
(see descriptions in \citet{2000A&AS..143....9W} and \citet{2000A&AS..143...23O}).
This work would have been impossible without the old ROSAT staff (H/W + S/W), 
who are too numerous to mention individually.
\end{acknowledgements}

\bibliography{aa25648-15}

\clearpage

\begin{appendix}

\section{Detection algorithm}\label{sec:DetectionAlgorithm}

\subsection{Source detection algorithm}\label{sec:SourceDetection}

The source detection was performed using a three-step approach 
and is described in Sect.~\ref{sec:SourceDetectionGeneral}.

One goal of 2RXS is the detection of point-like sources 
down to a likelihood limit of 6.5; 
very extended sources are not dealt with here;
the detailed discrimination is explained below in 
Sect.~\ref{sec:SourceDetectionMasking}.
The detection algorithm follows the standard scheme that was applied for the BSC and FSC  
\citep{Voges1999}, 
with important extensions.
The publicly available software packages 
ESO-MIDAS\footnote{ESO-MIDAS: http://www.eso.org/sci/software/esomidas/}
{version P03SEPpl1.2} \citep{Banse1992}
and
EXSAS\footnote{For the EXSAS users guide and additional information 
for the used commands see web page (Sect.~\ref{sec:WEB})}
{version 05APR\_EXP} \citep{Zimmermann1998}
were used on a Linux platform throughout this analysis.

\subsection{Energy bands}\label{sec:SourceDetectionBands}

The PSPC response varies significantly over the bandpass of $0.1 - 2.4$\,keV, with a
break at the carbon absorption edge (0.28 keV) ( cf. {\it heasarc.gsfc.nasa.gov/docs/pspc\_matrices.html}).
Several energy bands were constructed to determine the source properties.
These are expressed in 
pulse-height-invariant channels in units of approximately 10 eV,
broad (B): channels 11 -- 235 (0.1 -- 2.4 keV),
soft (S): 11 -- 41 (0.1 -- 0.4 keV), 
hard (H): 52 -- 201 (0.5 -- 2.0 keV), as in \citet{Voges1999}.
In addition, we performed source detection in the following bands:
medium (M): 52 -- 90 (0.5 -- 0.9 keV),
very hard (V): 91 -- 201 (0.9 -- 2.0 keV), and
wide (W): 11 -- 201 (0.1 -- 2.0 keV).

Source counts that were corrected for vignetting and dead-time were used to determine
hardness ratios HR according to $\rm HR = (hard - soft)/(hard + soft)$ with the bands
$$ 
 {\rm HR1} = \frac{H-S}{H+S}\;,\;\;
 {\rm HR2} = \frac{V-M}{V+M}\;,
$$
as in \citet{Voges1999}.

\subsection{Regions with very extended bright X-ray emission}\label{sec:SourceDetectionMasking}

The source detection algorithm is focused on point-like sources with an
extent smaller than 5\arcmin\ radius, therefore we excluded bright, extended emission
features such as the Vela or Kepler supernova remnants.
Thirty-two regions were manually identified and marked using polygon masks.
Sources detected in these regions were later automatically flagged and
excluded from the catalogue.
In Table~\ref{tab:AppExttab} we list the sky fields that have been identified
as such masked regions. These fields can be assessed through the 
the catalogue web page under {\tt cone search}.

\begin{figure}[htp]
  \centering
  \includegraphics[angle=-90,width=90mm,clip=]{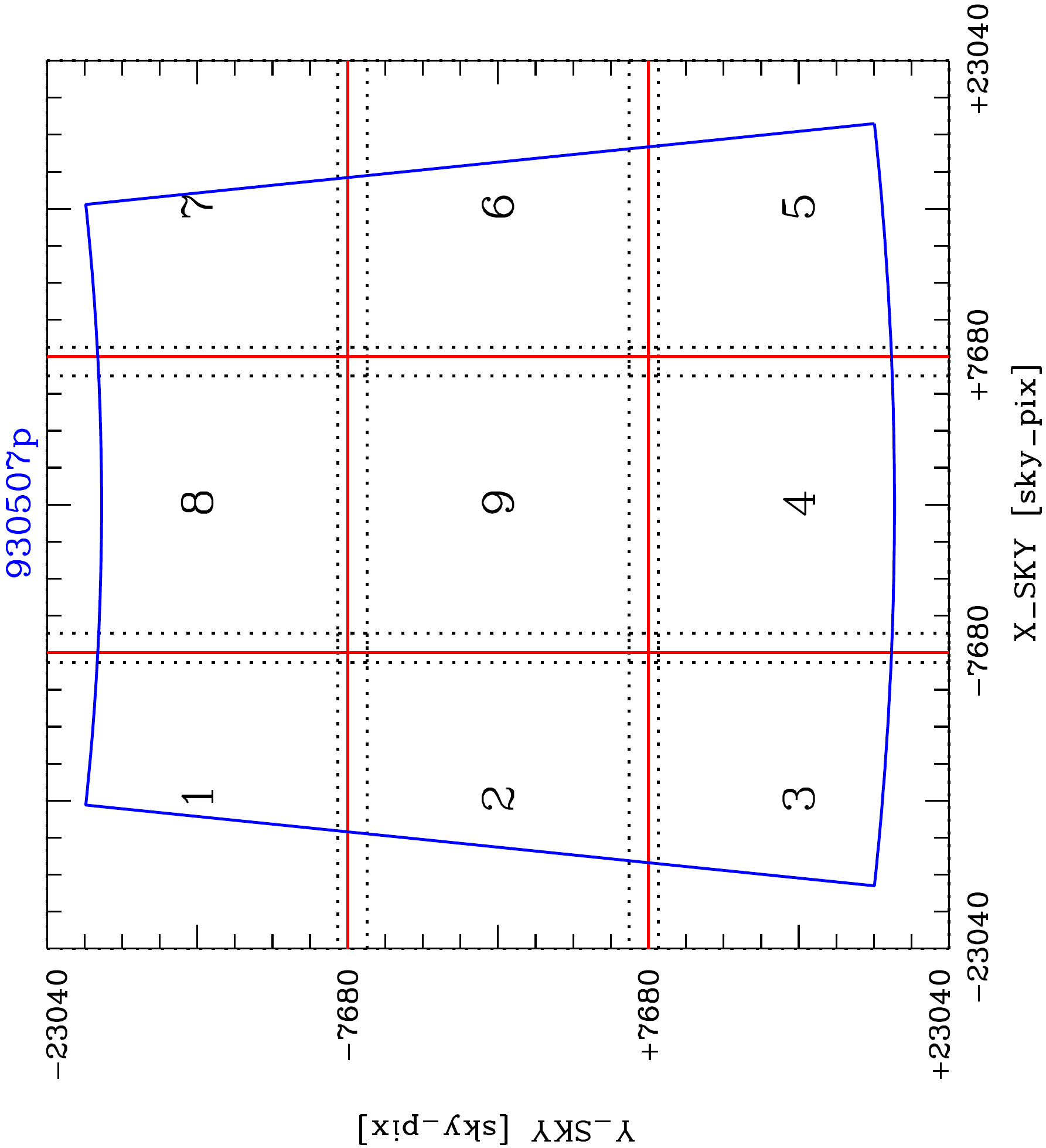}
  \caption{%
  Sub-images used in our detection scheme: the whole frame corresponds to a full
  $6.4\degr \times 6.4\degr$ RASS field, 
  the blue area denotes the unique region 
  (rectangle in equatorial coordinates, here for field 930507p). 
  The red lines illustrate the unique areas for each sub-image, while the
  dashed lines show the overlap among the sub-images. 
  All sub-images have the same size.
  {\tt sky\_pix} units are 0.5~\arcsec.
  The sub-image number {\tt USKY} as well as the 
  {\tt (X\_SKY,Y\_SKY)} coordinate values
  for each detection are stored in the catalogue.
  For details see text.
         }
   \label{plot_2rxs_subimages}
\end{figure}

\subsection{Local detection}\label{sec:SourceDetectionLdetect}

The first detection procedure is running a sliding window 
method\footnote{Using the command {\tt DETECT/LOCAL} of the {\tt EXSAS} package} 
on a binned image with 45\arcsec\ squared pixels\footnote{Produced with {\tt CREATE/SOURCE\_DETECT\_IMAGE}}. 
Like in \citet{Voges1999}, this is applied to an image with a size of 512 pixels $\times$ 512 pixels.
One important extension that was added to the previous analysis by us is a multiple 
detection run on a mosaic of $3\times 3$ overlapping sub-images of about 1/9 (12.6\%) in area where 
the remainder is masked and thus not considered (see Fig.\ref{plot_2rxs_subimages}).
This leads to a total of ten detection lists per energy band and field,
with a minimum existence likelihood of 5.5 for the sub-images and 6.0 
for the full image\footnote{Due to restrictions in the EXSAS 
package the likelihood limits were increased by 5\% until the number of 
detections per band and sub-image were less than 400}. %

\subsection{Creation of background map}

Detection lists from the previous step with a 
minimum likelihood of 8.0
are used as input to excise the brightest sources %
with a cut radius of 5 FWHM of the PSF.
A smoothed map is created
when filling the holes with a 2-d spline
fit\footnote{\tt CREATE/BG\_IMAGE}.
The multiple sub-image masking procedure is also applied to the background
map creation process. Thus, bright sources or sharp emission gradients in one
sub-image have no effects (over-shooting, etc.) on 
background maps of other sub-images.

\subsection{Detection using background map}\label{sec:SourceDetectionMdetect}

The sliding-window detection process is repeated using the smooth background 
maps\footnote{\tt DETECT/MAP}
with a minimum existence likelihood of 
5.5 for sub-regions, and 8.0 for the full image\footnote{Due to restrictions in the EXSAS package the likelihood 
limits were increased by 5\% until the number of detections per band and 
sub-image were less than 1000}.   %

\subsection{Source list merging (I)}

The source lists produced by the local and map detection runs 
(three energy bands S, H, and B for each of the two runs) were 
merged\footnote{\tt MERGE/SOURCE\_LIST} separately
for each of the ten images for each of the 1378 fields.
These lists were then used as input for the next step.
The ten detection lists for each of the remaining three bands (M, V, and W)
were kept for later individual inspection, 
but were not used explicitly in the catalogue production. 
Multi-band detection was performed sequentially and not simultaneously.

\subsection{Maximum likelihood detection}\label{sec:SourceDetectionMLdetect}

After the image-based detection steps described above were completed,
a photon-based algorithm
\citep{BoeseDoeb}
was applied to use the full spatial resolution. 
The ten merged lists were used as input for a maximum likelihood evaluation 
of the candidate X-ray sources. 
Each photon is properly weighted for vignetting and detector dead-time,
and the distribution is compared with a model of the PSF.
Finally, the maximum likelihood 
algorithm\footnote{\tt DETECT/MAXLIK}
yields source positions and count rates for the broad energy
band \footnote{Note that this ML algorithm can only return detections close to 
positions provided by the input lists. Therefore, care has to be taken that no 
proper source candidate is missed in previous detection steps.
}. 
The final detection list was built from sub-image entries down to a 
detection limit of 5.5 and with full-field-only detections 
added above a limit of 7.0 (with a flag {\tt USKY==0} to recognise them).

\subsection{Small-map source list merging (II)}

After the source detection process, the detection lists were unified.
First, the nine lists of the sub-regions were merged after limiting them 
to the unique areas in tangential plane-projected sky coordinates.
Then, additional sources from full-field detection (if not already 
contained in the sub-image analysis) were included.
Finally, detections were selected according to the rectangular 
equatorial coordinate ranges for each of the 1378 fields
(see Fig.\,\ref{Fig_sky_field}).

\subsection{Automatic flagging and rejection}

Detections inside masked regions were flagged and excluded from the catalogue.
Of the close multiple (mostly pairs) detections (within 10\arcsec\ ),
only the one with highest likelihood value was kept, while the others were flagged 
and excluded from the catalogue.
Later, a visual screening was applied and further detections were flagged and removed
(see Sect.~\ref{sec:Screening}).

\subsection{Point spread function and detection cell size}\label{sect_psf_cellsize}
The ROSAT PSPC survey point spread function (PSF) is almost energy independent
in the range of $60\%-90\%$ encircled energy fraction (EEF, see Fig.~19 in \citet{Boese2000}),
corresponding to an integration radius of $100 - 200$ \arcsec.
The effective detection cell size of our multi-stage detection scheme has to be
calibrated {\em \textup{a posteriori}} using simulations
(see also Sect.~5.7 in \citet{Zimmermann1998}).
In Fig.~\ref{Fig_spurious} we fit the likelihood function to the distribution
of simulated spurious sources and derived an average number of detection cells per field of 11000,
corresponding to a detection cell size of 
$116$ \arcsec, equivalent to an EEF of $64.2\%$.

\newpage

\section{2RXS sources with extreme variability}

\subsection{2RXS sources versus ROSAT pointed observations}
In Table~\ref{tab:2rxs_2rxp_50} we list 2RXS sources that were also detected 
in ROSAT pointed observations, which
show amplitude variations greater than 50.
Table~\ref{tab:2rxs_2rxp_50} also includes the count rates from 2RXS survey 
and 2RXP pointed observations, the source count ratio 
and the likely identification of the source.

\begin{table*}[htb]
\caption{Sources with amplitude variability greater than 50 for sources included 
in both the survey 2RXS and pointed 2RXP catalogues. The amplitude variability is
the ratio between the 2RXS and the 2RXP count rates.  
}\label{tab:2rxs_2rxp_50}  %
{\tiny
\begin{tabular}{lllccccrll}
\hline\hline
2RXS         & RA           &Dec               &\!\!\!Rate 2RXS\!\!\!& Error                &\!\!Rate 2RXP\!\!&  Error      & Ratio & ID \\
detection ID & J2000        &J2000             & $\rm counts\ s^{-1}$ &$\!\!\rm counts\,s^{-1}\!\!$ & $\rm counts\,s^{-1}$ &$\!\!\rm counts\,s^{-1}\!\!$&       &  \\
\noalign{\smallskip}\hline\noalign{\smallskip}
930722\_0269 &$13\;07\;53.74 $&$+53\;51\;37.0$&   1.975 & 0.063 &1.58e-03 &4.32e-04 &1243.2 & EV UMa, AM Her type CV \\ 
931752\_0042 &$19\;11\;16.01 $&$+00\;35\;05.5$&  12.756 & 0.184 &3.16e-02 &1.52e-03 & 403.2 & Aql X-1, LMXB \\
932602\_0089 &$00\;39\;15.57 $&$-51\;16\;59.9$&   1.005 & 0.067 &2.84e-03 &7.69e-04 & 353.7 & 2MASX J00391586-5117013, Seyfert 1\\
931863\_0079 &$23\;16\;03.76 $&$-05\;27\;13.9$&   1.216 & 0.061 &8.93e-03 &8.60e-04 & 136.1 & 2MASS J23160363-0527089, AM Her type CV\!\!\!\\
931231\_0204 &$12\;37\;41.39 $&$+26\;42\;30.2$&   5.303 & 0.105 &4.23e-02 &5.94e-03 & 125.2 & IC 3599\\
931715\_0125 &$05\;16\;11.56 $&$-00\;08\;04.3$&   2.388 & 0.077 &2.30e-02 &1.65e-03 & 103.6 & Ark 120 \\
931716\_0149 &$05\;40\;42.82 $&$-02\;05\;33.8$&   0.167 & 0.021 &1.77e-03 &4.00e-04 &  94.7 & 1RXS J054042.8-020533, in Orion\\
930725\_0204 &$15\;15\;23.25 $&$+55\;30\;57.8$&   0.349 & 0.019 &4.41e-03 &8.76e-04 &  79.2 & NGC 5905 \\
930902\_0146 &$00\;45\;29.24 $&$+42\;18\;55.4$&   0.055 & 0.012 &7.27e-04 &2.18e-04 &  75.5 & 1RXS J004528.7+421850, in M31\\
932331\_0076 &$13\;13\;17.08 $&$-32\;59\;13.8$&   1.968 & 0.082 &2.81e-02 &1.55e-03 &  70.0 & V1043 Cen, AM Her type\\
931406\_0050 &$02\;08\;11.95 $&$+15\;08\;39.0$&   0.103 & 0.021 &1.68e-03 &4.49e-04 &  61.6 & WW Ari\\
931737\_0095 &$13\;43\;25.26 $&$-00\;01\;11.8$&   0.034 & 0.013 &6.29e-04 &2.00e-04 &  55.1 & no counterpart within 40 arcsec\\
932319\_0009 &$08\;12\;28.92 $&$-31\;14\;55.4$&   0.331 & 0.029 &6.25e-03 &1.50e-05 &  52.9 & V572 Pup \\
930902\_0132 &$00\;47\;44.81 $&$+42\;38\;42.4$&   0.044 & 0.011 &8.81e-04 &2.53e-04 &  50.4 & 1RXS J004744.6+423842, in M31\\
\noalign{\smallskip}\hline
\end{tabular}
}
\end{table*}

\subsection{Normalised excess variance.}

In Table~\ref{tab:2rxs_exc_10} we list 2RXS sources that show 
normalised excess variance values above 10\,$\sigma$, which are
not listed in \citet{FUH}. 
\begin{table*}   %
\caption{Sources with normalised excess variance values above 10\,$\sigma$ and not 
listed in \citet{FUH}.}
\label{tab:2rxs_exc_10}
\begin{tabular}{lllccll}
\hline\hline\noalign{\smallskip}
2RXS        & RA              & Dec             & Excess  & Distance& Name       & Type               \\   
detection ID& J2000           & J2000           &$\sigma$ &\arcsec\ &            &                    \\
\noalign{\smallskip}\hline\noalign{\smallskip}
932022\_0081& $08\;15\;06.70$ & $-19\;03\;08.5$ & 30.87   & 12.6          & VV Pup           & AM Her type CV           \\ 
932822\_0060& $15\;28\;17.81$ & $-61\;52\;58.2$ & 25.43   &  4.3   & KY TrA      & LMXB                       \\
931146\_0057& $19\;58\;22.29$ & $+35\;12\;01.8$ & 25.05   & 10.9   & Cyg X-1    & HMXB                  \\
931301\_0035& $00\;11\;52.81$ & $+22\;59\;13.9$ & 14.83   &  9.6   & LP 348-40  & young low-mass star   \\
932910\_0097& $07\;48\;33.98$ & $-67\;45\;06.5$ & 13.00   &  7.7   & UY Vol     & LMXB                  \\
932437\_0026& $17\;03\;58.37$ & $-37\;50\;42.7$ & 12.41   & 20.3          & 4U 1700-37   & HMXB                  \\
\noalign{\smallskip}\hline
\end{tabular}
\end{table*}

Sources that posses normalised excess variance values greater than 20\,$\sigma$
and are listed in \citet{FUH}, but are not shown as graphical 
illustrations, are also listed in Table~\ref{tab:2rxs_exc_20}. 

\begin{table*}   %
\caption{Sources with normalised excess variance values above 20\,$\sigma$ and
listed in \citet{FUH} (FUH), but not shown as graphical illustration.}\label{tab:2rxs_exc_20}
\begin{tabular}{lllcccl}
\hline\hline\noalign{\smallskip}
2RXS         & RA             & Dec            & Excess     & Distance &  Seq  & Name       \\ 
detection ID & J2000          & J2000          & $\sigma$   & \arcsec\ &  FUH  &            \\
\noalign{\smallskip}\hline\noalign{\smallskip}
932907\_0149 &$05\;32\;50.29$ & $-66\;22\;11.9$&  77.50     &2.5       & 0281     & LMC X-4         \\
931139\_0078 &$16\;57\;49.93$ & $+35\;20\;25.3$&  72.53     &0.6       & 0862     & Her X-1         \\
932446\_0004 &$20\;47\;44.98$ & $-36\;35\;39.6$&  54.95     &0.4       & 1092     & BO Mic           \\
932907\_0187 &$05\;35\;41.24$ & $-66\;51\;52.5$&  49.12     &0.2       & 0289     & 1A 0535-66   \\
930311\_0231 &$15\;07\;57.88$ & $+76\;12\;15.1$&  29.58     &1.1       & 0755     & HD 135363  \\
932420\_0091 &$09\;02\;07.11$ & $-40\;33\;08.9$&  29.39     &0.5       & 0495     & GP Vel           \\
931837\_0161 &$13\;34\;43.60$ & $-08\;20\;34.8$&  27.15     &6.8       & 0676  & EQ Vir     \\ 
932902\_0046 &$01\;41\;00.76$ & $-67\;53\;31.9$&  25.92     &5.0       & 0063  & BL Hyi     \\
930944\_0108 &$21\;42\;43.25$ & $+43\;35\;09.5$&  24.96     &4.9       & 1134  & SS Cyg     \\
930627\_0065 &$19\;03\;18.14$ & $+63\;59\;34.0$&  24.36     &5.7       & 1004  & GJ 4094    \\
931352\_0057 &$20\;05\;41.93$ & $+22\;39\;55.3$&  22.28     &3.4       & 1061  & QQ Vul     \\
932808\_0259 &$05\;38\;56.67$ & $-64\;04\;58.6$&  22.15     &5.0       & 0293  & LMC X-3    \\
\noalign{\smallskip}\hline
\end{tabular}
\end{table*}

\subsection{Maximum amplitude variability}

In Table~\ref{tab:2rxs_amp_10} we list significant sources that have a maximum amplitude variability greater than 10\,$\sigma$. 
\begin{table*}   %
\caption{Sources with maximum amplitude variabilities above 10\,$\sigma$. 
The list is ordered by decreasing significance.
The ratio ($\rm ampl\_max$) and the significance ($\rm ampl\_sig$) are defined in Sect.~\ref{sec:variab}.
}
\label{tab:2rxs_amp_10}
\begin{tabular}{lllccrl}
\hline\hline\noalign{\smallskip}
2RXS         & RA             & Dec             & significance    & ratio& Distance     & ID \\   
detection ID & J2000          & J2000           &($\sigma$)       &                 & \arcsec\ &    \\   
\noalign{\smallskip}\hline\noalign{\smallskip}  
932446\_0004 &$20\;47\;44.98$ & $-36\;35\;39.6$ & 39.8 &  38.3 & 4.8    & BO Mic      \\
932022\_0081 &$08\;15\;06.70$ & $-19\;03\;08.5$ & 26.4 &  21.2 & 12.6   & VV Pup      \\
931139\_0078 &$16\;57\;49.93$ & $+35\;20\;25.3$ & 21.2 &  57.7 & 10.7   & Her X-1     \\
932348\_0008 &$20\;45\;09.38$ & $-31\;20\;22.9$ & 20.1 &  15.2 & 1.5    & AU Mic      \\
932808\_0259 &$05\;38\;56.67$ & $-64\;04\;58.6$ & 20.0 &  20.3 & 6.5    & LMC X-3     \\
932910\_0097 &$07\;48\;33.98$ & $-67\;45\;06.5$ & 19.8 &  21.7 & 9.0    & UY Vol      \\
930311\_0231 &$15\;07\;57.88$ & $+76\;12\;15.1$ & 19.2 &  13.0 & 14.0   & HD 135363   \\
931352\_0057 &$20\;05\;41.93$ & $+22\;39\;55.3$ & 18.8 &  24.6 & 6.5    & QQ Vul      \\
932420\_0091 &$09\;02\;07.11$ & $-40\;33\;08.9$ & 18.1 &  13.6 & 11.7   & GP Vel      \\
932822\_0060 &$15\;28\;17.81$ & $-61\;52\;58.2$ & 18.1 &  14.6 & 4.3    & KY TrA      \\
930946\_0100 &$22\;46\;50.70$ & $+44\;20\;07.3$ & 17.9 &  12.3 & 11.5   & EV Lac      \\ 
932907\_0149 &$05\;32\;50.29$ & $-66\;22\;11.9$ & 16.6 &  18.9 & 4.0    & LMC X-4     \\
931229\_0073 &$11\;49\;55.53$ & $+28\;45\;09.1$ & 15.6 &   6.8 & 4.1    & EU UMa      \\
931531\_0062 &$11\;32\;49.27$ & $+12\;10\;27.4$ & 13.8 &   5.4 & 3.8    & BD+12 2343  \\
930627\_0065 &$19\;03\;18.14$ & $+63\;59\;34.0$ & 13.1 &   4.6 & 2.8    & GJ 4094     \\ 
931118\_0143 &$07\;34\;37.16$ & $+31\;52\;23.3$ & 12.6 &   6.5 & 10.5   & $\alpha$ Gem C \\  
932540\_0085 &$19\;38\;35.79$ & $-46\;12\;58.7$ & 12.6 &  11.0 & 1.8    & QS Tel      \\ 
931134\_0075 &$14\;42\;07.57$ & $+35\;26\;27.7$ & 12.3 &   6.2 & 8.7    & Mrk 478     \\ 
931146\_0057 &$19\;58\;22.29$ & $+35\;12\;01.8$ & 12.3 &  40.3 & 10.9   & Cyg X-1     \\ 
932907\_0187 &$05\;35\;41.24$ & $-66\;51\;52.5$ & 12.1 &   4.5 & 2.6    & 1A 0535-668 \\ 
931852\_0036 &$19\;18\;47.81$ & $-05\;14\;14.8$ & 12.0 &   5.0 & 5.9    & 4U 1916-053 \\ 
931448\_0046 &$18\;02\;06.59$ & $+18\;04\;36.2$ & 11.9 &   3.9 & 11.2   & V884 Her    \\
930718\_0324 &$10\;51\;35.33$ & $+54\;04\;37.3$ & 11.4 &   3.7 & 5.7    & EK UMa      \\          
932902\_0046 &$01\;41\;00.76$ & $-67\;53\;31.9$ & 10.9 &  12.5 & 8.1    & BL Hyi      \\
930945\_0061 &$22\;08\;40.84$ & $+45\;44\;31.0$ & 10.8 &   7.4 & 3.0    & AR Lac      \\
930314\_0063 &$20\;30\;09.90$ & $+79\;50\;42.4$ & 10.8 &   3.8 & 8.6    & GSC 04593-01344 \\
931713\_0128 &$04\;37\;37.20$ & $-02\;29\;41.5$ & 10.4 &   6.1 & 13.6   & GJ 3305     \\
932437\_0026 &$17\;03\;58.37$ & $-37\;50\;42.7$ & 10.1 &   4.9 & 19.3   & HD 153919   \\
932907\_0068 &$05\;28\;44.94$ & $-65\;26\;56.4$ & 10.0 &   9.7 &  1.0   & RST 137     \\
\noalign{\smallskip}\hline
\end{tabular}
\end{table*}

\clearpage

\section{Extended X-ray emission regions with embedded point sources in the 
ROSAT all-sky survey}\label{sec:AppExt}

\begin{table*}
 \caption{Sources added manually to 1RXS BSC, see text for details.}
 \label{tab:1rxs_flatsch}
\begin{tabular}{lcll}
\hline\hline
1RXS name             &  SeqID  & counterpart        & 2RXS comment \\
\hline
1RXS J021756.0+624359 &  930604 & SNR G132.4+02.2    & masked       \\
1RXS J045851.3+515028 &  930809 & SNR G156.2+05.7    & masked       \\
1RXS J050036.1+461859 &  930910 & SNR G160.4+02.8    & masked       \\
1RXS J205118.2+310312 &  931148 & SNR Cygnus Loop    & masked       \\
1RXS J161954.9-153542 &  932043 & Sco X-1            &  not observed (detector off) \\
1RXS J181545.4-164043 &  932048 & SNR G014.1-00.1    & masked       \\
1RXS J171333.1-394544 &  932438 & SNR G347.3-00.5    & masked       \\
1RXS J083520.6-451035 &  932518 & SNR Vela           & Vela pulsar automatically detected   \\
1RXS J112441.5-682621 &  932914 & artifact           & --           \\
\hline
\end{tabular}
\end{table*}

\begin{table*}[thbp]
\caption{\label{tab:AppExttab} Large regions with extended X-ray emission in the ROSAT all-sky survey, masked in standard detection run.}
\centering
\begin{tabular}{lllll}
\hline\hline\noalign{\smallskip}
SeqID  &RA        &Dec         &Count rate             &Identification \\  
      &J2000      &J2000       &$\rm counts\ s^{-1}$   &\\  
\noalign{\smallskip}\hline\noalign{\smallskip}
930604 & $02\;18\;39$ & $+62\;49\;40$ & 5.48              & SNR G132.4+02.2 \\  
930738 & $23\;01\;11$ & $+58\;51\;32$ & 9.67    & SNR G109.1-01.0 \\
930739 & $23\;23\;24$ & $+58\;48\;54$ & 51.66           & SNR Cas-A\\ %
930809 & $04\;58\;03$ & $+51\;46\;56$ & 18.56            & SNR G156.2+05.7 \\
930910 & $04\;59\;06$ & $+46\;20\;33$ & 9.78              & SNR G160.9+2.6 \\
930910 & $04\;50\;17$ & $+45\;02\;17$ & 0.97              & 3C 129.1, Cluster of Galaxies\\
930911 & $05\;26\;30$ & $+42\;56\;00$ & 2.16              & SNR G166.0+04.3 \\
930939 & $19\;21\;07$ & $+43\;58\;43$ & 4.24              & Abell 2319\\
931007 & $02\;54\;14$ & $+41\;35\;30$ & 4.88              & 2A 0251+413, AWM 7 \\
931008 & $03\;19\;47$ & $+41\;30\;52$ & 34.02            & Perseus Cluster \\
931036 & $16\;28\;37$ & $+39\;32\;49$ & 7.65              & Abell 2199\\
931148 & $20\;48\;30$ & $+31\;16\;20$ & 2200              & Cygnus Loop \\    %
931315 & $05\;34\;32$ & $+22\;00\;52$ & 913               & Crab Nebula/Pulsar\\
931330 & $11\;44\;48$ & $+19\;39\;27$ & 4.51              & Abell 1367\\
931443 & $16\;02\;26$ & $+15\;58\;53$ & 3.28              & ACO 2147, Cluster of Galaxies\\
931452 & $19\;38\;16$ & $+17\;18\;50$ & 1.68              & SNR G053.6-02.2\\
932034 & $12\;57\;12$ & $-17\;24\;34$ & 1.61              & ACO 1644, Cluster of Galaxies\\
932048 & $18\;15\;45$ & $-16\;40\;43$ & 3.93              & SNR G014.1-00.1\\
932144 & $17\;12\;25$ & $-23\;21\;01$ & 9.74              & Ophiuchus Cluster, Cluster of Galaxies\\
932146 & $18\;01\;23$ & $-23\;17\;20$ & 7.41              & SNR G006.4-00.1 \\
932147 & $18\;07\;22$ & $-23\;17\;48$ & 5.95              & NGC 6546, Cluster of Stars \\
932309 & $03\;38\;30$ & $-35\;27\;18$ & 0.84              & Fornax Cluster, Cluster of Galaxies \\
932309 & $03\;38\;52$ & $-35\;35\;40$ & 0.30              & NGC 1404, Group of Galaxies\\
932325 & $10\;30\;00$ & $-35\;19\;35$ & 2.90              & ACO S 636, Cluster of Galaxies\\
932438 & $17\;13\;33$ & $-39\;45\;44$ & 8.5                & SNR G347.3-00.5 \\
932518 & $08\;35\;20$ & $-45\;10\;33$ & 2810, (510)     & Vela SNR (Puppis A),  also in 932419, 932517, 932519, 932615, 932616\\ %
932622 & $12\;10\;00$ & $-52\;26\;31$ & 11.7              & SNR G295.5+09.7 \\
932726 & $15\;53\;00$ & $-56\;10\;00$ & 1.9                & SNR G326.3-01.8\\
932815 & $10\;45\;08$ & $-59\;53\;04$ & 15.1              & Carinae Nebula including OB associations Car OB1 and Car OB2 \\
932817 & $11\;50\;43$ & $-62\;43\;42$ & 8.1                & SNR G296.1-00.5 \\
932821 & $14\;43\;24$ & $-62\;27\;42$ & 21.9       & SNR G315.0-02.3 (Cen X-1)\\
932823 & $16\;14\;22$ & $-60\;52\;07$ & 8.1                & Abell 3627\\
932914 & $11\;23\;32$ & $-68\;24\;56$ & $-$                & false entry in BSC, non-celestial (instrumental artifact)\\
\noalign{\smallskip}\hline
\end{tabular}
\tablefoot{When possible, we have assigned individual objects to the regions. 
The analysis of the point source component and the diffuse component will be 
performed in a subsequent paper. 
In Col. 1 we list the field number, Cols. 2 and 3 give the equatorial coordinates
of the extended region centre, 
Col. 4 the estimated total count rate, and in the last column we provide an 
identification, whenever possible. 
}
\end{table*}

 \begin{figure}[thb]
  \centering
  \resizebox{0.99\hsize}{!}{\includegraphics[clip=]{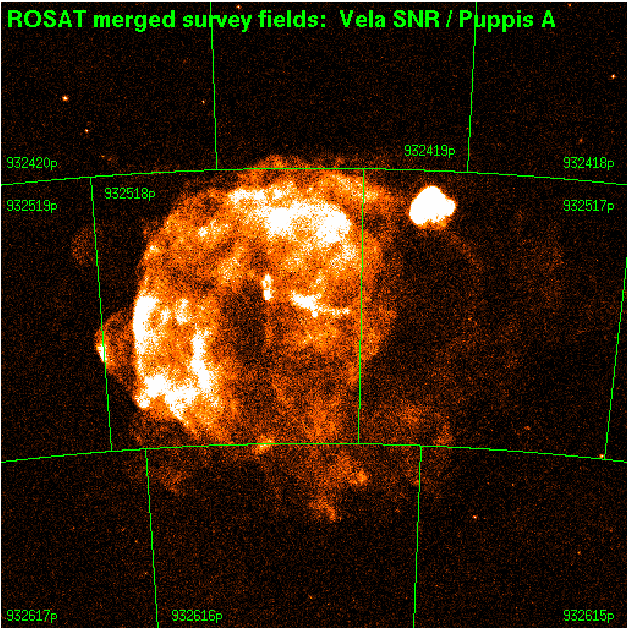}}
  \caption{ROSAT all-sky image of the Vela SNR. In the upper right corner the SNR Puppis A is visible.
We have merged the event files from several sky fields to produce the image. 
The size of the image is 10.5\degr\ $\times$ 10.5\degr. 
Details will be reported in a subsequent paper.  
}              \label{Vela}
  \end{figure}

For the 1RXS BSC catalogue, nine extended sources had been added manually
as listed in Table~\ref{tab:1rxs_flatsch}. No source detection numbers,
detection likelihoods, or (e.g.) count rates are available for them there.
Six of them are very extended supernova remnants, which we have masked
from our standard source detection procedure.
Sco X-1 has never been observed with the ROSAT PSPC: during the survey, the PSPC had been switched off before scanning Sco X-1, and been switched on again afterwards.
In the survey completion phase (Feb.\ 1997) most of this unobserved
region has been filled up with pointed observations;
also a lunar occultation observation of Sco X-1 had been performed (Feb.\ 1998). 
This will be described in \cite{Freyberg2016a}.
The Vela SNR had been manually
included in 1RXS with the coordinates of the Vela pulsar. 
We have detected the Vela pulsar
automatically and included it in 2RXS 
(together with two other sources inside the Vela SNR region).
The last entry in Table~\ref{tab:1rxs_flatsch} is due to single reflections
from Nova Muscae in January 1991, and appeared as an extended emission region
in the survey images. We did not add this to 2RXS.

We have identified 
32 
large extended regions with diffuse 
emission and embedded point sources, as well as large, brighter, and extended 
sources, such as\ clusters of galaxies or supernova remnants. 
These regions are listed in Table~\ref{tab:AppExttab}. 
For three of these extended regions the central source is included in the present 
2RXS catalogue. 
We interactively determined the count rate
of these regions and listed the source identification. 
The point source content of these regions will be published 
as additional survey sources in a separate paper \citep{Freyberg2016b}.

The mask in field 930738 corresponds to SNR G109.1-01.1 (CTB 109), with the central point source being 1E 2225.0+5837.  
For field 931232 the coordinates were obtained from the 2RXS processing 
for source 86
from central sub-image.
In field 931316 two supernova remnants were detected, SNR G189.1+02.9 and SNR IC 433.
Figure~\ref{Vela} depicts the ROSAT all-sky survey image of the Vela SNR. 

\clearpage
\newpage

\end{appendix}

\vfill

\end{document}